\documentclass[a4paper,10pt]{article}
\pdfoutput=1

\usepackage{jheppub} %
\usepackage{bold-extra}
\usepackage{cleveref}

\usepackage{mathtools}
\usepackage{xspace}
\RequirePackage{snapshot}

\usepackage{algpseudocode,algorithm}

\usepackage{amsmath}
\usepackage{etoolbox}

\usepackage[margin=0pt,font+=smaller,labelformat=parens,labelsep=space,skip=6pt,list=false,hypcap=false]{subcaption}

\preprint{MCnet-24-14}

\title{\boldmath KrkNLO matching for colour-singlet processes}

\author[a]{Pratixan Sarmah,}
\author[a]{Andrzej Siódmok,}
\author[a,b]{James Whitehead}

\affiliation[a]{Jagiellonian University, \\
ul.\ prof.\ Stanis\l{}awa \L{}ojasiewicza 11, 30-348 Krak\'{o}w, Poland}
\affiliation[b]{Institute of Nuclear Physics, Polish Academy of Sciences,\\
ul.\ Radzikowskiego 152, 31-342 Krak\'{o}w, Poland}

\emailAdd{pratixan.sarmah@doctoral.uj.edu.pl}
\emailAdd{andrzej.siodmok@uj.edu.pl}
\emailAdd{james.whitehead@uj.edu.pl}

\abstract{Matched calculations combining perturbative QCD with parton showers are an
indispensable tool for LHC physics.
Two methods for NLO matching are in widespread use: \mcatnlo and \powheg.
We describe an alternative, \krknlo, reformulated to be easily applicable to any colour-singlet process.
The primary distinguishing characteristic of \krknlo is its use of an alternative factorisation scheme,
the `Krk' scheme, to achieve NLO accuracy.
We describe the general implementation of KrkNLO in \herwigseven,
using diphoton production as a test process.
We systematically compare its predictions to those produced by \mcatnlo
with several different choices of shower scale,
both truncated to one-emission and with the shower running to completion,
and to \atlas data from LHC Run 2.
}


\DeclareRobustCommand{\ensuremathrm}[1]{\ensuremath{\mathrm{#1}}\xspace}

\newcommand{\order}[1]{\ensuremath{{O}\left(#1\right)}}

\newcommand{\lo}{\mathrm{LO}}
\newcommand{\nlo}{\mathrm{NLO}}

\newcommand{\cut}{\mathrm{cut}}

\newcommand{\dd}{\mathrm{d}}
\newcommand{\dv}[2]{\frac{\dd #1}{\dd #2}}

\newcommand{\qbar}{\bar{q}}

\newcommand{\alphas}{\alpha_\mathrm{s}}

\newcommand{\msbar}{\ensuremath{{\overline{\mathrm{MS}}}}\xspace}
\newcommand{\dis}{\ensuremath{{\mathrm{DIS}}}\xspace}

\newcommand{\FS}{\mathrm{FS}}

\newcommand{\II}{\mathrm{II}}

\newcommand{\rB}{\mathrm{B}}
\newcommand{\rC}{\mathrm{C}}
\newcommand{\rD}{\mathrm{D}}
\newcommand{\rF}{\mathrm{F}}
\newcommand{\rFS}{\mathrm{FS}}
\newcommand{\rI}{\mathrm{I}}
\newcommand{\rK}{\mathrm{K}}

\newcommand{\rP}{\mathrm{P}}
\newcommand{\rR}{\mathrm{R}}
\newcommand{\rPS}{\mathrm{PS}}
\newcommand{\rS}{\mathrm{S}}
\newcommand{\rT}{\mathrm{T}}
\newcommand{\rV}{\mathrm{V}}

\newcommand{\calO}{\mathcal{O}}

\newcommand{\pt}{p_\rT}

\DeclareRobustCommand{\cut}{\ensuremathrm{cut}\xspace}

\newcommand{\muf}{\mu_\rF}
\newcommand{\mur}{\mu_\rR}

\newcommand{\cf}{C_\mathrm{F}}

\newcommand{\tR}{T_\mathrm{R}}
\newcommand{\ca}{C_\mathrm{A}}

\newcommand{\nf}{n_f}

\newcommand{\krk}{\ensuremath{\mathrm{Krk}}\xspace}
\newcommand{\krknlo}{\ensuremath{\mathrm{KrkNLO}}\xspace}
\newcommand{\mcatnlo}{\ensuremath{\mathrm{\textsc{Mc@Nlo}}}\xspace}
\newcommand{\powheg}{\ensuremath{\mathrm{\textsc{Powheg}}}\xspace}

\newcommand{\nnlops}{\ensuremath{\mathrm{\textsc{NNLOPS}}}\xspace}
\newcommand{\unnlops}{\ensuremath{\mathrm{\textsc{UN}^2\textsc{LOPS}}}\xspace}
\newcommand{\geneva}{\ensuremath{\mathrm{\textsc{Geneva}}}\xspace}
\newcommand{\minnlops}{\ensuremath{\mathrm{\textsc{MiNNLO}_{PS}}}\xspace}

\newcommand{\herwig}{\textsf{Herwig}\xspace}
\newcommand{\herwigseven}{\textsf{Herwig 7}\xspace}
\newcommand{\matchbox}{\textsf{Matchbox}\xspace}
\newcommand{\madgraph}{\textsf{MadGraph}\xspace}
\newcommand{\openloops}{\textsf{OpenLoops}\xspace}
\newcommand{\gosam}{\textsf{GoSam}\xspace}

\newcommand{\rivet}{\textsf{Rivet}\xspace}

\newcommand{\vegas}{\textsf{Vegas}\xspace}

\newcommand{\atlas}{\textsc{Atlas}\xspace}

\DeclareRobustCommand{\GeV}{\ensuremathrm{GeV}\xspace}
\DeclareRobustCommand{\TeV}{\ensuremathrm{TeV}\xspace}

\DeclareRobustCommand{\pt}{\ensuremath{p_\rT}\xspace}
\DeclareRobustCommand{\ptcut}{\ensuremath{p_\rT^\cut}\xspace}
\DeclareRobustCommand{\ptof}[1]{\ensuremath{p_{\rT,{#1}}}\xspace}

\DeclareRobustCommand{\ptg}[1]{\ensuremath{p_{\rT}^{\gamma_{#1}}}\xspace}

\DeclareRobustCommand{\ptj}[1]{\ensuremath{p_{\rT}^{j_{#1}}}\xspace}

\DeclareRobustCommand{\Etiso}{\ensuremath{E_{\rT}^{\text{iso}}}\xspace}
\DeclareRobustCommand{\Etisopart}{\ensuremath{E_{\rT}^{\text{iso,part}}}\xspace}

\DeclareRobustCommand{\dR}{\ensuremath{\Delta R}\xspace}

\DeclareRobustCommand{\dRgg}{\ensuremath{\Delta R_{\gamma\gamma}}\xspace}

\DeclareRobustCommand{\absetaj}{\ensuremath{\left| \eta^{j} \right| }\xspace}
\DeclareRobustCommand{\absyg}{\ensuremath{\left| y^{\gamma} \right| }\xspace}

\DeclareRobustCommand{\Mgg}{\ensuremath{M_{\gamma\gamma}}\xspace}

\DeclareRobustCommand{\dphigg}{\ensuremath{\Delta\phi_{\gamma\gamma}}\xspace}

\newcommand{\lnomxoomxplus}[1]{{\left[\frac{\log^{#1} (1-x)}{1-x}\right]_+}}

\makeatletter
\newlength{\negph@wd}
\DeclareRobustCommand{\negphantom}[1]{%
  \ifmmode
    \mathpalette\negph@math{#1}%
  \else
    \negph@do{#1}%
  \fi
}
\newcommand{\negph@math}[2]{\negph@do{$\m@th#1#2$}}
\newcommand{\negph@do}[1]{%
  \settowidth{\negph@wd}{#1}%
  \hspace*{-\negph@wd}%
}

\def\negstrip#1 #2\relax{-#1}

\newcommand*{\negthick}{%
	\mkern-\thickmuskip
}

\makeatother

\begin{document}
\maketitle
\flushbottom

\section{Introduction}
\label{sec:intro}

In the precision era of LHC physics, `matched' calculations
combining next-to-leading-order (NLO) perturbative accuracy with the all-orders logarithmic resummation
provided by parton shower algorithms remain an indispensable
workhorse for LHC phenomenology.
Two methods for matching at NLO, the \mcatnlo \cite{Frixione:2002ik}
and \powheg methods \cite{Nason:2004rx,Frixione:2007vw,Alioli:2010xd},
have been widely used for LHC physics 
and the associated matching uncertainties investigated in detail for a wide variety
of processes and for a number of independent implementations in
\cite{Alioli:2008gx,Alioli:2008tz,Alioli:2009je,Hamilton:2010mb,Platzer:2011bc,Hoeche:2011fd,Nason:2012pr,Frederix:2012dh,Heinrich:2017kxx,Jones:2017giv,Cormier:2018tog,Jager:2020hkz,ATL-PHYS-PUB-2023-029}.

The differences between results generated by alternative matching schemes
may be considered `matching uncertainties', the full general extent of which is unclear, but which comprise formally
higher-order terms
introduced beyond NLO by the choice of matching method.

In order to achieve the goal of a fully-general matching method to combine
parton showers with NNLO fixed-order calculations, it is necessary
to look beyond established matching methods to alternatives
which may more readily admit generalisation beyond NLO.
Early efforts in the direction of general NNLO matching include
\nnlops \cite{Hamilton:2013fea},
\unnlops and similar \cite{Platzer:2012bs,Lonnblad:2012ix,Hoche:2014dla,Bellm:2017ktr},
\geneva \cite{Alioli:2012fc,Alioli:2013hqa,Alioli:2015toa}
and
\minnlops \cite{Monni:2019whf}.
Simultaneously, exploring alternative schemes at NLO allows a better assessment
of the magnitude and significance of matching uncertainties for matched
NLO+PS predictions.

The \krknlo method for parton shower matching \cite{Jadach:2011cr,Jadach:2015mza,Jadach:2016qti}
is one such alternative.
Its key feature is to exploit the general freedom (at NLO and beyond) to choose a PDF factorisation scheme for the calculation.
The factorisation scheme employed (the `\krk scheme' \cite{Jadach:2016acv}) is defined
by the requirement that it remove the double-counting between
real- and shower-emissions which lies at the heart of the NLO matching problem.

The \krknlo method has previously been formulated and implemented for a simplified version of
the Drell-Yan process \cite{Jadach:2015mza}
and for Higgs production via gluon fusion in the infinite-top-mass limit \cite{Jadach:2016qti}.

In this work, we refresh and extend the formulation to the production 
of arbitrary colour-singlet final-states
which proceed via $q\qbar$-annihilation at leading-order.
As an initial test process to study matching uncertainty we consider 
diphoton production, where the absence
of a resonance peak and the characteristic falling spectrum offer
a useful sandbox for the identification of different contributions.

This is the first process calculated independently of the derivation of the \krk
factorisation scheme and is an intermediate step towards a fully-automated calculation
for arbitrary colour-singlet final-states,
which will be the subject of a future publication.

In \cref{sec:nlomatching} we introduce and re-formulate the \krknlo
method, alongside the \mcatnlo method in comparable notation;
in \cref{sec:implementation} we describe its implementation in \herwigseven.

Interested readers who wish to bypass the technicalities of \cref{sec:nlomatching}
on first reading
may prefer to begin with \cref{sec:matchingcomparisons,sec:pheno}.
In \cref{sec:matchingcomparisons} we compare the method with the
\mcatnlo method, both after a single emission
and after the parton shower has been allowed to run to its conclusion.
In \cref{sec:pheno} we compare the alternative methods with
LHC Run II data from \atlas.

\section{NLO matching with KrkNLO}
\label{sec:nlomatching}

The \krknlo method was introduced in 
\cite{Jadach:2015mza,Jadach:2016qti}
following preliminary work in
\cite{Jadach:2011cr,Jadach:2012vs};
the \krk factorisation scheme, which will be introduced below,
was the subject of dedicated discussion in~\cite{Jadach:2016acv,Jadach:2020xfl}.

In \cref{sec:nlomatching_notationdefinitions}
we establish notation for NLO fixed-order calculations, parton showers, and
NLO matching and review the \mcatnlo method in this notation.
In \cref{sec:nlomatching_krknlo} we outline the \krknlo method for a general process
$p p \to X$
with no QCD partons in the final state.
In \cref{sec:nlomatching_krkfs} we motivate and summarise the necessary factorisation scheme 
transformation into the \krk scheme for the NLO matching condition to be satisfied.
Finally in \cref{sec:nlomatching_krkvsothers} we summarise the difference in higher-order terms
between \krknlo and \mcatnlo that can be expected to contribute
to the differences observed in practice in \cref{sec:matchingcomparisons}.

\subsection{Notation and definitions}
\label{sec:nlomatching_notationdefinitions}

\subsubsection*{QCD at NLO}
\label{sec:nlomatching_notationdefinitions_fixedorder}

We write the differential hadronic cross-section for
$2\to m$ process $A + B \to X$, for
an (IRC-safe) observable's measurement function $\mathcal{O}$, 
within the framework of collinear factorisation as
\begin{align}
	\label{eq:nlomatching_notationdefinitions_fixedorder_dsigmaAB}
    \dd {\sigma}_{AB} [\calO]
    (P_1, P_2) = 
    \dd \xi_1 \, \dd \xi_2
    \;
    \sum_{a,b} f_a^A(\xi_1, \muf) \, f_b^B(\xi_2, \muf) \;
    \dd \hat{\sigma}_{ab}[\calO] (\xi_1 P_1, \xi_2 P_2)
\end{align}
for partonic flavours $a,b$,
\msbar parton distribution functions $f_a^A,f_b^B$,
incoming hadronic momenta $P_1, P_2$
and collinear momentum fractions
$\xi_{1,2}$, and
where implicitly we further allow the factorisation scale $\muf$ to be an infrared-safe function of
the phase-space kinematics.
The partonic differential cross-section may be expanded perturbatively
to next-to-leading-order,
\begin{align}
	\label{eq:nlomatching_notationdefinitions_fixedorder_dsigma_expansion}
	\dd \hat{\sigma}_{ab}[\calO] (\xi_1 P_1, \xi_2 P_2)
	=
	\left(\frac{\alphas (\mur)}{2\pi}\right)^k
	\left[
	\dd \hat{\sigma}_{ab}^{(0)}[\calO] (\xi_1 P_1, \xi_2 P_2)
	+
	\left(\frac{\alphas (\mur)}{2\pi}\right)
	\dd \hat{\sigma}_{ab}^{(1)}[\calO] (\xi_1 P_1, \xi_2 P_2; \muf, \mur)
	+
	\order{\alphas^2}
	\right],
\end{align}
where $\mur$ is the renormalisation scale and for the processes we consider subsequently, $k=0$.

The leading-order contribution 
$\dd \hat{\sigma}_{ab}^{(0)}$
is given by
\begin{align}
    \dd \hat{\sigma}_{ab}^{(0)}[\calO] (\xi_1 P_1, \xi_2 P_2)
    &{} =
    \dd \Phi_m (\xi_1 P_1, \xi_2 P_2) \;
    \frac{1}{2\hat{s}_{12}} \;
    \rB_{ab} (\Phi_m) \;
    \Theta_{\cut} \left[\Phi_m\right] \;
    \calO(\Phi_m)
\end{align}
where $\dd \Phi_m$ denotes the $m$-particle (Born) Lorentz-invariant phase-space measure,
$\rB_{ab}$ the Born matrix-element for the partonic channel $ab$,
and $\Theta_{\cut} \left[\Phi_m\right]$
a theta-function imposing `generator' cuts on the Born phase-space;
fiducial cuts where different are included within the definition of $\calO$.%
\footnote{In particular, where the Born phase-space contains limits
in which the Born matrix-element diverges, $\Theta_{\cut}$
must
be chosen to remove them by, at the very least,
imposing a technical cut vetoing the singular regions.
}
The partonic flux factor $1/(2\hat{s}_{12})$ contains the momentum invariant between the incoming partons,
\begin{align}
	\hat{s}_{12} = (p_1 + p_2)^2 = 2 p_1 \cdot p_2 
\end{align}
For brevity we suppress the arguments of $\dd \Phi_m$ where unambiguous.

The next-to-leading-order (NLO) contribution may be written using the subtraction formalism
as
\begin{align}
    \notag
    \dd \hat{\sigma}_{ab}^{(1)}[\calO] (\xi_1 P_1, \xi_2 P_2)
    &{} =
    \dd \Phi_m \;
    \frac{1}{2\hat{s}_{12}} \;
    \Biggl[ \rV_{ab} (\Phi_m) 
    + \sum_{(\alpha)} \rB_{ab}^{(\alpha)} (\Phi_m)
    	\int \dd \Phi_{+1}^{(\alpha)} \; x \, \rS^{(\alpha)} (\Phi_{+1}^{(\alpha)})
    \Biggr] \;
    \Theta_{\cut} \left[\Phi_m\right] \;
    \calO(\Phi_m) \\ \notag
    &{} + \dd \eta_1 \, \dd \eta_2 \;
    \dd \Phi_m(\eta_1 \xi_1 P_1, \eta_2 \xi_2 P_2) \;
    \frac{1}{2\hat{s}_{12}} \;
    \rC^{\rFS}_{ab}(\eta_1, \eta_2; \muf; \Phi_m) \;
    \Theta_{\cut} \left[\Phi_m\right] \;
    \calO(\Phi_m) \\
    &{} + 
    \dd \Phi_{m+1} \;
    \frac{1}{2\hat{s}_{12}} \;
    \Biggl[ \rR_{ab} (\Phi_{m+1}) \;
    \Theta_{\cut} \left[\Phi_{m+1}\right] \;
    \calO(\Phi_{m+1})
    \\ \notag
    &{} \hphantom{{} + \Phi_{m+1} \; \frac{1}{2\hat{s}_{12}} \; \Biggl[}
    - \sum_{(\alpha)} \rS^{(\alpha)} (\tilde{\Phi}_{+1}^{(\alpha)}) \; \rB^{(\alpha)} (\tilde{\Phi}_m^{(\alpha)}) \;
    \Theta_{\cut} \left[ \tilde{\Phi}_m^{(\alpha)} \right] \;
    \calO(\tilde{\Phi}_m^{(\alpha)})
    \Biggr],
\end{align}
for virtual matrix-elements $\rV$, real matrix-elements $\rR$,
mass-factorisation counterterms $\rC^{\rFS}$,
subtraction kernels $\rS^{(\alpha)}$
with corresponding phase-space mappings $\tilde{\Phi}_m^{(\alpha)}(\Phi_{m+1})$
and
$\tilde{\Phi}_{+1}^{(\alpha)}(\Phi_{m+1})$
and reduced matrix-elements $\rB^{(\alpha)}$,
all indexed by generalised index $(\alpha)$.
We suppress an implicit summation over colour- and spin-indices
between the subtraction kernels $\rS^{(\alpha)}$
and the corresponding reduced-matrix-elements $\rB^{(\alpha)}$.

Here we adopt the momentum mappings and subtraction terms of 
Catani--Seymour dipole subtraction \cite{Catani:1996vz}.
For our purposes, initial-initial dipoles suffice, so 
we employ the notation%
\footnote{The corresponding mapping in which the second
incoming parton is the emitter, $\II_2$,
may be obtained by relabelling.}
\begin{align}
\begin{rcases*}
p_1 = \xi_1 {P}_1 & emitter \\
p_2 = \xi_2 {P}_2 & spectator \\
q & emission \\
\negphantom{\bigl\{}\bigl\{ {q}_i \bigl\}_{i=1}^{m} & FS particles
\end{rcases*} = \Phi_{m+1}(p_1,p_2)
\longmapsto
\tilde{\Phi}^{\II_1}_{m} (\Phi_{m+1}) = 
\begin{cases*}
{p}^{\II_1}_1 = x \, p_1 = x \, \xi_1 {P}_1 & \\
{p}^{\II_1}_2 = p_2 = \xi_2 {P}_2 & \\
{q}^{\II_1}_i = \Lambda({p}_1+{p}_2-q, {p}^{\II_1}_1+{p}^{\II_1}_2) \, {q}_i & \\
\end{cases*}
\label{eq:nlomatching_notationdefinitions_IIdipolemapping_sub}
\end{align}
where 
\begin{align}
 x = x^{\II_1}(\Phi_{m+1}) = 1 - \frac{q \cdot (p_{1} + p_{2})}{p_{1} \cdot p_{2}}
\equiv x^{\II_2}(\Phi_{m+1})
\end{align}
is the collinear momentum fraction of the splitting
and $\Lambda$ a Lorentz transformation restoring overall momentum conservation
by mapping ${p}_1+{p}_2-q$ onto ${p}^{\II_1}_1+{p}^{\II_1}_2$.
We require only 
$(\alpha) \in \{ (q_{a}g_{i},b), (g_{a} q_{i},b) \}$ in the notation of \cite{Catani:1996vz}, with
\begin{align}
	\rS^{(\alpha)} ( x ) & {} =
	\frac{8 \pi}{x s_{ai}}
	\rD^{(\alpha)} (x)
\end{align}
where the momentum invariant $s_{ai}$ is between the
emitter and the emission,
and
\begin{align}
	\rD^{(q_{a}g_{i},b)} (x) & {} =
	\cf \frac{1+x^2}{1-x}
	&
	\rD^{(g_aq_i,b)} (x) & {} =
	\tR \left(x^2 + (1-x)^2\right).
	\label{eq:nlomatching_notationdefinitions_dipolefunctions_sub}
\end{align}
We omit the spectator from the label $(\alpha)$ where unambiguous.
We further write the integrated dipoles 
\begin{align}
	\rB^{(\alpha)} (\Phi_m) \int \dd \Phi_{+1}^{(\alpha)} \; x \, \rS^{(\alpha)} (\Phi_{+1}^{(\alpha)})
\end{align}
as the sum of a fully-integrated part in the Born phase-space $\rI^{(\alpha)}(\Phi_m)$
and a collinear contribution which we write schematically as $\rK^{(\alpha)} (x) \, \rB^{(\alpha)} (\Phi_m)$ and
combine with the mass-factorisation counterterms
$\rC$, which comprise the DGLAP splitting kernels.

For colour-singlet final-states, we may write the simplified expression for each $q\qbar$ channel as
\begin{align}
	\label{eq:nlomatching_notationdefinitions_qcdatnlo_dsigmaqqb_nlo_1_singlet}
    \dd \hat{\sigma}_{q\qbar}^{(1)}[\calO] (\xi_1 P_1, \xi_2 P_2)
    &{} =
    \dd \Phi_m \;
    \frac{1}{2\hat{s}_{12}} \;
    \Biggl[ \rV_{q\qbar} (\Phi_m) 
    + \rI_{q\qbar} (\Phi_m)
    \Biggr] \;
    \calO(\Phi_m) \\ \notag
    {} + \dd \eta_1 \, \dd \eta_2 \; &
    \dd \Phi_m(\eta_1 \xi_1 P_1, \eta_2 \xi_2 P_2) \;
    \frac{1}{2\hat{s}_{12}} \;
    \Bigl[
    \left(\rP(\muf) +  \rK^{\rFS}\right)_{qq} \negthick (\eta_1) \; \rB_{q\qbar}(\Phi_m) \, \delta^1_{\eta_2}
    \\ \notag
    & \hphantom{\dd \Phi_m(\eta_1 \xi_1 P_1, \eta_2 \xi_2 P_2) \;
    	\frac{1}{2\hat{s}_{12}} \;
    	\Bigl[}
     + \delta^1_{\eta_1} \, \rB_{q\qbar}(\Phi_m) \left(\rP(\muf) +  \rK^\rFS\right)_{qq} \negthick (\eta_2)
     \Bigr]
    \calO(\Phi_m) \\ 
    \notag
    {} + 
    \dd \Phi_{m+1} \; &
    \frac{1}{2\hat{s}_{12}} \;
    \Biggl[ \rR_{q\qbar} (\Phi_{m+1}) \;
    \calO(\Phi_{m+1})
    \\ \notag
    &{}
    - \frac{8 \pi}{x} \, \rD^{(qg)}(x) \, \left[ 
    \frac{1}{s_{qg}} 
    \rB_{q\qbar}(\tilde{\Phi}^{\II_1}_m) \;
    \calO(\tilde{\Phi}_m^{\II_1})
     + \frac{1}{s_{\qbar g}} 
     \rB_{q\qbar}(\tilde{\Phi}^{\II_2}_m)  \,
     \calO(\tilde{\Phi}_m^{\II_2}) \right]
    \Biggr],
\end{align}
dropping the $\Theta_{\cut}$ factors for brevity,
and writing $\delta^1_{\eta_i}$ as a shorthand for the Dirac delta function $\delta(1-\eta_i)$.
Explicit expressions for the $\rP$ and $\rK$ contributions are given in \cref{sec:appendix_cssummary}.
Likewise for each $qg$-type channel,
\begin{align}
	\label{eq:nlomatching_notationdefinitions_qcdatnlo_dsigmaqg_nlo_1_singlet}
    \dd \hat{\sigma}_{qg}^{(1)}[\calO] (\xi_1 P_1, \xi_2 P_2)
    &{} =
    \dd \eta_1 \, \dd \eta_2 \;
    \dd \Phi_m(\eta_1 \xi_1 P_1, \eta_2 \xi_2 P_2) \;
    \frac{1}{2\hat{s}_{12}} \;
    \delta^1_{\eta_1} \,
    \rB_{q\qbar}(\Phi_m) \left(\rP(\muf)	 +  \rK^{\rFS}\right)_{gq} \negthick (\eta_2)  \,
    \calO(\Phi_m)  \notag \\
    &{} + 
    \dd \Phi_{m+1} \;
    \frac{1}{2\hat{s}_{12}} \;
    \Biggl[ \rR_{qg} (\Phi_{m+1}) \;
    \calO(\Phi_{m+1})
    -
    \rS^{(gq)}(x) \, 
    \rB_{q\qbar}(\tilde{\Phi}^{\II_2}_m)  \,
    \calO(\tilde{\Phi}_m^{\II_2})
    \Biggr],
\end{align}
and similarly for $\dd \hat{\sigma}_{g\overline{q}}^{(1)}$.

\subsubsection*{Parton showers}
\label{sec:nlomatching_notationdefinitions_partonshower}

A parton shower 
typically uses the veto algorithm
\cite{Seymour:1994df,Sjostrand:2006za,Platzer:2011dq,Lonnblad:2012hz,Kleiss:2016esx}
to sample successive splitting scales from the
so-called Sudakov distribution characterising the probability
of generating a splitting of type
$i$ at a scale $t \in (t_0,t_1)$,
\begin{align}
    p_i(t; t_0, t_1) =  
    \Theta\left[ t_0 < t < t_1 \right] 
        \; P_i(t) \;
        \Delta\bigr\vert_{t}^{t_1},
\label{eq:nlomatching_notationdefinitions_Suddist}
\end{align}
where
$P_i(t)$ represents the conditional probability density that a splitting of type $i$ occurs at scale $t$ given that no splitting has occurred at a higher scale,
and 
\begin{align}
	\label{eq:nlomatching_notationdefinitions_SudFF}
	\Delta\bigr\vert_{t}^{t_1} = \prod_i \Delta_{i}\bigr\vert_{t}^{t_1}
\end{align}
represents the Sudakov factor describing the `survival' probability that no splitting
is generated between scales $t$ and $t_1$, i.e.
concretely
\begin{align}
	p_\varnothing(t; t_0, t_1) = \delta(t-0) \; \Delta\bigr\vert_{t_0}^{t_1}.
\end{align}
Solving the resulting differential equations,
\begin{align}
	\dv{}{t} \, \Delta_i\bigr\vert_{t}^{t_1}
	=
	\Delta_i\bigr\vert_{t}^{t_1} P_i(t),
\end{align}
for a single splitting, and
\begin{align}
	\dv{}{t} \, \Delta\bigr\vert_{t}^{t_1} = \sum_i p_i(t;t_0,t_1) = \Delta\bigr\vert_{t}^{t_1} \sum_i P_i(t),
\end{align}
for all spittings,
in the possible-splitting region, gives
the closed-form expressions for the Sudakov factor associated with each splitting,
\begin{align}
	\label{eq:nlomatching_notationdefinitions_partonshower_sudexp}
    \Delta_i \bigr\vert_{t}^{t_1}
    & {} =
    \exp \left[
    - \int_t^{t_1} \dd t' \; P_i(t')
    \right],
\end{align}
and that associated with all splittings,
\begin{align}
	\label{eq:nlomatching_notationdefinitions_SudFF_closedform}
	\Delta\bigr\vert_{t}^{t_1} = \prod_i \Delta_{i}\bigr\vert_{t}^{t_1}
	= 
	\exp \left[
	- \int_t^{t_1} \dd t' \; \sum_i P_i(t')
	\right].
\end{align}

For each scale sampled from this distribution the shower generates
a higher-multiplicity phase-space configuration
in which an additional resolved emission has the generated scale, with kinematics distributed
in phase-space according to some kernel related to the density by
\begin{align}
	\label{eq:nlomatching_notationdefinitions_partonshower_showerkernel}
    P_i(t) = \int \dd \Phi_{+1} \; \frac{\alphas(\mu(\Phi_{+1}))}{2\pi} P_i(\Phi_{+1})
    \;
    \delta(t - t(\Phi_{+1})).
\end{align}

In practice, the possible splittings and splitting phase-space available to be considered for each
splitting depend further on the initial phase-space configuration, allowing us to express the action of
the parton shower formally as an iterative operator,
\begin{align}
	\label{eq:nlomatching_notationdefinitions_partonshower_psoperator}
    \rPS \left[ \mathcal{O} \right] \left(\Phi_m; \boldsymbol{\xi} \right)
    = {}
    & \Delta \bigr\vert_{t_0}^{t_1(\Phi_m)}(\Phi_m; \boldsymbol{\xi}) \;
    \calO (\Phi_m)
    \\ \notag {} + {}
    \sum_{(\alpha) \in \Phi_m}
    & \dd \Phi_{+1}^{(\alpha)} \;
    \Theta \Bigl[ t_0 < t(\Phi_{m+1}^{(\alpha)}) < t_1(\Phi_m) \Bigr]
    \;
    \left(
    \frac{\alphas(\mu^{(\alpha)}(\Phi_{+1}^{(\alpha)}))}{2\pi}
    \; P_m^{(\alpha)} (\Phi_{+1}^{(\alpha)}; \boldsymbol{\xi})
    \right)
    \\
    & \qquad \times 
    \;
    \Delta \bigr\vert_{t(\Phi_{m+1}^{(\alpha)})}^{t_1(\Phi_m)} (\Phi_m; \boldsymbol{\xi})
    \;
    \rPS\left[ \calO \right] (\Phi^{(\alpha)}_{m+1}; \boldsymbol{\xi}^{(\alpha)}),
\end{align}
where we replace index $i$ with 
a generalised index $(\alpha)$ to represent the possible splittings
from momentum configuration $\Phi_m$,
$m$ to indicate the final-state multiplicity,
and $\boldsymbol{\xi} = (\xi_1, \xi_2)$ denotes
the incoming partonic momentum-fractions.
For brevity we will write
\begin{align}
	\Theta_{t_0}^{t_1(\Phi_m)} 
	\equiv
	\Theta \Bigl[ &t_0 < t(\Phi_{m+1}) < t_1(\Phi_m) \Bigr]
\end{align}
when the argument of the $\Theta$-function is unambiguous.
Assembling the components, the Sudakov factor $\Delta \bigr\vert_{t_0}^{t_1(\Phi_m)}(\Phi_m; \boldsymbol{\xi})$
may be expanded perturbatively as
\begin{align}
	\Delta \bigr\vert_{t_0}^{t_1(\Phi_m)}(\Phi_m; \boldsymbol{\xi})
	& {} \equiv
	\prod_{(\alpha) \in \Phi_m}
	\Delta_{(\alpha)} \bigr\vert_{t_0}^{t_1(\Phi_m)}(\Phi_m; \boldsymbol{\xi})
	\\ \label{eq:nlomatching_notationdefinitions_partonshower_sudakovnloexpansion}
	& {} = 1 
	- 
	\frac{\alphas}{2\pi}
	\sum_{(\alpha) \in \Phi_m}
        \int
	\dd \Phi_{+1}^{(\alpha)} \;
	\Theta_{t_0}^{t_1(\Phi_m)} 
	\; P_m^{(\alpha)} (\Phi_{+1}^{(\alpha)}; \boldsymbol{\xi}^{(\alpha)})
	+ \order{\alphas^2}.
\end{align}
For brevity one or both of the phase-space
and partonic momentum-fraction
arguments of $\Delta$ will typically be omitted where unambiguous.
In full generality the cut-off $t_0$ may be a function of the underlying phase-space,
$t_0(\Phi_m)$; here we restrict to the case where it is a constant.%
\footnote{In this context, the shower cut-off is typically tuned alongside
	hadronisation model parameters to give good agreement with data.
	Although we will not discuss hadronisation, this fixed shower cut-off is
	in practice used  as a scale governing the transition from the parton
	shower model to the non-perturbative hadronisation model.}

The $\rPS$ operator acts as a functional on the measurement function of IRC-safe observable $\calO$ and represents
a single iteration of the shower algorithm: 
either a splitting is selected and performed (the second term),
or no resolvable emission is generated above the cut-off scale, in which case the observable is evaluated on the input phase-space configuration and the iteration is terminated (the first term).

The case when $\calO$ consists only of a product of theta-functions and $\calO (\Phi_m) \equiv \calO(\Phi_n)$ for all $n \geqslant m$
corresponds to a calculation of the inclusive cross-section, by unitarity,
`integrating out' all possible shower emissions.
Unitarity can be verified at NLO by comparing the first-order expansion of the Sudakov factor
\cref{eq:nlomatching_notationdefinitions_partonshower_sudakovnloexpansion}
to the corresponding differential contributions in
\cref{eq:nlomatching_notationdefinitions_partonshower_psoperator}.

The choices of $t(\Phi_m)$, $P_m^{(\alpha)}$ and $\Phi_{+1}^{(\alpha)}$ define the parton shower algorithm.
We will focus upon the (Catani--Seymour) dipole shower 
\cite{Nagy:2005aa,Nagy:2006kb,Dinsdale:2007mf,Schumann:2007mg,Platzer:2009jq}
and concretely, its implementation in \herwigseven \cite{Platzer:2011bc,Bellm:2015jjp,Bewick:2023tfi}%
, in which the
ordering variable $t$ is the transverse momentum of the emitted
parton relative to the emitter-spectator pair in its rest-frame,
the splitting densities
$P_m^{(\alpha)}$
are chosen to be Catani--Seymour dipole functions \cite{Catani:1996vz,Catani:2002hc},
and the phase-space factorisation implicit in \cref{eq:nlomatching_notationdefinitions_partonshower_psoperator}
is provided by the corresponding 
Catani--Seymour momentum mappings.

The dipole functions relevant to the NLO matching of 
processes considered in the present work are
the same polarisation-averaged initial--initial splitting dipoles
used for dipole subtraction in \cref{eq:nlomatching_notationdefinitions_dipolefunctions_sub}
\cite{Schumann:2007mg}.
The initial--initial massless momentum mapping is given, without loss of generality labelling $p_1$ as the emitter, by
\begin{align}
\begin{rcases*}
p_1 = \xi_1 {P}_1 & emitter \\
p_2 = \xi_2 {P}_2 & spectator \\
\negphantom{\bigl\{}\bigl\{ {q}_i \bigl\}_{i=1}^m & FS
\end{rcases*} = \Phi_m(p_1,p_2)
\xmapsto{\Phi_{+1}}
{\Phi}^{\II_1}_{m+1} = 
\begin{cases*}
{p}^{\II_1}_1 = x^{-1} p_1 = x^{-1} \xi_1 {P}_1 & \\
{p}^{\II_1}_2 = p_2 = \xi_2 {P}_2 & \\
{q}^{\II_1}_i = \Lambda({p}_1+{p}_2, {p}^{\II_1}_1+{p}^{\II_1}_2-{q}^{\II_1}_{m+1}) \, {q}_i & \\
{q}^{\II_1}_{m+1} = (1-x-v) {p}^{\II_1}_1 + v {p}_2 + k_\rT
\end{cases*}
\label{eq:nlomatching_notationdefinitions_IIdipolemapping_split}
\end{align}
where $\Lambda$ is a Lorentz transformation restoring overall momentum conservation.
This is the inverse mapping of \cref{eq:nlomatching_notationdefinitions_IIdipolemapping_sub} and we will
implicitly use that
\begin{align}
    \tilde{\Phi}^{(\alpha)}_m( \Phi_{m+1}^{(\alpha)}(\Phi_m,\Phi_{+1})) & {} = \Phi_m,
    &
    \tilde{\Phi}^{(\alpha)}_{+1}( \Phi_{m+1}^{(\alpha)}(\Phi_m,\Phi_{+1})) & {} = \Phi_{+1},
\end{align}
and $x( \Phi_{m+1}^{(\alpha)}(\Phi_m,\Phi_{+1})) = x$.

The transverse momentum of the generated splitting in
\cref{eq:nlomatching_notationdefinitions_IIdipolemapping_split} is governed by
$k_\rT$, a massless four-vector generated with uniformly-distributed azimuth around the emitter-spectator axis
in the rest frame of the emitter-spectator pair, with spatial magnitude given by
\begin{align}
    \lVert \mathbf{k}_\rT \rVert^2 = \frac{1}{x} v \, (1 - x - v) \; \hat{s}_{12} .
\end{align}
We will denote the transverse momentum of the $i$\textsuperscript{th} generated splitting,
with respect to its emitter-spectator pair in the dipole rest frame, as $\ptof{i}$.

The three random variables which parametrise the three additional dimensions
required to describe an $({m+1})$-particle phase-space from an $m$-particle phase-space
are therefore $x, v$ and $\phi$, and give rise to a convolution over the collinear momentum fraction $x$,
\begin{align}
    \dd \Phi_{m+1} (p_1, p_2)
    =
    \int_0^1
    \dd x'
    \ 
    \dd \Phi_m ( x' p_1, p_2 )
    \; \dd \Phi_{+1} (x'; p_1, p_2)
\end{align}
where the radiation phase-space $\dd \Phi_{+1}$ is
\begin{align}
    \dd \Phi_{+1} (x'; p_1, p_2)
    =
    \frac{2 p_1 \cdot p_2}{16 \pi^2} \; \dd x \, \delta(x-x') \; \dd v \; \frac{\dd \phi}{2\pi}
    \;
    \Theta\left[ 0 < x < 1 \right]
    \,
    \Theta\left[ 0 < v < 1-x \right].
\end{align}
Further details may be found in \cite{Dinsdale:2007mf,Schumann:2007mg,Platzer:2010ppa}.

For initial--initial splittings the shower proceeds from the hard process
by `backwards evolution' \cite{Sjostrand:1985xi}, identifying in each splitting a higher-scale parton within
the parent hadron. 
Accordingly, the shower densities depend on both the splitting dipoles
and upon the relevant PDFs, evaluated at their respective momentum fractions:
\begin{align}
	\label{eq:nlomatching_notationdefinitions_partonshower_Pmab}
    P_m^{(ab)} (\Phi_{+1}; \xi_1, \xi_2) = 
    \frac{f_{a} (\frac{\xi_1}{x} , \mu^{(ab)})}{f_{b} (\xi_1, \mu^{(ab)}) }
    \;
    \rS^{ab} (x) 
.
\end{align}

For a $p p \to X$
process whose only Born partonic subprocess 
is $q \qbar \to X$,
the
`one-emission' expansion of the shower applied to the Born matrix-element $\rB_{q\qbar}$
is therefore%
\footnote{Note that these are not strictly `purely'-partonic differential cross-sections, despite the $\dd \hat{\sigma}$ notation,
as the initial-state backward-evolution of the parton-shower induces
an explicit dependence on the partonic momentum fractions through
the ratios of PDFs within the splitting densities.
}
{ \footnotesize
\begin{align}
\dd \hat{\sigma}^{\lo+\rPS_1}_{q\qbar} [\calO] (\boldsymbol{\xi}) = {} &
    \dd \Phi_m \; 
    \frac{1}{2\hat{s}_{12}} \rB_{q \qbar}(\Phi_m) \Biggl\{
    \Delta \bigr\vert_{\ptcut}^{Q(\Phi_m)} \;
    \calO (\Phi_m)
    \\ \notag {} +
    \dd \Phi_{+1}^{\II_1} \;
    \Theta \bigl[ \ptcut < \ptof{1} & < Q(\Phi_m) \bigr]
    \; \Delta \bigr\vert_{\ptof{1}}^{Q(\Phi_m)} 
    (\boldsymbol{\xi}) \;
    \Delta \bigr\vert_{\ptcut}^{\ptof{1}} 
    \left(\Phi^{\II_1}_{m+1}; \frac{\xi_1}{x}, \xi_2\right)
    \left[ 
    \frac{\alphas(\mu^{\II_1})}{2\pi}
    \sum_{a \in \{q,g\}}
    \frac{f_{a} (\frac{\xi_1}{x} , \mu^{\II_1})}{f_{q} (\xi_1, \mu^{\II_1}) }
    \;
    \rS^{qa} (x) \;
    \calO  (\Phi^{\II_1}_{m+1}) \right]
    \\ \notag {} +
    \dd \Phi_{+1}^{\II_2} \;
    \Theta \bigl[ \ptcut < \ptof{1} & < Q(\Phi_m) \bigr]
    \; \Delta \bigr\vert_{\ptof{1}}^{Q(\Phi_m)}
        (\boldsymbol{\xi}) \;
    \Delta \bigr\vert_{\ptcut}^{\ptof{1}}
    \left(\Phi^{\II_2}_{m+1}; \xi_1, \frac{\xi_2}{x} \right)
    \left[ 
    \frac{\alphas(\mu^{\II_2})}{2\pi}
    \sum_{b\negphantom{b}\hphantom{a}\in \{q,g\}}
    \frac{f_{b\negphantom{b}\hphantom{a}} (\frac{\xi_2}{x} , \mu^{\II_2})}{f_{\qbar} (\xi_2, \mu^{\II_2}) }
    \;
    \rS^{qb\negphantom{b}\hphantom{a}} (x) \;
    \calO  (\Phi^{\II_2}_{m+1}) \right]
    \Biggr\},
\end{align}
}
where
$\ptcut$ is the shower cut-off scale,
$\ptof{1}$ the transverse momentum of the generated emission,
and $Q(\Phi_m)$ denotes the shower starting-scale.
Note that if the shower algorithm is allowed to proceed,
subsequent emissions are only generated at higher orders of $\alphas$,
so this is also the perturbative expansion of the full shower to NLO.

For consistency with subsequent sections, we redistribute these terms
according to their initial-state partonic channel after the shower emission,
and expand in $\alphas$:%
\footnote{The $\order{\alphas^0}$ contribution to the Sudakov factor,
	$
	\Delta^{(0)} \bigr\vert_{t_0}^{t_1} (\Phi; \boldsymbol{\xi}),
	$ 
	is retained to allow the convenient deduction of contributions 
	arising at higher orders, but as implied by 
	\cref{eq:nlomatching_notationdefinitions_partonshower_sudexp}
	is exactly 1.}
{ \footnotesize
\begin{align}
	\dd \hat{\sigma}^{\lo+\rPS_1(0)}_{q\qbar} [\calO] (\boldsymbol{\xi}) = {} &
	\dd \Phi_m \; 
	\frac{1}{2\hat{s}_{12}} \;
	\Biggl[
	\rB_{q \qbar}(\Phi_m) 
	\Biggr]
	\Theta_{\cut} \left[\Phi_m\right] \;
	\Delta^{(0)} \bigr\vert_{\ptcut}^{Q(\Phi_m)} \;
	\calO (\Phi_m)
	\\
	\dd \hat{\sigma}^{\lo+\rPS_1(1)}_{q\qbar} [\calO]  (\boldsymbol{\xi}) = {} &
	\dd \Phi_m \; 
	\frac{1}{2\hat{s}_{12}} \;
	\Biggl[
	\rB_{q \qbar}(\Phi_m) 
	\Biggr]
	\Theta_{\cut} \left[\Phi_m\right] \;
	\Delta^{(1)} \bigr\vert_{\ptcut}^{Q(\Phi_m)} \;
	\calO (\Phi_m)
	\\ \notag {} + {}
	\dd \Phi_{m+1} 
	\frac{1}{2 \hat{s}_{12}} \Biggl[
	& 
	\sum_{i=1}^2
	\rB_{q \qbar}(\tilde{\Phi}^{\II_i}_m) \;
	\Theta_{\cut} \left[\tilde{\Phi}_m^{\II_i}\right] \;
	\Theta_{\ptcut}^{Q(\tilde{\Phi}_m^{\II_i})} (\tilde{\Phi}_m^{\II_i})
	\;
	\rS^{q_ig} (x) \;
	\Delta^{(0)} \bigr\vert_{\ptof{1}}^{Q(\tilde{\Phi}_m^{\II_i})}
	( \tilde{\Phi}^{\II_i}_m; \boldsymbol{\xi}^{\II_i} )
	\Biggr]
	\Delta^{(0)} \bigr\vert_{\ptcut}^{\ptof{1}}  %
	\;
	\calO  (\Phi_{m+1})
	\\
	\dd \hat{\sigma}^{\lo+\rPS_1(1)}_{qg} [\calO] (\boldsymbol{\xi}) = {} &
	\dd \Phi_{m+1} \; 
	\frac{1}{2 \hat{s}_{12}} \;
	\rB_{q \qbar}(\tilde{\Phi}^{\II_2}_m) \;
	\Theta_{\cut} \left[\tilde{\Phi}^{\II_2}_m\right] \;
	\\ \notag {}
	& 
	\hphantom{
		\dd \Phi_{m+1} \; 
		\frac{1}{2 \hat{s}_{12}} \;}
	\times
	\Theta_{\ptcut}^{Q(\tilde{\Phi}_m^{\II_2})} \;
	\rS^{gq} (x) \;
	\Delta^{(0)} \bigr\vert_{\ptof{1}}^{Q(\tilde{\Phi}^{\II_2}_m)}
	( \tilde{\Phi}^{\II_2}_m; \xi_1, x \xi_2 )
	\;
	\Delta^{(0)} \bigr\vert_{\ptcut}^{\ptof{1}} \;
	\calO  (\Phi_{m+1})
\end{align}

}

Here and in subsequent similar expressions, we omit factors of the form 
\begin{align}
	\label{eq:nlomatching_notationdefinitions_PDFratio_order_alphas2}
	\frac{f_{q}(\xi_1 ,  {\muf}(\tilde{\Phi}_m^{\II_1}))}{f_{q}(\xi_1, \mu^{\II_1})}
	\frac{f_{q}(\frac{\xi_1}{x}, \mu^{\II_1})}{f_{q}(\frac{\xi_1}{x}, \muf({\Phi}_{m+1}))}
	\frac{f_{\qbar}(\xi_2, {\muf}(\tilde{\Phi}_m^{\II_1}))}{f_{\qbar}(\xi_2, \muf(\Phi_{m+1}))}
	= 1 + \order{\alphas}
\end{align}
or similar, arising from the initial generation of the Born process
and the PDF ratio within the shower emission \cref{eq:nlomatching_notationdefinitions_partonshower_Pmab}
and 
\begin{align}
	\label{eq:nlomatching_notationdefinitions_alphas_order_alphas2}
	\frac{\alphas(\mu_1)}{\alphas(\mu_2)}
	=
	1 
	+ \order{\alphas}
\end{align}
arising from the difference in strong-coupling scale between the expansion convention of
\cref{eq:nlomatching_notationdefinitions_fixedorder_dsigma_expansion}
and that chosen for the shower kernels \cref{eq:nlomatching_notationdefinitions_partonshower_showerkernel}.

\subsubsection*{NLO matching%
	\footnote{This section is based on ongoing work in collaboration with Simon Pl\"atzer \cite{matchboxmatchingtoappear}
	and may be read in companion with the corresponding section
	of \cite{herwigmanualtoappear}.}
}
\label{sec:nlomatching_notationdefinitions_nlomatching}

The objective of NLO matching is to retain perturbative accuracy but combine it with the logarithmic resummation of a parton shower.
This problem has previously been solved in general by the \mcatnlo \cite{Frixione:2002ik} and \powheg \cite{Nason:2004rx,Frixione:2007vw,Alioli:2010xd} methods.

Concretely, we require that for an infrared-safe observable $\mathcal{O}$ and a suitable choice of generation cuts,
matched predictions for its expectation value
differ from their NLO fixed-order counterpart by terms that are higher-order in $\alphas$,
or by terms which give power-corrections (i.e.\ non-logarithmic) in the cut-off scale upon integration.
That is, for the expectation-value of $\calO$ calculated via some matching scheme, 
we adopt as an `NLO matching condition' that
\begin{align}
    \sigma^{\nlo+\rPS(0)} [\calO] (P_1, P_2) &= \sigma^{(0)} [\calO] (P_1, P_2)
    \\ 
    \label{eq:nlo_match_1}
    \sigma^{\nlo+\rPS(1)} [\calO] (P_1, P_2) &= \sigma^{(1)} [\calO] (P_1, P_2)
    + \order{\left(\frac{\ptcut}{\sqrt{s}}\right)^2}
\end{align}
where the expectation value for $\calO$ implied by a given differential cross-section is
\begin{align}
    \langle \calO \rangle \equiv \sigma[\calO] (P_1, P_2) \equiv \int \dd {\sigma}[\calO] (P_1, P_2).
\end{align}

The \mcatnlo method (`subtractive matching') uses the shower approximation to the real matrix-elements itself
as the subtraction term, except below a cutoff
where it switches to the exact dipole subtraction to ensure
exact cancellation of the infrared divergences of the real-emission
matrix-element in its singular limits.
In our case, using dipole subtraction and the dipole shower,
and for colour-singlet processes,
the contribution generated by the shower coincides with the subtraction term.

The subtracted real events generated from the $\Phi_{m+1}$ phase-space (`H-events')
are used as an initial condition for the shower evolution separately from the 
combined Born, virtual, and integrated subtraction-term events generated from the
$\Phi_m$ phase-space (`S-events'), i.e.:

{ \footnotesize
	\begin{align}
		\label{eq:nlomatching_notationdefinitions_nlomatching_dsigmaab_mcatnlo_0}
		\dd \hat{\sigma}_{ab}^{\mcatnlo} & [\calO] (\boldsymbol{\xi}) %
		{} =
		\dd \Phi_m \;
		\frac{1}{2\hat{s}_{12}} \;
		\Biggl[
		\rB_{ab} (\Phi_m)
		\Biggr] \;
		\Theta_{\cut} \left[\Phi_m\right] \;
		\rPS\left[ \calO \right] (\Phi_m; \boldsymbol{\xi})
		\\
		\notag
		& 
		+ \left(\frac{\alphas(\mur)}{2\pi}\right) \Biggl\{
		\dd \Phi_m \;
		\frac{1}{2\hat{s}_{12}} \;
		\Biggl[ \rV_{ab} (\Phi_m) 
		+ \sum_{(\alpha)} \rB_{ab}^{(\alpha)} (\Phi_m)
		\int \dd \Phi_{+1}^{(\alpha)} \; 
		\Theta_{t_0}^{t_1({\Phi}_m)}
		P_m^{(\alpha)} (\Phi_{+1}^{(\alpha)}; \boldsymbol{\xi}^{(\alpha)})
		\\ \notag
		&{}
		\hphantom{{} = \dd \Phi_m \; \frac{1}{2\hat{s}_{12}} \; \Biggl[}
		+ \sum_{(\alpha)} \rB_{ab}^{(\alpha)} (\Phi_m)
		\int \dd \Phi_{+1}^{(\alpha)} \; 
		\Theta_{0}^{t_0} \;
		x\, \rS^{(\alpha)} (\Phi_{+1}^{(\alpha)})
		\Biggr] \;
		\Theta_{\cut} \left[\Phi_m\right] \;
		\rPS\left[ \calO \right] (\Phi_m; \boldsymbol{\xi})
		\\ \notag
		&{} + \dd \eta_1 \, \dd \eta_2 \;
		\dd \Phi_m(\eta_1 \xi_1 P_1, \eta_2 \xi_2 P_2) \;
		\frac{1}{2\hat{s}_{12}} \;
		\rC^{\rFS}_{ab}(\eta_1, \eta_2; \muf; \Phi_m) \;
		\Theta_{\cut} \left[\Phi_m\right] \;
		\rPS\left[ \calO \right] (\Phi_m; \eta_1 \xi_1, \eta_2 \xi_2)
		\\ \notag
		&{} + 
		\dd \Phi_{m+1} \;
		\frac{1}{2\hat{s}_{12}} \;
		\Biggl[ \rR_{ab} (\Phi_{m+1}) \;
		\Theta_{\cut} \left[\Phi_{m+1}\right] \;
		\\ \notag
		&{} \hphantom{{} + \Phi_{m+1} \; \frac{1}{2\hat{s}_{12}} \; \Biggl[}
		- \sum_{(\alpha)} 
		\Theta_{t_0}^{t_1(\tilde{\Phi}^{(\alpha)}_m)}
		P_m^{(\alpha)} (\Phi_{m+1}; \boldsymbol{\xi}) \; \rB^{(\alpha)} (\tilde{\Phi}_m^{(\alpha)}) \;
		\Theta_{\cut} \left[ \tilde{\Phi}_m^{(\alpha)} \right] \;
		\\ \notag
		&{} \hphantom{{} + \Phi_{m+1} \; \frac{1}{2\hat{s}_{12}} \; \Biggl[}
		- \sum_{(\alpha)}
		\Theta_{0}^{t_0} \;
		\rS^{(\alpha)} (\tilde{\Phi}_{+1}^{(\alpha)}) \; \rB^{(\alpha)} (\tilde{\Phi}_m^{(\alpha)}) \;
		\Theta_{\cut} \left[ \tilde{\Phi}_m^{(\alpha)} \right] \;
		\Biggr]
		\rPS\left[ \calO \right] (\Phi_{m+1}; \boldsymbol{\xi})
		\Biggr\}
		\\ \notag
		& \hphantom{\calO} + \order{\alphas^2}.
	\end{align}
}

The effects of this in practice have been explored in detail in
\cite{Hoeche:2011fd,Nason:2012pr,Jones:2017giv,Cormier:2018tog,Jadach:2011cr}
and applied for phenomenological studies throughout the operation of the LHC.

Reorganised by order in $\alphas$ and by their contribution to $\calO(\Phi_m)$ and
$\calO(\Phi_{m+1})$, the contributions arising up to NLO within $\mcatnlo$
for colour-singlet processes are:

{ \footnotesize

\begin{align}
	\label{eq:nlomatching_notationdefinitions_nlomatching_dsigmaqqb_mcatnlo_0_singlet}
	\dd \hat{\sigma}^{\mcatnlo(0)}_{q\qbar} [\calO] = {} &
	\dd \Phi_m \; 
	\frac{1}{2\hat{s}_{12}} \;
	\Biggl[
	\rB_{q \qbar}(\Phi_m) 
	\Biggr]
	\Theta_{\cut} \left[\Phi_m\right] \;
	\Delta^{(0)} \bigr\vert_{\ptcut}^{Q(\Phi_m)} \;
	\calO (\Phi_m)
	\\
	\label{eq:nlomatching_notationdefinitions_nlomatching_dsigmaqqb_mcatnlo_1_singlet}
	\dd \hat{\sigma}^{\mcatnlo(1)}_{q\qbar} [\calO] = {} &
	\dd \Phi_m \; 
	\frac{1}{2\hat{s}_{12}} \;
	\Biggl[
	\rB_{q \qbar}(\Phi_m) \;
	\Delta^{(1)} \bigr\vert_{\ptcut}^{Q(\Phi_m)} 
    + \biggl\{ \rV_{q\qbar} (\Phi_m; \mur) 
		+ \rI_{q\qbar} (\Phi_m; \mur)
	\\ \notag 
	& \hphantom{\dd \Phi_m \; 
		\frac{1}{2\hat{s}_{12}} \;
		\Biggl[ }
    - \rB_{q \qbar}(\Phi_m) \sum_{(\alpha)} \int \dd \Phi_{+1}^{(\alpha)} \;
    \Theta_{Q(\Phi_m)}^{\infty} \;
    x\, \rS^{(\alpha)} (\Phi_{+1}^{(\alpha)})
	{}  \biggr\} \;
		\Delta^{(0)} \bigr\vert_{\ptcut}^{Q(\Phi_m)} 
	\Biggr]
	\Theta_{\cut} \left[\Phi_m\right] \;
	\calO (\Phi_m)
	\\ \notag
	{} + \dd \eta_1 \, \dd \eta_2 \; &
	\dd \Phi_m(\eta_1 \xi_1 P_1, \eta_2 \xi_2 P_2) \;
	\frac{1}{2\hat{s}_{12}} \;
	\Biggl[
	\left(\rP(\muf) +  \rK^{\rFS}\right)_{qq}  \negthick (\eta_1) \; \rB_{q\qbar}(\Phi_m) \; \delta^1_{\eta_2}
	\\ \notag
	& \hphantom{\dd \Phi_m \;
		\frac{1}{2\hat{s}_{12}} \;
		\Bigl[}
	+ \delta^1_{\eta_1} \, \rB_{q\qbar}(\Phi_m)
	 \left(\rP(\muf) +  \rK^\rFS\right)_{qq} \negthick (\eta_2)
	\Biggr] \;
	\Theta_{\cut} \left[\Phi_m\right] \;
	\Delta^{(0)} \bigr\vert_{\ptcut}^{Q(\Phi_m)} \;
	\calO(\Phi_m)
	\\ \notag 
	{} + {} & \dd \Phi_{m+1} 
	\frac{1}{2 \hat{s}_{12}} \Biggl[
	\rR_{q \qbar} (\Phi_{m+1}) \;
	\Theta_{\cut} \left[\Phi_{m+1}\right] \;
	\Delta^{(0)} \bigr\vert_{\ptcut}^{Q({\Phi}_{m+1})}
	\\ \notag
	&
	\hphantom{\dd \Phi_{m+1} \; 
		\frac{1}{2 \hat{s}_{12}}  \Biggl[}
	{} + 	\sum_{i=1}^2
    \biggl(	
	\Theta_{\ptcut}^{Q(\tilde{\Phi}_m^{\II_i})} \;
	\Delta^{(0)} \bigr\vert_{\ptof{1}}^{Q(\tilde{\Phi}_m^{\II_i})} \;
	\Delta^{(0)} \bigr\vert_{\ptcut}^{\ptof{1}} (\Phi_{m+1})
	\\ \notag
	& \hphantom{\dd \Phi_{m+1} \; 
		\frac{1}{2 \hat{s}_{12}}  \Biggl[ +  + \biggl(}
	{} - 
	 \Theta_{0}^{Q(\tilde{\Phi}_m^{\II_i})} \;
  \Delta^{(0)} \bigr\vert_{\ptcut}^{Q({\Phi}_{m+1})}
	 \biggr) \;
    \Theta_{\cut} \left[\tilde{\Phi}_m^{\II_i}\right] \;
	\rB_{q \qbar}(\tilde{\Phi}^{\II_i}_m) \;
	\rS^{q_ig} (x)
	\Biggr]
	\calO  (\Phi_{m+1})
	\\
	\label{eq:nlomatching_notationdefinitions_nlomatching_dsigmaqg_mcatnlo_1_singlet}
	\dd \hat{\sigma}^{\mcatnlo(1)}_{qg} [\calO] = {} &
	\dd \eta_2 \;
	\dd \Phi_m(\xi_1 P_1, \eta_2 \xi_2 P_2) \;
	\frac{1}{2\hat{s}_{12}} \;
	\rB_{q\qbar}(\Phi_m) \left(\rP(\muf)	 +  \rK^{\rFS}\right)_{gq} \negthick (\eta_2)  \,
	\\ \notag & \hspace{2in} \times
	\Theta_{\cut} \left[\Phi_m\right] \;
	\Delta^{(0)} \bigr\vert_{\ptcut}^{Q(\Phi_m)} \;
	\calO(\Phi_m)
	\\ \notag & {} + 
	\dd \Phi_{m+1} \; 
	\frac{1}{2 \hat{s}_{12}}  \Biggl[
	\rR_{q g} (\Phi_{m+1}) \;
	\Theta_{\cut} \left[\Phi_{m+1}\right] \,
	\Delta^{(0)} \bigr\vert_{\ptcut}^{Q({\Phi}_{m+1})}
	\\ \notag
	&
	\hphantom{\dd \Phi_{m+1} \; 
		\frac{1}{2 \hat{s}_{12}}  \Biggl[}
	{} +  
	 \biggl(
    \Theta_{\ptcut}^{Q(\tilde{\Phi}_m^{\II_2})} \;
	\Delta^{(0)} \bigr\vert_{\ptof{1}}^{Q(\tilde{\Phi}_m^{\II_2})} \;
	\Delta^{(0)} \bigr\vert_{\ptcut}^{\ptof{1}} (\Phi_{m+1})
	\\ \notag
	& 
	\hphantom{\dd \Phi_{m+1} \; 
		\frac{1}{2 \hat{s}_{12}}  \Biggl[ +  + \biggl(}
	{} -
    \Theta_{0}^{Q(\tilde{\Phi}_m^{\II_2})} \;
	\Delta^{(0)} \bigr\vert_{\ptcut}^{Q({\Phi}_{m+1})}
	\biggr) \;
    \Theta_{\cut} \left[\tilde{\Phi}_m^{\II_2}\right] \,
	\rB_{q \qbar}(\tilde{\Phi}^{\II_2}_m) \;
	\rS^{gq} (x)
	\Biggr]
	\calO  (\Phi_{m+1})
\end{align}
}
where we have again omitted ratios of PDFs and of $\alphas$ evaluated at different scales.

This can be verified to satisfy the NLO matching condition:
the error relative to \cref{eq:nlomatching_notationdefinitions_qcdatnlo_dsigmaqqb_nlo_1_singlet}
is
\begin{align}
	\label{eq:nlomatching_notationdefinitions_nlomatching_mcatnlo_powcorrRmS}
	\Theta^{\ptcut}_0 \times \biggl[
    \sum_{(\alpha)}
	\rB_{q \qbar}(\tilde{\Phi}^{(\alpha)}_m) \;
	\rS^{(\alpha)} (x)
	\Bigl( \calO(\Phi_{m+1}) - \calO(\tilde{\Phi}^{(\alpha)}_m) \Bigr)
	\Biggr],
\end{align}
which gives a non-logarithmic correction upon integration for IR-safe observable $\calO$.
The precise power of the power-correction in \cref{eq:nlo_match_1}
depends on the rate of convergence of the observable in the unresolved limit.
For cross-section-type observables, in which $\calO(\Phi) \equiv 1$ (or a product of theta-functions
implementing fiducial cuts), this vanishes for sufficiently small cut-off scales,
in which case the NLO-matching is exact.

In the interest of pedagogical clarity, we give a brief guided tour of the key features of
\cref{eq:nlomatching_notationdefinitions_nlomatching_dsigmaqqb_mcatnlo_0_singlet,eq:nlomatching_notationdefinitions_nlomatching_dsigmaqqb_mcatnlo_1_singlet,eq:nlomatching_notationdefinitions_nlomatching_dsigmaqg_mcatnlo_1_singlet}
germane to a comparison with
the corresponding dipole-subtracted NLO expressions of
\cref{eq:nlomatching_notationdefinitions_qcdatnlo_dsigmaqqb_nlo_1_singlet,eq:nlomatching_notationdefinitions_qcdatnlo_dsigmaqg_nlo_1_singlet}.
The leading-order terms in the expansion of the Sudakov factors, $\Delta^{(0)} \bigr\vert_{\ptcut}^{Q(\Phi_m)}$,
are 1 and, for strict fixed-order comparisons, may be ignored.  They are retained purely to facilitate
subsequent comparisons between matching schemes, in which the exponential suppression arising from 
the inclusion of the higher-order terms within \cref{eq:nlomatching_notationdefinitions_partonshower_sudexp}
becomes relevant.
The $\Theta_{\cut}$ factors may also, in practice, be considered to be identically 1
as long as the generator cuts are appropriately chosen for $\calO$
(they may only be zero if $\calO$ is also zero);
they are retained here to allow terms to be identified directly as either `H'-events, generated from $\Phi_{m+1}$
and therefore accompanied by $\Theta_{\cut}[\Phi_{m+1}]$, or as `S'-events generated from $\Phi_m$.
This division allows the required cancellations between contributions to be easily identified.
Finally, we note that $\Delta^{(1)} \bigr\vert_{\ptcut}^{Q(\Phi_m)}$ contains contributions
corresponding to \cref{eq:nlomatching_notationdefinitions_partonshower_Pmab} for all possible splittings,
and therefore includes terms, proportional to the gluon PDF $f_g$, which cancel against
the (differential) splitting term generated by the parton shower within
\cref{eq:nlomatching_notationdefinitions_nlomatching_dsigmaqg_mcatnlo_1_singlet}.

\subsection{The KrkNLO method}
\label{sec:nlomatching_krknlo}

The \krknlo method generates all events from the Born phase-space $\Phi_m$ and uses the shower algorithm
itself to generate real-type events with $\Phi_{m+1}$ kinematics.
According to the decision tree implied by the shower algorithm, events are reweighted multiplicatively to introduce
the real and virtual matrix-elements, and satisfy the matching condition.

The method may be summarised succinctly as:
\begin{algorithm}
	\begin{algorithmic}
		\ForAll{Born events} {shower}
		\If {first emission generated, from kernel $(\alpha)$}
		\State $w\gets w \times \frac{R(\Phi_{m+1})}{P_m^{(\alpha)}(\Phi_{m+1})}$
		\EndIf
		\State $w \gets w \times
		\left[1 + \frac{\alphas(\mur)}{2\pi}\left(\frac{\rV(\Phi_m;\, \mur)}{\rB(\Phi_m)} + \frac{\rI(\Phi_m;\, \tilde{\mu}_\rR)}{\rB (\Phi_m)} + \Delta_0^\rFS \right)\right]$
		\EndFor
	\end{algorithmic}
\end{algorithm}

This is possible only because of the \krk factorisation scheme, which will be discussed
further in \cref{sec:nlomatching_krkfs}; as a consequence of this,
no additional collinear convolutions
are required at the level of the hard cross-section,
and NLO accuracy can be achieved through multiplicative reweighting alone.
Within \krknlo, $\Delta_0^{\rFS}$ is a factorisation-scheme-dependent constant,
as will be discussed
further in \cref{sec:nlomatching_krkfs}.

To $\order{\alphas}$, the \krknlo algorithm generates the following contributions:
{ \footnotesize

\begin{align}
	\label{eq:nlomatching_notationdefinitions_nlomatching_dsigmaqqb_krknlo_0_singlet}
	\dd \hat{\sigma}^{\krknlo(0)}_{q\qbar} [\calO] = {} &
	\dd \Phi_m \; 
	\frac{1}{2\hat{s}_{12}} \;
	\Biggl[
	\rB_{q \qbar}(\Phi_m) 
	\Biggr]
	\Theta_{\cut} \left[\Phi_m\right] \;
	\Delta^{(0)} \bigr\vert_{\ptcut}^{Q(\Phi_m)} \;
	\calO (\Phi_m)
	\\
	\label{eq:nlomatching_notationdefinitions_nlomatching_dsigmaqqb_krknlo_1_singlet}
	\dd \hat{\sigma}^{\krknlo(1)}_{q\qbar} [\calO] = {} &
	\dd \Phi_m \;
	\frac{1}{2\hat{s}_{12}} \;
	\Biggl[
	\rB_{q \qbar}(\Phi_m) \;
	\Delta^{(1)} \bigr\vert_{\ptcut}^{Q(\Phi_m)} 
	\\ \notag
	& \hphantom{\dd \Phi_m \;
		\frac{1}{2\hat{s}_{12}} \;
		\Bigl[}
	{} + \biggl\{ \rV_{q\qbar} (\Phi_m;  \mur) 
	+ \rI_{q\qbar} (\Phi_m; \mur) 
	+ \Delta_0^{\krk} \; \rB_{q\qbar}(\Phi_m)
	\biggr\} 
	\Delta^{(0)} \bigr\vert_{\ptcut}^{Q(\Phi_m)} 
	\Biggr]
	\Theta_{\cut} \left[\Phi_m\right] \;
	\calO (\Phi_m)
	\\ \notag
	{} + \dd \eta_1 \, \dd \eta_2 \;
	\dd \Phi_m & (\eta_1 \xi_1 P_1, \eta_2 \xi_2 P_2) \;
	\frac{1}{2\hat{s}_{12}} \;
	\Bigl[
	 \left( \rP(\muf) + \rK^{\msbar\to\krk}\right)_{qq}  \negthick (\eta_1) \;
	\rB_{q\qbar}(\Phi_m) \; \delta^1_{\eta_2}
	\\ \notag
	& \hphantom{ (\eta_1 \xi_1 P_1, \eta_2 \xi_2 P_2) \;
		\frac{1}{2\hat{s}_{12}} \;
		\Bigl[}
	+ \delta^1_{\eta_1} \, \rB_{q\qbar}(\Phi_m) 
	 \left( \rP(\muf) + \rK^{\msbar\to\krk} \right)_{qq} \negthick (\eta_2) 
		\Bigr]
	\Theta_{\cut} \left[\Phi_m\right] \;
	\Delta^{(0)} \bigr\vert_{\ptcut}^{Q(\Phi_m)} \;
	\calO(\Phi_m)
	\\ \notag {} + {}
	\dd \Phi_{m+1} 
	\frac{1}{2 \hat{s}_{12}} \Biggl[
	& 
	\frac{\rR_{q \qbar} (\Phi_{m+1})}{	
		\rB_{q \qbar}(\tilde{\Phi}^{\II_1}_m) \; 
		\rS^{q_1g} (x) \;
		\Theta_{\cut}\left[ \tilde{\Phi}^{\II_1}_m \right]
		+
		\rB_{q \qbar}(\tilde{\Phi}^{\II_2}_m) \; 
		\rS^{q_2g} (x) \;
		\Theta_{\cut}\left[ \tilde{\Phi}^{\II_2}_m \right]	}
	\\ \notag
	{} \times & \biggl( \sum_{i=1}^2	
	\Theta_{\cut} \left[\tilde{\Phi}_m^{\II_i}\right] \;
	 \Theta_{\ptcut}^{Q(\tilde{\Phi}_m^{\II_i})} (\tilde{\Phi}_m^{\II_i}) \;
	\Delta^{(0)} \bigr\vert_{\ptof{1}}^{Q(\tilde{\Phi}_m^{\II_i})} \;
	\Delta^{(0)} \bigr\vert_{\ptcut}^{\ptof{1}} (\Phi_{m+1}) \;
	\rB_{q \qbar}(\tilde{\Phi}^{\II_i}_m) \;
	\rS^{q_ig} (x)
	\biggr)
	\Biggr]
	\;
	\calO  (\Phi_{m+1})
	\\
	\label{eq:nlomatching_notationdefinitions_nlomatching_dsigmaqg_krknlo_1_singlet}
	\notag
	\dd \hat{\sigma}^{\krknlo(1)}_{qg} [\calO] = {} &
	\dd \eta_2 \;
	\dd \Phi_m (\xi_1 P_1, \eta_2 \xi_2 P_2) \;
	\frac{1}{2\hat{s}_{12}} \;
	\rB_{q\qbar}(\Phi_m)
	\left(\rP_{gq}(\muf) + \rK_{qg}^{\msbar\to\krk} \right)
	 (\eta_2)  \;
	\Theta_{\cut} \left[\Phi_m\right] \;
	\Delta^{(0)} \bigr\vert_{\ptcut}^{Q(\Phi_m)} \;
	\calO(\Phi_m)
	\\ {} + {}
	\dd \Phi_{m+1} \; 
	\frac{1}{2 \hat{s}_{12}}
	& \rR_{q g} (\Phi_{m+1}) \;
	\Theta_{\cut} \left[\tilde{\Phi}_m^{\II_2}\right] \;
	\Theta_{\ptcut}^{Q(\tilde{\Phi}_m^{\II_2})} (\tilde{\Phi}_m^{\II_2}) \;
	\Delta^{(0)} \bigr\vert_{\ptof{1}}^{Q(\tilde{\Phi}_m^{\II_2})} \;
	\Delta^{(0)} \bigr\vert_{\ptcut}^{\ptof{1}} (\Phi_{m+1}) \;
	\calO  (\Phi_{m+1})
\end{align}
}

Note that here, for consistency with the other sections, we continue with the convention
used in \cref{eq:nlomatching_notationdefinitions_fixedorder_dsigmaAB} whereby we
write the partonic cross-section in the \msbar scheme,
for convolution with \msbar PDFs.
In practice, as will be discussed further in \cref{sec:nlomatching_krkfs},
the \krknlo method uses its own factorisation scheme,
defined by the $\rK^{\msbar\to\krk}_{ab}$ terms above,
to achieve the NLO matching condition.

For numerical stability we have here chosen the real weight to be
\begin{align}
	\label{eq:nlomatching_notationdefinitions_nlomatching_krknlo_wR_sym}
		\frac{\rR_{q \qbar} (\Phi_{m+1})}{	
		\rB_{q \qbar}(\tilde{\Phi}^{\II_1}_m) \; 
		\rS^{q_1g} (x) \;
		\Theta_{\cut}\left[ \tilde{\Phi}^{\II_1}_m \right]
		+
		\rB_{q \qbar}(\tilde{\Phi}^{\II_2}_m) \; 
		\rS^{q_2g} (x) \;
		\Theta_{\cut}\left[ \tilde{\Phi}^{\II_2}_m \right]	
		}
\end{align}
rather than simply partitioning the real contribution and reweighting each dipole separately,
\begin{align}
	\label{eq:nlomatching_notationdefinitions_nlomatching_krknlo_wR_unsym}
		\frac{1}{2}
		\frac{\rR_{q \qbar} (\Phi_{m+1})}{	
		\rB_{q \qbar}(\tilde{\Phi}^{\II_1}_m) \; 
		\rS^{q_1g} (x)}, \qquad
		\frac{1}{2}
		\frac{\rR_{q \qbar} (\Phi_{m+1})}{
		\rB_{q \qbar}(\tilde{\Phi}^{\II_2}_m) \;
		\rS^{q_2g} (x)}
\end{align}
for each generated shower emission respectively.
In the statistical limit they are equivalent;
the inclusion of the second dipole in the denominator regulates the weight
distribution when the emitting dipole kernel is small.

From
\cref{eq:nlomatching_notationdefinitions_nlomatching_dsigmaqqb_krknlo_1_singlet,eq:nlomatching_notationdefinitions_nlomatching_dsigmaqg_krknlo_1_singlet}
it is clear that for NLO accuracy we require from $\Theta_{\ptcut}^{Q(\tilde{\Phi}_m)}$
that the shower starting-scale be unrestricted (i.e.\ a `power' shower \cite{Plehn:2005cq,Corke:2010zj});
otherwise high-$\pt$ regions of the real-emission phase-space $\Phi_{m+1}$ are excluded.
This corresponds to $Q(\Phi_m) \equiv \infty$
(in practice we will use $Q_{\max{}}(\Phi_m)$ to denote the upper-bound
arising from kinematic restrictions%
\footnote{The kinematic upper-bound differs between the two II-type dipoles
but may be expressed concretely as
$$
Q_{\max{}}(\Phi_{m}) = \frac{1-\xi_{\min{}}}{2\sqrt{\xi_{\min{}}}} \sqrt{\hat{s}_{12}}
$$
where $\xi_{\min{}} = \min\{\xi_1,\xi_2\}$.}).

For soft emissions, relative to the fixed-order expression
\cref{eq:nlomatching_notationdefinitions_qcdatnlo_dsigmaqqb_nlo_1_singlet,eq:nlomatching_notationdefinitions_qcdatnlo_dsigmaqg_nlo_1_singlet}, 
we are also excluding events in which the hardest emission is softer than $\ptcut$.
Assuming this is chosen to be sufficiently small in relation to the relevant hard scales of the process,
it acts as a slicing parameter, and
below this threshold we are in the singular region in which
the dipole factorisation yields a good approximation
for the corresponding real matrix element and for any IRC-safe observable 
$\calO(\Phi_{m+1}) = \calO(\tilde{\Phi}^{(\alpha)}_m)$.
The `missing' events with an emission softer than $\ptcut$
are restored by the mismatch
in integration domains
between the integrated dipoles 
and the implicit integration within the shower Sudakov factor, with an error
given by
\begin{align}
	\label{eq:nlomatching_notationdefinitions_nlomatching_krknlo_powcorrRmS}
	\Theta^{\ptcut}_0 \times \biggl[
	\Bigl( \rR (\Phi_{m+1}) 
	- {} & \sum_{(\alpha)} \rB^{(\alpha)}_{q\qbar}(\tilde{\Phi}^{(\alpha)}_m) \; \rS^{(\alpha)} (x) \Bigr)
		 \; \calO(\Phi_{m+1}) 
	\\ \notag
	{} + {}
	& \sum_{(\alpha)} \rB^{(\alpha)}_{q\qbar}(\tilde{\Phi}^{(\alpha)}_m) \; \rS^{(\alpha)} (x)
	\Bigl( \calO(\Phi_{m+1}) - \calO(\tilde{\Phi}^{(\alpha)}_m) \Bigr)
	\biggr].
\end{align}
The former term in
\cref{eq:nlomatching_notationdefinitions_nlomatching_krknlo_powcorrRmS}
is the difference between the real-emission
matrix element $\rR$ and its Catani--Seymour dipole-subtraction counterterm, within the singular region regulated by $\ptcut$.
By construction the IR-divergent terms therefore cancel and this term can contribute at most a power-correction (i.e.\@, not logarithmically-enhanced).
The latter term
is the difference between the measurement
function 
$\calO$ for an IRC-safe observable
applied to $\Phi_{m+1}$ and to a reduced phase-space configuration in which the 
unresolved particle is mapped away $\tilde{\Phi}^{(\alpha)}_m$, in the unresolved region,
and therefore also contributes at most a power-correction \cite{Platzer:2010ppa}.

In order to satisfy the NLO matching condition, it is necessary
to correctly reproduce the NLO cross-section when setting $\calO(\Phi)$
equal to a product of theta-functions implementing fiducial cuts.
This requires the factorisation scheme kernels $\rK^{\msbar\to\rFS}_{ab}$
to cancel the collinear contributions generated by the integral of the shower
kernels over the radiative phase-space
within 
$\Delta^{(1)} \bigr\vert_{\ptcut}^{Q(\Phi_m)}.$%
\footnote{Although we present this for the dipole shower, taking advantage of the explicit
	phase-space integrations performed for dipole subtraction in \cite{Catani:1996vz},
	it can in principle be extended to other shower algorithms with the integration performed
	numerically where necessary.}
This is sufficient for \krknlo to satisfy the NLO matching condition for general $\calO$.
The corresponding factorisation scheme, called the \krk scheme%
\footnote{The same factorisation scheme was formerly named the `Monte Carlo' scheme in \cite{Jadach:2015mza,Jadach:2016acv,Jadach:2016qti}.}
\cite{Jadach:2016acv,Jadach:2020xfl},
is introduced in further detail in \cref{sec:nlomatching_krkfs}
and \cref{sec:appendix_krkpdfs}.

Finally, we note that terms contributing to $ \calO(\Phi_{m+1}) $ are generated
with identical `Born' Sudakov factors
corresponding to the parton-shower probability of generating an emission at that
scale, starting from the relevant underlying Born kinematics.
Whilst at the order shown these Sudakov factors are 1, the exponential behaviour of
\cref{eq:nlomatching_notationdefinitions_partonshower_sudexp}
makes the higher-order contributions of these factors non-negligible, especially where 
$\ptof{1} \ll Q_{\max{}}(\tilde{\Phi}_m^{(\alpha)})$.
In the high-$\pt$ tail, the \krknlo and fixed-order predictions should coincide.

\subsection{The Krk factorisation scheme}
\label{sec:nlomatching_krkfs}

As described in \cref{sec:nlomatching_krknlo}, the \krk scheme is defined by the requirement 
that it cancel the collinear terms arising from the integral over the shower emission kernels
within the shower Sudakov factor. Full expressions for the
factorisation scheme transformation are given in \cref{sec:appendix_krkpdfs}.

Given a choice of factorisation scheme with PDFs related to those in the \msbar scheme by
\begin{align}
	\mathbf{f}^{\rFS}
	&=
	\mathbb{K}^{\msbar \to \rFS}
	\otimes
	\mathbf{f}^{\msbar},
\end{align}
where explicitly
\begin{align}
	\label{eq:nlomatching_krkfs_CS_faFS}
	f^{\rFS}_a (x, \muf)
	&=
	\sum_b
	\int_x^1
	\frac{\dd z}{z}
	\;
	\mathbb{K}^{\msbar \to \rFS}_{ab} \left(z, \muf\right) \ 
	f_b^{\msbar} \left(\frac{x}{z}, \muf\right)
\end{align}
and the matrix of convolution kernels $\mathbb{K}^{\msbar \to \rFS}_{ab}$ has perturbative expansion
\begin{align}
	\mathbb{K}^{\msbar \to \rFS}_{ab} \left(z, \mu\right)
	=
	\delta_{ab} \; \delta(1-z)
	+
	\frac{\alphas(\mu)}{2\pi}
	\;
	\rK^{\msbar \to \rFS}_{ab}(z) + \order{\alphas^2},
\end{align}
the NLO partonic cross-section must be transformed correspondingly into the
new factorisation scheme to retain NLO accuracy,
\begin{align}
	\dd \hat{\sigma}_{ab}^{\rFS(1)}	(\xi_1 P_1, \xi_2 P_2)
	=
	\dd \hat{\sigma}_{ab}^{\msbar(1)}
	-
	\dd \eta_1 \, \dd \eta_2 \;
	\sum_{c}
	\Bigl[ 
	\rK^{\msbar\to\rFS}_{ca}(\eta_1) \;
	& \dd \hat{\sigma}_{cb}^{\msbar(0)}(\eta_1 \xi_1, \eta_2 \xi_2) 
	\;
	\delta_{\eta_2}^1
	\\ \notag
	{} + 
	\delta_{\eta_1}^1
	\;
	& \dd \hat{\sigma}_{ac}^{\msbar(0)}(\eta_1 \xi_1, \eta_2 \xi_2)
	\;
	\rK^{\msbar\to\rFS}_{cb}(\eta_2)
	\Bigr].
\end{align}
For \cref{eq:nlomatching_notationdefinitions_nlomatching_dsigmaqqb_krknlo_1_singlet,eq:nlomatching_notationdefinitions_nlomatching_dsigmaqg_krknlo_1_singlet}
the adoption of the \krk scheme 
therefore amounts to the cancellation of the
$\rK_{ab}^{\msbar\to\krk}$ convolution terms within the perturbative cross-section,
and the convolution of the resulting expressions
for 
$ \dd \hat{\sigma}^{\krk} $
with \krk-scheme PDFs, for which the 
scheme-transformation \cref{eq:nlomatching_krkfs_CS_faFS} has been pre-calculated.

The additional choice of $\muf^2 = \hat{s}_{12}$ removes the 
mass-factorisation terms contained within the $\rP$ operator,
since these terms 
(\cref{eq:app_CS_Pop})
are proportional to $\log \muf^2 / \hat{s}_{12}$.
We make the central choice $\muf^2 = \hat{s}_{12}$ throughout.
Factorisation scale variation by a constant factor 
(e.g.\@ following the conventional scale-variation heuristic for estimation of missing-higher-order uncertainties),
or by a function of $x$,
is equivalent in this case to a variation of the factorisation scheme
and would here arise as a factorisation-scheme uncertainty.%
\footnote{Like the choice of renormalisation or factorisation scale,
	the choice of factorisation scheme is unphysical.
	An all-order calculation should therefore be independent of the scheme used.
	However, at any fixed order of perturbation theory, higher-order terms beyond the
	claimed level of perturbative accuracy are scheme-dependent.
	The envelope of predictions arising from alternative choices of factorisation scheme
	may therefore be considered a theory uncertainty,
	`factorisation-scheme uncertainty',
	akin to factorisation- and renormalisation-scale uncertainty.
	There is no clear metric by which one scheme may be considered intrinsically superior to any other
	save perhaps perturbative convergence, which may be altered by the addition or removal of (perhaps-large) logarithmic terms
	from the perturbative cross-section.
	We defer a detailed discussion of factorisation-scheme uncertainty, both for the \krk scheme
	and for other proposed alternatives to the `default' \msbar-scheme, to future work.
}
Since the impact of factorisation-scale variation is well-understood we do not explore this further here.

By comparison of 
\cref{eq:nlomatching_notationdefinitions_nlomatching_dsigmaqqb_krknlo_1_singlet,eq:nlomatching_notationdefinitions_nlomatching_dsigmaqg_krknlo_1_singlet}
with
\cref{eq:nlomatching_notationdefinitions_nlomatching_dsigmaqqb_mcatnlo_1_singlet,eq:nlomatching_notationdefinitions_nlomatching_dsigmaqg_mcatnlo_1_singlet}
it is clear that NLO accuracy is achieved provided that 
\begin{align}
	\label{eq:nlomatching_krkfs_Deltaeq_Kqq}
	\rK^{\msbar\to\krk}_{qq} (x)
	= {} & 
	\rK^{\msbar}_{qq} (x)
	-
	\frac{1}{2} \, \Delta_0^{\FS} \, \delta(1-x). \\
	\label{eq:nlomatching_krkfs_Deltaeq_Kqg}
	\rK^{\msbar\to\krk}_{qg} (x)
	= {} & 
	\rK^{\msbar}_{gq} (x)
\end{align}
That is, the contribution to $\rK_{qc}^{\msbar\to\krk}$ with non-trivial 
$x$-dependence is fully determined by the integrated Catani--Seymour dipoles,
 whilst for the flavour-diagonal transition,
`virtual' terms proportional to $\delta(1-x)$,
i.e.\@ associated solely with the unconvoluted Born phase-space $\Phi_m(\xi_1 P_1,\xi_2 P_2)$,
are unconstrained by the requirement
that the collinear contributions from $\rK^{\FS}$ are absorbed into the factorisation-scheme transformation.
Alternative choices of
$\Delta_0^{\FS}$ therefore define a one-parameter family of factorisation schemes
all of which give NLO accuracy.
To fix it concretely, for the `full' \krk scheme we follow
\cite{Jadach:2016acv}
and impose 
\begin{align}
	\label{eq:nlomatching_krkfs_PDFmomsumrule}
	\sum_a \int_0^1 \dd \xi \; \xi \, f_{a}^{\krk} (\xi; \muf) = 1
\end{align}
as a momentum sum rule \cite{Collins:1981uw} on the transformed \krk-scheme PDFs (an earlier `DY' scheme used a flavour sum rule \cite{Jadach:2015mza});
this is achieved by modifying the flavour-diagonal `virtual' corrections,
i.e.\@ fixing $\Delta_0^{\FS}$,
to impose for each flavour $a$
\begin{align}
	\sum_{b} \int_0^1 \dd z \; z \, \rK^{\msbar\to\krk}_{ab} (z) = 0.
\end{align}
The `full' \krk scheme further includes an additional redefinition of the gluon PDF
derived from $gg$-channel processes
(see \cref{sec:appendix_krkpdfs} for details); since this entails a modification
of the gluon PDF from the $\msbar$ PDF 
at order $\alphas$ only, this contributes only $\order{\alphas^2}$
terms to the class of processes considered here and is therefore irrelevant to
the NLO matching condition.

As a result of the factorisation scheme and factorisation scale choices,
there are no remaining collinear terms, allowing
NLO accuracy to be achieved
within the hard process solely by multiplicative reweighting
with no additional collinear convolutions.

Explicit expressions for the PDF convolution transformation
into the \krk scheme are given in \cref{sec:appendix_krkpdfs}
together with a description of the way the transformation is performed in practice;
the numerical effect of the transformation of \msbar PDFs into the \krk
factorisation scheme is shown in \cref{fig:appendix_krkvsmsbarpdf}.

\subsection{Comparison with \mcatnlo}
\label{sec:nlomatching_krkvsothers}

Comparison of the $\order{\alphas}$ contributions to the matched \krknlo
differential cross-section (\cref{eq:nlomatching_notationdefinitions_nlomatching_dsigmaqqb_krknlo_1_singlet,eq:nlomatching_notationdefinitions_nlomatching_dsigmaqg_krknlo_1_singlet})
with their \mcatnlo counterparts
(\cref{eq:nlomatching_notationdefinitions_nlomatching_dsigmaqqb_mcatnlo_1_singlet,eq:nlomatching_notationdefinitions_nlomatching_dsigmaqg_mcatnlo_1_singlet})
is instructive and illustrates the underlying cause of some of the numerical differences between the matching schemes
which will emerge at a formally higher-order in $\alphas$
despite both satisfying the NLO matching condition and using identical
shower algorithms, physics parameters, and underlying \msbar PDF sets.
We briefly highlight these here.

The \krknlo method requires the phase-space for the first parton shower emission 
to be unrestricted (a `power' shower),
i.e.\@ uses a starting-scale $Q_{\max{}}(\Phi_m)$ restricted only by the possible kinematics of the splitting.
In $\mcatnlo$ the shower starting-scale
is typically chosen as a characteristic scale related to the underlying hard process;
as usual with such choices, this may be varied as a heuristic to estimate shower uncertainties 
\cite{Gieseke:2004tc,Bellm:2016voq,Bothmann:2016nao,Mrenna:2016sih,Bellm:2016rhh,Cormier:2018tog}.
This difference in $Q(\Phi_m)$ choice between \krknlo and \mcatnlo
generates different Sudakov factors associated with emissions from the Born phase-space.
This difference will be largest when $Q(\Phi_m) \ll Q_{\max{}}(\Phi_m)$,
with additional suppression for \krknlo arising from the integral over the shower phase-space excluded within \mcatnlo.

The \mcatnlo method further requires the additional choice $Q(\Phi_{m+1})$ for the starting-scale
for `H'-events, in which the shower starts from $\Phi_{m+1}$.
This is the scale that defines the Sudakov factor for `H'-events and bounds the phase-space
for the second and subsequent emissions (in our case, $\ptof{2},\ptof{3}$ etc.).
Where $Q(\Phi_{m+1})$ differs significantly from $\ptj{1}$,%
\footnote{Note that for the processes we consider here, the $\pt$
relative to the beam axis is the same as the `shower $\pt$' relative to the
emitter--spectator axis in the dipole rest frame,
and so $Q(\Phi_{m+1}) = \ptj{1}$ gives a shower phase-space 
for the second-emission identical to that arising from
the shower $\pt$-ordering in the \krknlo case.  For a general process
with contributions from IF/FI/FF dipoles,
the $\pt$ relative to the beam axis will differ from that relative to a parent dipole.} 
the phase-space available for the second parton-shower emission
(and consequently, the associated Sudakov factor)
will differ substantially between the two methods.

For \mcatnlo, where $\ptof{1} > Q(\Phi_{m}^{\II_i})$,
i.e.\ the real-emission phase-space inaccessible by the parton shower from the underlying Born,
the `pure' real-emission matrix-element
is unmodified by shower-subtraction terms or the real-emission generated from `S'-events.
Notably, for the `power'-shower choice of $Q(\Phi_m)$ within \mcatnlo, this never occurs.
For the phase-space accessible by the shower from the underlying Born,
at this order the contributions from the `H' and `S' events cancel, but at higher-orders
would be governed by the hierarchy of scales between $Q(\Phi_m)$ and $\ptof{1}$
(for the Born Sudakov applied to the `S'-events, absent from the `H'-events)
and between $Q(\Phi_{m+1})$ and $\ptof{1}$ (for the `real-emission' Sudakov).

Within \krknlo the $\calO(\Phi_{m+1})$ contribution receives a single
unified Sudakov factor associated with emissions from the underlying Born 
process.
In \mcatnlo contributions are associated with different Sudakov factors
according to the underlying division between `H'- and `S'-events,
which is arbitrary and unphysical \cite{Frixione:2002ik,Schumann:2007mg}.
One consequence of this is a reduced Sudakov suppression for \mcatnlo
where $\ptof{1} \ll Q(\Phi_m^{\II_i})$ and, for example, 
especially where $\ptof{1} \sim \ptcut$.

The \krknlo method adopts $\muf(\Phi_m)^2 = \hat{s}_{12}$, as described
in \cref{sec:nlomatching_krkfs}, while the functional form for the factorisation scale
in \mcatnlo may be chosen freely,
and is typically chosen as a characteristic scale related to the underlying hard process.
This may be expected to contribute
to formally-higher-order deviations in regions of phase-space where the ratio
$\muf^2 / \hat{s}_{12}$ is large. This is likely to be especially significant
where $\muf$ is chosen as a $\pt$-based scale, and for instance in
the extremes of rapidity distributions \cite{Gehrmann:2020oec}.

This factorisation scale is used only for the Born process; since the real-emission
is generated by the parton shower, the scales used by the PDFs accompanying the real-emission
matrix-element $\rR$ are $\muf(\Phi_m)$ for the spectator PDF, and $\mu^{\II_i}$
for the emitter PDF.  Relative to the \mcatnlo method there are therefore factors
of the form of \cref{eq:nlomatching_notationdefinitions_PDFratio_order_alphas2,eq:nlomatching_notationdefinitions_alphas_order_alphas2}.
In addition, due to the adoption of the \krk scheme for the PDFs used within the parton
shower kernels of \cref{eq:nlomatching_notationdefinitions_partonshower_Pmab}
(necessary for the first emission to satisfy the matching condition)
further modifies the results relative to the ratio of \msbar PDFs that arise within
the \mcatnlo method.

The \krknlo method generates factors of $\alphas$ 
at NLO and beyond within the shower,
which use the renormalisation scale 
$\mur = \mu^{(\alpha)}(\Phi_{+1}^{(\alpha)})$
(by default, within the Herwig dipole shower $\mu^{(\alpha)}(\Phi_{+1}^{(\alpha)})$ is the 
transverse momentum of the splitting with respect to its parent dipole, though
this is modified according to the CMW scheme if enabled \cite{Catani:1990rr,Bellm:2017bvx,Bellm:2019zci}). 
Since the real-emission matrix-element is generated as a reweighted shower-emission,
this is accompanied by a factor of $\alphas(\mu^{(\alpha)}(\Phi_{m+1}^{(\alpha)}))$
unless explicitly reweighted.
Within $\mcatnlo$, the factors of $\alphas$ associated with $\rR$ are typically
all assigned a renormalisation scale $\mur(\Phi_{m+1})$.

The adoption of the \krk factorisation scheme for the \krknlo calculation is motivated
by the NLO matching condition as outlined in \cref{sec:nlomatching_krkfs}.  Beyond NLO,
however, the transformed PDFs defined to satisfy the NLO matching condition are also applied
to the real and virtual contributions where they 
generate additional contributions of $\order{\alphas^2}$.

The \mcatnlo populates the real-emission phase-space directly
and may therefore be combined with standard Monte Carlo variance-reduction techniques
to efficiently sample the real-emission phase-space.
The \krknlo method uses the shower algorithm to sample from the real-emission
phase-space, and so relatively under-samples the high-$\pt$ tail where the 
emission kernels, derived to be good approximations only in the factorisable singular limit,
underestimate the true real matrix-element.

The latter could be resolved, along with the relaxation of the requirement for full phase-space coverage
by the parton shower, by a hybrid method KrkMC@NLO (see also \cite{Nason:2021xke}) in which
the two methods are patched together along a hypersurface within phase-space.
This amounts to the insertion of a $\Theta$-function, and its complementary $\Theta$-function,
into the two methods, and summing the result.
In particular, if this were chosen to remove from the \mcatnlo contribution
the region of phase-space in which the \mcatnlo expressions \cref{eq:nlomatching_notationdefinitions_nlomatching_dsigmaqqb_mcatnlo_1_singlet,eq:nlomatching_notationdefinitions_nlomatching_dsigmaqg_mcatnlo_1_singlet}
amount to an over-subtraction, it would fix the problem of negative weights arising within \mcatnlo
by assigning regions generating negativity to \krknlo (in which all weights are positive).
The matching uncertainty resulting from the above differences could then be
assessed as a function of the choice of hypersurface, e.g.\@ 
by defining a one-parameter family smoothly interpolating between the two methods.
We defer this to future studies.

\section{Implementation}
\label{sec:implementation}

The algorithm described above has been implemented within \herwigseven \cite{Bellm:2015jjp,Bellm:2017bvx,Bellm:2019zci,Bewick:2023tfi}, using the \herwig dipole shower.
In \cref{sec:imp_krknloH7} we outline the general features of the implementation
used for the results presented in \cref{sec:matchingcomparisons,sec:pheno}.
We postpone a detailed discussion for an anticipated future public release of the \krknlo
code, for general colour-singlet processes, within \herwig \textsf{7.4}.
In \cref{sec:imp_validation} we summarise the validation of the implementation
of the diphoton production process,
which is the specific process chosen for the present work
and the subject of the matching studies in \cref{sec:matchingcomparisons,sec:pheno}.
The implementation of the Drell-Yan process described in \cite{Jadach:2015mza}
was derived by factorising production from decay and therefore unsuitable
to generalise to arbitrary processes.
We performed an independent derivation for $pp\to\gamma\gamma$ as a
stepping-stone on the road to the forthcoming fully general implementation.

The diphoton process is the simplest colour-singlet process with no intrinsic mass-scale
or resonance,
so can usefully elucidate the differing behaviours of the alternative parton-shower matching
methods without being obfuscated by convolution with a resonant matrix-element.

\subsection{KrkNLO in \herwigseven}
\label{sec:imp_krknloH7}

Within the \herwig dipole shower, the \krknlo algorithm is implemented in practice via the abstract \texttt{DipoleEventReweight} class \cite{Bellm:2017bvx},
by providing a concrete implementation of its two methods, \texttt{weight} and \texttt{weightCascade}:
\begin{description}
	\item[\texttt{weight}] applies a multiplicative weight each time an emission is generated.
	In our case, this weight is non-trivial only for the first emission,
	in which case it reweights the parton-shower dipole emission kernel
	according to
	\cref{eq:nlomatching_notationdefinitions_nlomatching_krknlo_wR_sym}
	or
	\cref{eq:nlomatching_notationdefinitions_nlomatching_krknlo_wR_unsym}.
	\item[\texttt{weightCascade}] applies a multiplicative weight to the
	entire shower cascade derived from the Born kinematics $\Phi_m$.
	This is applied to every event, when the shower algorithm reaches the
	cutoff scale $t_0$, and is therefore applied exactly once (unless the
	shower is forced to terminate after \texttt{NEmissions}).
	In our case, this calculates and applies the virtual weight implied by
	\cref{eq:nlomatching_notationdefinitions_nlomatching_dsigmaqqb_krknlo_1_singlet},
	\begin{align}
		1 + \frac{\alphas(\mur(\Phi_m))}{2\pi}
		\frac{1}{\rB_{q\qbar}(\Phi_m)}
		\biggl\{
		 \rV_{q\qbar} (\Phi_m;  \mur) 
		+ \rI_{q\qbar} (\Phi_m; \mur) 
		+ \Delta_0^{\rFS} \; \rB_{q\qbar}(\Phi_m) \biggr\},
	\end{align}
	for the $q\qbar$-channel only.
\end{description}
For testing and validation purposes, the contributions 
of $\rB$, $\rR$, $\rV$, $\rI$ and $\Delta_0^{\rFS}$
can each be enabled or disabled independently.

Both \texttt{weight} and \texttt{weightCascade} allow the renormalisation scales
used for the $\order{\alphas}$ contributions
to be set flexibly and independently, or (in the real emission case)
to be left at the scale generated by the shower.

The value of $\Delta_0^{\rFS}$ is set according to the \texttt{KrkNLOEventReweight:PDF}
parameter, which should reflect the factorisation scheme of the PDF set chosen for the hard process.
This allows the calculation to be performed in the \krk scheme
(both `full' \cite{Jadach:2016qti,Jadach:2016acv} and `DY' \cite{Jadach:2015mza})
and for validation purposes also allows the calculation to be performed in
the \msbar scheme (i.e.\@ $\Delta_0^{\rFS}=0$), in which case external implementations of the contributions provided by the $\rP$ and $\rK$
convolution terms in
\cref{eq:nlomatching_notationdefinitions_nlomatching_dsigmaqqb_krknlo_1_singlet,eq:nlomatching_notationdefinitions_nlomatching_dsigmaqg_krknlo_1_singlet}
are required for NLO accuracy.

\subsection{Validation}
\label{sec:imp_validation}

To validate the implementation of the \krknlo method within \herwigseven
for the diphoton process,
we have tested each component independently
to verify that it behaves as anticipated and, in the relevant limits,
correctly reproduces the required fixed-order components
for NLO accuracy.

\subsubsection{Real matrix elements}
\label{sec:imp_validation_R}

Within the \krknlo implementation, the hard-coded real-emission
matrix elements have been tested phase-space-point by phase-space-point
against those provided by \openloops \cite{Buccioni:2019sur}
and \madgraph \cite{Alwall:2014hca},
with machine-precision accuracy.

For \krknlo to achieve NLO accuracy, the weights
\cref{eq:nlomatching_notationdefinitions_nlomatching_krknlo_wR_sym,eq:nlomatching_notationdefinitions_nlomatching_krknlo_wR_unsym}
must further correspond exactly to the the momentum mappings and
splitting kernels used within the parton shower algorithm,
and the underlying Born matrix element used for the
LO event generation,
in order to correctly reproduce the real matrix element.

To verify this numerically within the \krknlo code, 
the shower is truncated to provide exactly one emission
and events in which no emissions are generated are vetoed.
The relevant Sudakov factor 
$\Delta \bigr\vert_{\ptof{1}}^{Q(\Phi_m)} (\Phi_m)$
is calculated for each splitting
by numerical integration (with \vegas \cite{Lepage:1977sw}) 
of the splitting kernels considered within the shower algorithm
over the splitting phase-space
between the scale of the splitting and the starting-scale used within the shower algorithm,
using identical kernels, kinematic limits, and scales (both for the PDF ratios
and $\alphas$) as are used in the parton shower algorithm.
Each event is then reweighted by 
$1 / \Delta \bigr\vert_{\ptof{1}}^{Q(\Phi_m)} (\Phi_m)$.

By comparison of
\cref{eq:nlomatching_notationdefinitions_nlomatching_dsigmaqqb_krknlo_1_singlet,eq:nlomatching_notationdefinitions_nlomatching_dsigmaqg_krknlo_1_singlet}
with
\cref{eq:nlomatching_notationdefinitions_qcdatnlo_dsigmaqqb_nlo_1_singlet,eq:nlomatching_notationdefinitions_qcdatnlo_dsigmaqg_nlo_1_singlet}
it is clear that this should lead to agreement,
for any observable,
between the differential cross-sections calculated by this procedure
and those arising from the real-emission contribution of a fixed-order NLO computation
(or equivalently, the LO diphoton-plus-jet process).

This has been verified against \matchbox (with matrix elements from \madgraph)
for the partonic $q\qbar$- and $qg$-channels,
and for the physical hadronic $pp$ process in which all partonic channels are summed over.
A selection of plots indicating excellent agreement is provided
in \cref{fig:validation_real}, \cref{sec:appendix_validation}.

\subsubsection{Virtual matrix elements}
\label{sec:imp_validation_V}

Within the \krknlo code, the hard-coded
$\varepsilon$-pole and finite remainders of the one-loop
matrix elements have been tested phase-space-point by phase-space-point
against those provided by \openloops, with machine-precision accuracy for arbitrary $\mur$.
The $\rI$ term has additionally been checked
against the implementation of the
Catani--Seymour $\mathbf{I}$-operator automatically generated by \matchbox,
and the poles in the dimensional regulator $\varepsilon$
have been verified to cancel against those within $\rV$
rendering the sum IR-finite.
The virtual matrix elements may be isolated for validation purposes
within the \krknlo implementation
by setting the shower cut-off $t_0$ sufficiently high as to
ensure $t_0 > \max_{\Phi_m} Q(\Phi_m)$, thus prohibiting any parton-shower
radiation.
The radiative phase-space is then empty and the
shower algorithm is effectively bypassed,
giving a Sudakov factor identically equal to 1 for all events
(i.e.\@ the $\order{\alphas}$ contribution 
$\Delta^{(1)} \bigr\vert_{\ptcut}^{Q(\Phi_m)} (\Phi_m)$,
and all higher-order contributions, vanish).
By enabling the $\rV$ and $\rI$ contributions
and disabling the Born and $\Delta_0^{\rFS}$ contributions,
 as described in \cref{sec:imp_krknloH7},
 the components contributing to $\calO(\Phi_m)$ in 
\cref{eq:nlomatching_notationdefinitions_nlomatching_dsigmaqqb_krknlo_1_singlet}
can be tested numerically
at the level of differential cross-sections
against those calculated using the automated \matchbox implementation within \herwig,
using one-loop matrix elements as given by either \openloops or \madgraph
(or any other supported one-loop provider).
A selection of plots indicating excellent agreement is provided
in \cref{fig:validation_virtual}, \cref{sec:appendix_validation}.

\subsubsection{Krk factorisation scheme}
\label{sec:imp_validation_FS}

PDFs generated by the convolution code used for \cite{Jadach:2015mza,Jadach:2016acv} 
have been checked against an independent implementation of the same convolution transformations in a separate code.
The Born differential cross-section convoluted with \krk-scheme PDFs has been verified
to agree numerically with the collinear contribution defined by the Catani--Seymour $\rP$ and $\rK$ operators%
\footnote{As discussed in \cref{sec:nlomatching_krkfs}, the contribution from the
	mass-factorisation terms given by the Catani--Seymour $\rP$ operator
    at $\muf = \hat{s}_{12}$ is in fact zero, but the \matchbox code explicitly
    calculates both operator contributions, as it would for any other scale choice.}
as implemented in \matchbox within \herwig, i.e.\@ verifying

{ \footnotesize
\begin{align}
	\label{eq:krkconvvalidation_krk}
	\sum_{f} f^{\krk}_{q_f} (\hat{s}_{12})
	 & \otimes_{\xi_1} 
  	\dd \Phi_m \;
    \frac{1}{2\hat{s}_{12}} \;
    \rB_{q_f \qbar_f}(\Phi_m) 
    \left(1 + \frac{\alphas({\hat{s}_{12}})}{2\pi} \Delta_{0}^{\rFS} \right) 
    \calO(\Phi_m)
 	\otimes_{\xi_2} 
 	f^{\krk}_{\qbar_f} (\hat{s}_{12})
 	\\ \label{eq:krkconvvalidation_msbar}
	& {} = 
	\sum_f
	f^{\msbar}_{q_f} (\hat{s}_{12})
	\otimes_{\xi_1} 
	\dd \Phi_m(\eta_1 \xi_1 P_1, \eta_2 \xi_2 P_2) \;
	\frac{1}{2\hat{s}_{12}} \;
	\rB_{q_f \qbar_f}(\Phi_m) 
	\\ \notag
	& \qquad \Biggl(1 
	+ \dd \eta_1 \, \dd \eta_2 \; \frac{\alphas(\hat{s}_{12})}{2\pi} \Biggl[
		\left(\rP(\muf) +  \rK^{\rFS}\right)_{qq}  \negthick (\eta_1) \; \delta^1_{\eta_2}
		+ \delta^1_{\eta_1} \,
		\left(\rP(\muf) +  \rK^\rFS\right)_{qq} \negthick (\eta_2)
		\Biggr] \Biggr) \; 
	\calO(\Phi_m)
	\\ \notag
	& \qquad \qquad
	\otimes_{\xi_2} 
	f^{\msbar}_{\qbar_f} (\hat{s}_{12}) 
	 + \order{\alphas^2}
\end{align}
}

Plots testing the validity of this relationship numerically are shown in 
\cref{fig:validation_PK} of \cref{sec:appendix_validation}.
This is a non-trivial cross-check of the interplay between the numerical
pre-convolution used for the calculation of the \krk-scheme PDFs, and
the equivalent numerical convolution operators used in \matchbox.
Overall there is good agreement with a difference of 5--10\% emerging at
high-$\Mgg$.  Within predictions made by the \krknlo method this may
be considered as a component of the overall `missing-higher-order'
perturbative uncertainty.

\section{Analysis of NLO matching uncertainty}
\label{sec:matchingcomparisons}

In this section we explore the consequences of using alternative matching methods for precision phenomenology.
As outlined in \cref{sec:nlomatching}, NLO accuracy is achieved by several different methods
which, even using the same NLO hard-process ingredients
and the same parton shower algorithm with the same parameters,
will nevertheless give predictions which differ.
These differences arise formally at a higher-order in the perturbative expansion,
but are not guaranteed to be numerically small.
Understanding the magnitude of this `matching uncertainty' is a central requirement for the application of 
matched NLO + parton shower predictions to precision phenomenology.
We focus on the similarities and differences between \mcatnlo and \krknlo.

Prior studies comparing alternative matching methods have focused primarily on neutral-current Drell--Yan
and related processes involving the production of a massive vector boson or other massive final-states
\cite{Nason:2006hfa,Hamilton:2008pd,Alioli:2008gx,Hoeche:2011fd,Jones:2017giv,Cormier:2018tog,Jager:2020hkz,FebresCordero:2021kcc}%
.
The boson mass scale and the associated resonance entering into such processes inevitably
dominate their phenomenology, making a clean comparison of matching and shower effects difficult.
We choose to focus instead upon the diphoton production process,
as the simplest Standard Model colour-singlet process without
an intrinsic mass-scale and lacking a vector-boson resonance.

Concretely, throughout this section we present results with fiducial cuts close%
\footnote{Most notably, the rapidity interval $1.37 \leqslant \absyg \leqslant 1.52$, corresponding
	to the transition region between the barrel and end-cap calorimeters, is excluded from the \atlas
	fiducial cuts (summarised in \cref{eqn:ATLAScuts}) but is included here.}
to those used by \atlas for LHC Run II at 13 TeV \cite{ATLAS:2021mbt}:
\begin{subequations}
	\label{eqn:pseudoATLAScuts}
	\begin{align}
		\ptg{1} &> 40 \;\GeV \,, 	&	\ptg{2} &> 30 \;\GeV \,,  \\
		\dRgg &> 0.4 \,, & \absyg &\in \left[ 0, 2.5 \right),\hphantom{\cup \left( 1.52, 2.37 \right)} \\
		\Etiso (r) &< 0.1 \, \ptg{} \; \chi(r; R)  & \text{ within cone } r &\leqslant R = 0.4
	\end{align}
\end{subequations}
and generator cuts
\begin{align}
	\label{eqn:rungeneratorcuts}
	\hphantom{\Etiso (r)} \negphantom{\ptg{}}
	\ptg{} &> 5 \;\GeV \, , 
	\hphantom{ 0.1 \, \ptg{} \; \chi(r; R) } \negphantom{30 \; \GeV \, ,} 
	& 
	\hphantom{\text{ within cone } r} \negphantom{\absyg}
	\absyg &< 25 \negphantom{25} \hphantom{\left[ 0, 2.5 \right), \cup \left( 1.52, 2.37 \right)}
\end{align}
consistently for all predictions.%
\footnote{The relatively low choice of generator cut for $\ptg{}$ is required for \krknlo as it governs
	access to the real-emission phase-space by `migration' upon the generation of the first parton-shower emission.  Simultaneously, however, this choice leads to the large deviation of the `power'-shower \mcatnlo prediction from the others, due to
	its large generator-cut dependence (attributable to larger migration in photon-$\pt$ arising from the emission of harder jets). This can be tamed by a choice of $\ptg{}$-cut closer to the fiducial cut.
}

In place of the experimental photon isolation we use
smooth-cone (`Frixione') isolation \cite{Frixione:1998jh}
with the `tight' isolation parameters from the 2013 Les Houches Accords \cite{Andersen:2014efa},
which corresponds to
\begin{align}
	\chi(r; R) 
	= \left(  \frac{1 - \cos r}{1 - \cos R}  \right)
	\equiv \left( \frac{\sin \frac{1}{2} r}{\sin \frac{1}{2} R}  \right)^2
\end{align}
where $\Etiso (r)$ is the cumulative transverse isolation energy within 
(rapidity-azimuth) radius $r$,
calculated
as the transverse magnitude of the total momentum of all
non-photon particles within a cone of radius $r$.

Results corresponding to the exact experimental fiducial cuts (summarised in \cref{eqn:ATLAScuts})
 are presented in \cref{sec:pheno}.
Where jet distributions are presented we use the 
anti-$k_\rT$ algorithm \cite{Cacciari:2008gp} with clustering radius 0.4 and a
$\pt$ cut of 1 GeV.
 
Throughout we use \texttt{CT18NLO} PDFs \cite{Hou:2019efy}, either in the \msbar scheme
or transformed into the \krk scheme as described in \cref{sec:appendix_krkpdfs}.
Accordingly, we adopt $\alphas(M_Z) = 0.118$ as the input to the running of the strong coupling
throughout the hard process, shower, and the \krknlo code.
We use the \herwigseven default dipole-shower cut-off scale $\ptcut = 1 \, \GeV$.
Within \herwig, we disable 
both hadronisation and the \texttt{RemnantDecayer} so the final-state of the hard-process is the only
source of final-state QCD partons and the hard-process is the only input into the parton shower
initial-conditions.

We consistently adopt the renormalisation and factorisation scales
\begin{align}
	\mur(\Phi_m) = \sqrt{\hat{s}_{12}} & {} \equiv \Mgg = \muf(\Phi_m) \\
	\mur(\Phi_{m+1}) & {} = M_{\gamma\gamma} = \muf(\Phi_{m+1}).
\end{align}
Note that as discussed in \cref{sec:nlomatching_krknlo}, within the
\krknlo method, the factorisation scale to be used is fixed by the choice of 
convolution defining
the PDF transformation from the \msbar to the \krk factorisation scheme.
This motivates our adoption of this scale consistently for both \krknlo and \mcatnlo.
Within the shower we use the scale $\mu^{(\alpha)} = \lVert \mathbf{k}_\rT^{(\alpha)} \rVert$,
the transverse-momentum of the generated splitting relative to the emitter-spectator dipole,
for both the ratios of PDFs and for the running of $\alphas$.

For \mcatnlo we consider several comparators with varying choices of shower starting scale,
concretely:
\begin{itemize}
	\item a `power-shower' with $Q(\Phi_m) = Q_{\max{}}(\Phi_m)$ and $Q(\Phi_{m+1}) = Q_{\max{}}(\Phi_{m+1})$;
	\item a `default' shower with $Q(\Phi_m) = \sqrt{\hat{s}_{12}} \equiv \Mgg$ and $Q(\Phi_{m+1}) = \ptj{1}$, and 
	\item a `DGLAP-inspired' choice in which the shower starting-scale consistently matches the factorisation scale, here $Q(\Phi_m) = M_{\gamma\gamma}$ and $Q(\Phi_{m+1}) = M_{\gamma\gamma}$.
\end{itemize}
For the \krknlo method, the shower starting-scale is fixed to $Q_{\max{}}$ as discussed in \cref{sec:nlomatching_krknlo}.
Note that the `power-shower' choice for \mcatnlo is generally not recommended \cite{Bellm:2016rhh},
and is included to enable a direct comparison with \krknlo.

In \cref{sec:matchingcomparisons_1em} we consistently truncate the shower
in the $\Phi_{m+1}$ phase-space (i.e.\@ for diphoton production, $\Phi_3$);
in the terminology of \mcatnlo, this corresponds to `H'-events with no shower emissions
and `S' events with up to one shower emission.
At this point the matching between the hard-process and the shower is complete
and the subsequent evolution of each event is handled entirely by the parton-shower
algorithm, which is the same in both cases.%
\footnote{This is subject to the caveat that the shower starting-scale for the
	first `post-matching' emission, $Q(\Phi_{m+1})$ within \mcatnlo,
	may not match the shower starting-scale for the continuation of the shower, $\ptof{1}$ within \krknlo.}

In \cref{sec:matchingcomparisons_fullshower} we allow the shower to run
to its final cut-off scale, fully populating the emission phase-space.
The same predictions, with the unsimplified \atlas cuts (and combined with the formally-NNLO gluon-gluon box contribution) are compared to experimental data
in \cref{sec:pheno}.

\subsection{First-emission only}
\label{sec:matchingcomparisons_1em}

As discussed in \cref{sec:nlomatching}, the first-emission truncation of alternative matching schemes
allows their effect to be compared at the level of the initial conditions
each method supplies to the parton shower for subsequent evolution
(this is often referred to as a `parton-level' comparison, e.g.\@ in \cite{Hoeche:2011fd}).
In this section we augment the analytical comparison of
\cref{sec:nlomatching_krkvsothers}
between the one-emission-truncated \mcatnlo and \krknlo methods with a
numerical comparison.

In the one-emission case, the scale for the emission of the first shower emission from `H'-events, $Q(\Phi_{m+1})$,
does not enter the calculation,
so the second (`default') and third (`DGLAP')
choices outlined above are formally identical.
We therefore combine them here.

\begin{figure}[p]
	\centering
	\includegraphics[width=.4\textwidth]{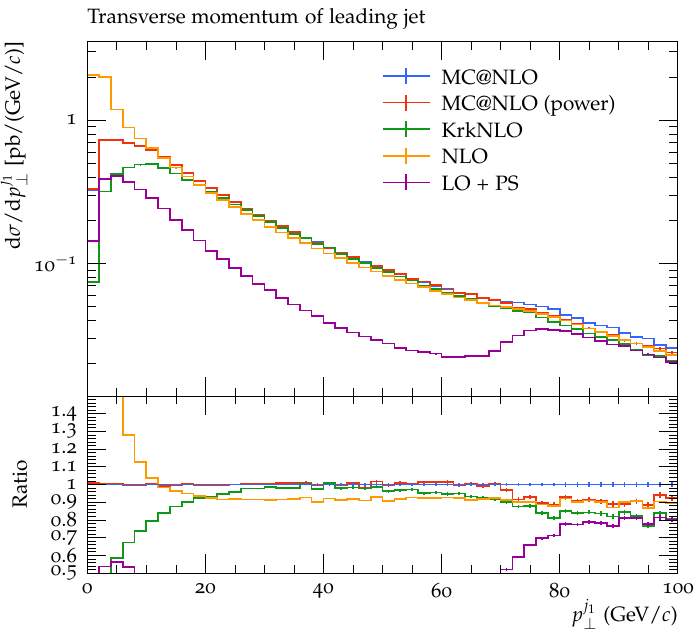}
	\qquad
	\includegraphics[width=.4\textwidth]{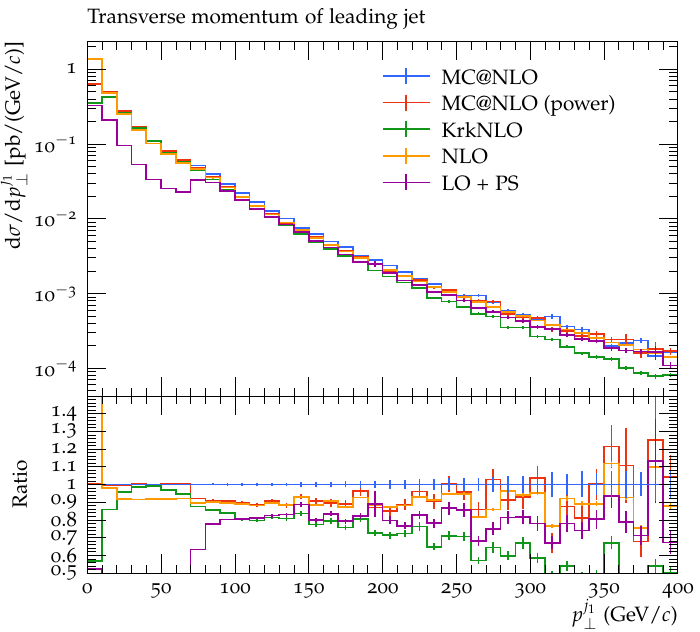}
	\includegraphics[width=.4\textwidth]{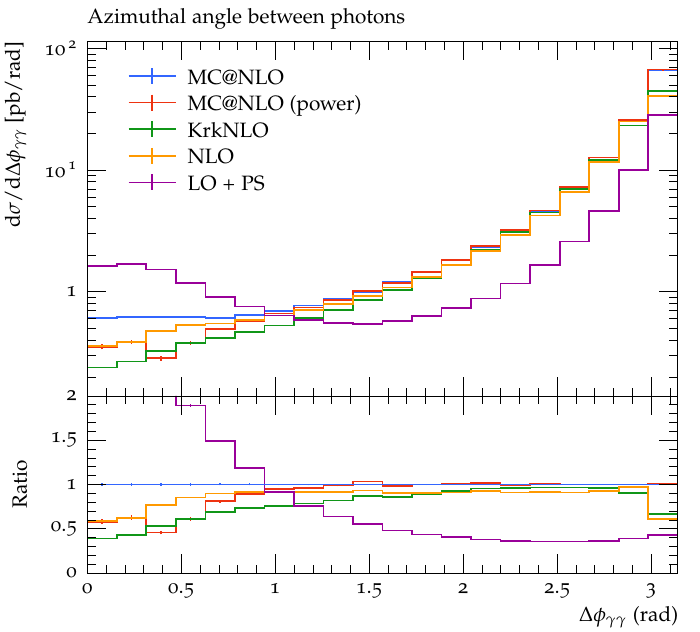}
	\qquad
	\includegraphics[width=.4\textwidth]{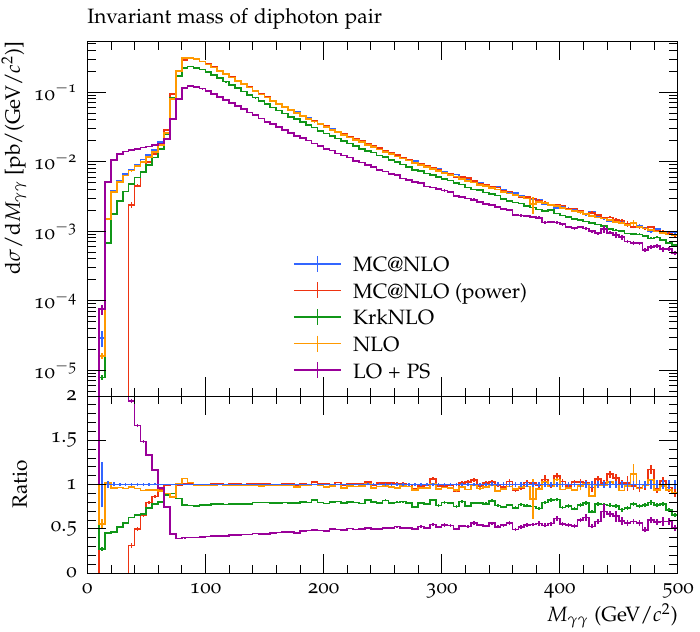}
	\includegraphics[width=.4\textwidth]{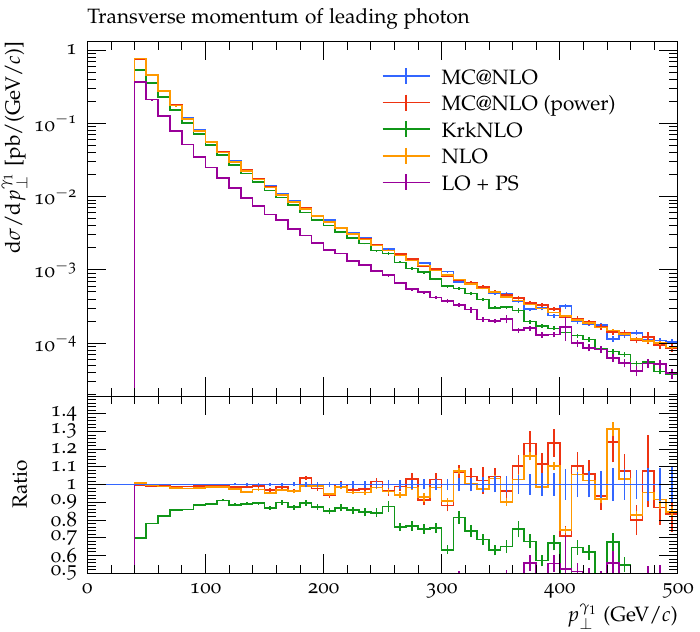}
	\qquad
	\includegraphics[width=.4\textwidth]{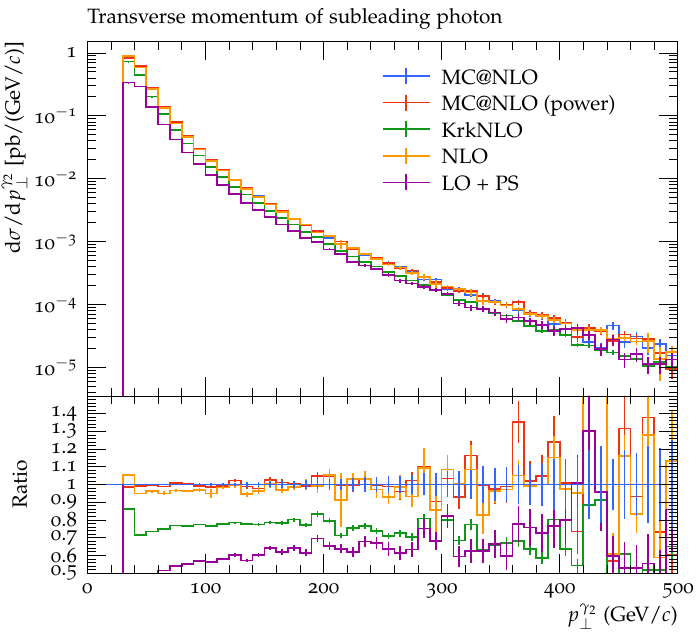}
	\includegraphics[width=.4\textwidth]{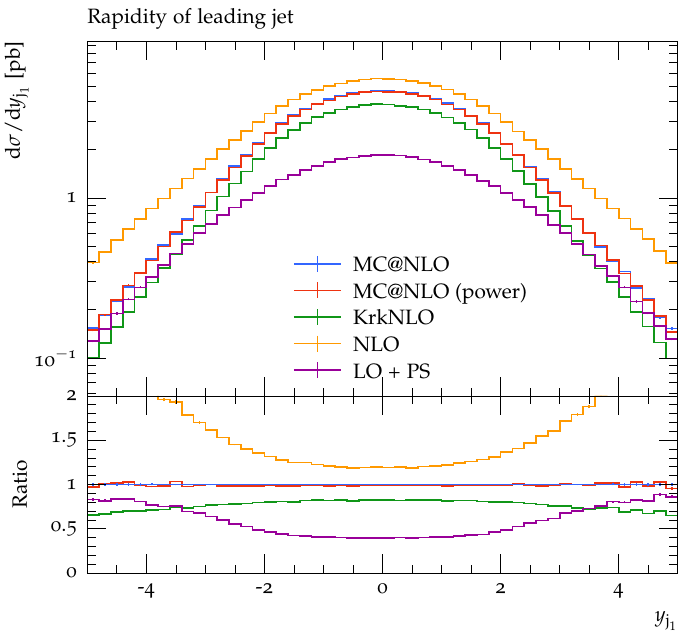}
	\qquad
	\includegraphics[width=.4\textwidth]{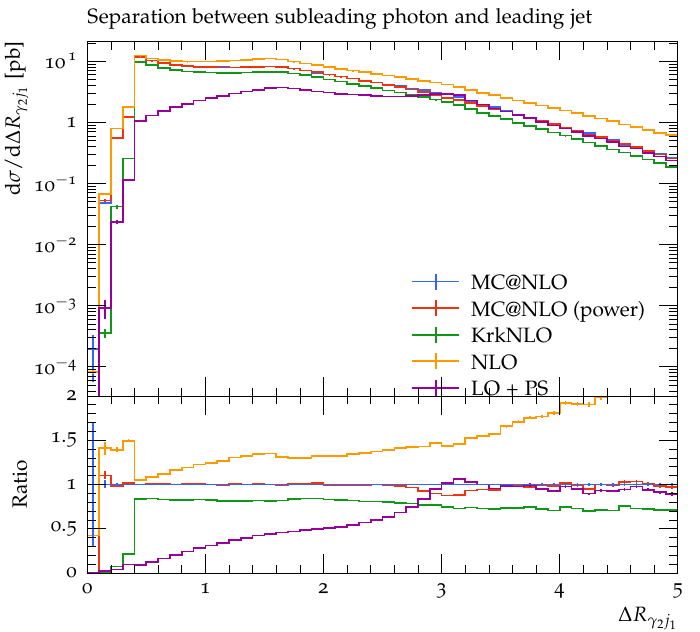}
	\caption{
		`Parton level' (first-emission) comparison of KrkNLO with MC@NLO, NLO fixed-order,
		and the corresponding first-emission distributions generated by the
		parton shower from a leading-order calculation.
		The shower in each case is a `power shower', i.e.\ with no phase-space restrictions.
		\label{fig:matchingcomparisons_1em}}
\end{figure}

We turn concretely to the distributions shown
in \cref{fig:matchingcomparisons_1em}.
As expected from the reasoning of \cref{sec:nlomatching_krkvsothers},
the predictions converge
with the fixed-order calculation
in the large-$\ptj{1}$ limit where the Sudakov factor is small and the
contribution to $\calO(\Phi_{m+1})$
is dominated in both cases
by the real-emission matrix-element.
This is a desired feature of an NLO matching method
(and stands in contrast with the behaviour of \powheg
due to its exponentiation of $\rR/\rB$
in the hard region
\cite{Hoeche:2011fd}).
In the low-$\ptj{1}$ region, we see the
expected additional
Sudakov suppression of the real-emission matrix-element
within \krknlo relative to \mcatnlo.
This leads to a substantial difference in the
amount of soft radiation
generated in close-to-Born kinematic configurations.

\begin{figure}[p]
	\centering
	\subcaptionbox{The transverse-momentum distribution of the hardest jet (here, also parton), $\dd \sigma / \dd \ptj{1}$.
		\label{fig:matchingcomparisons_1em_ptj_dphi}}[\textwidth]{

\makebox[\textwidth][c]{
		\includegraphics[width=.36\textwidth]{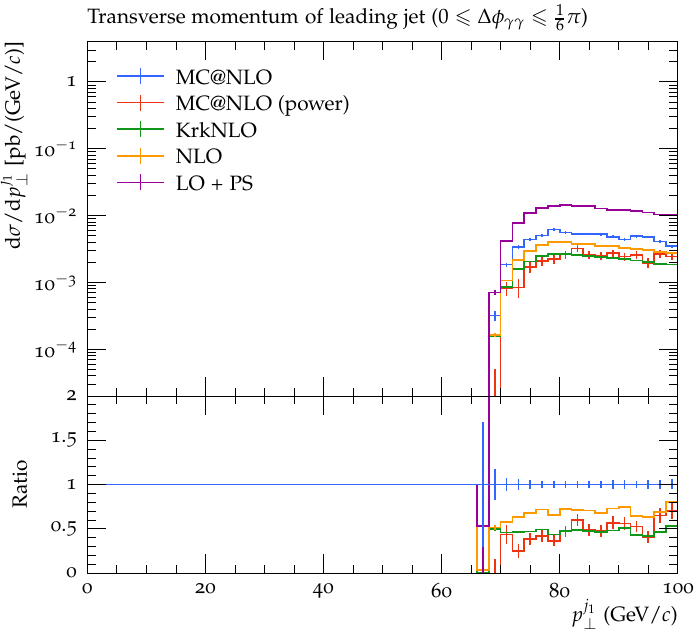}
		\includegraphics[width=.36\textwidth]{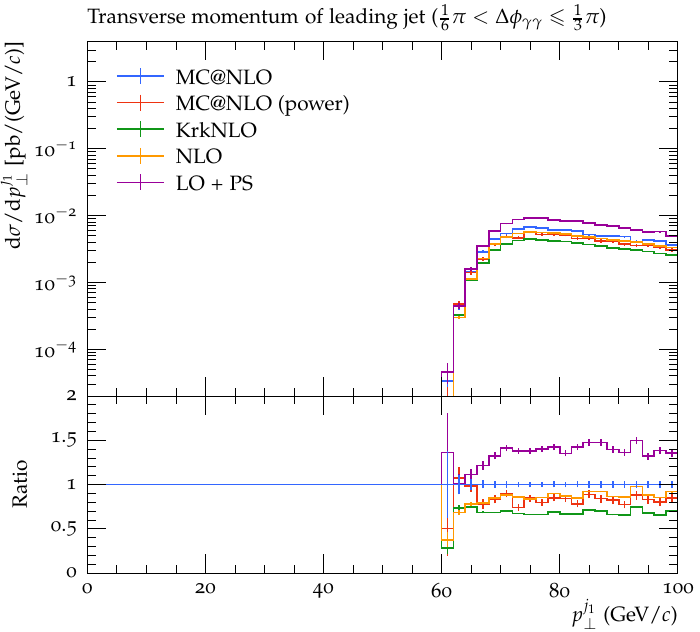}
		\includegraphics[width=.36\textwidth]{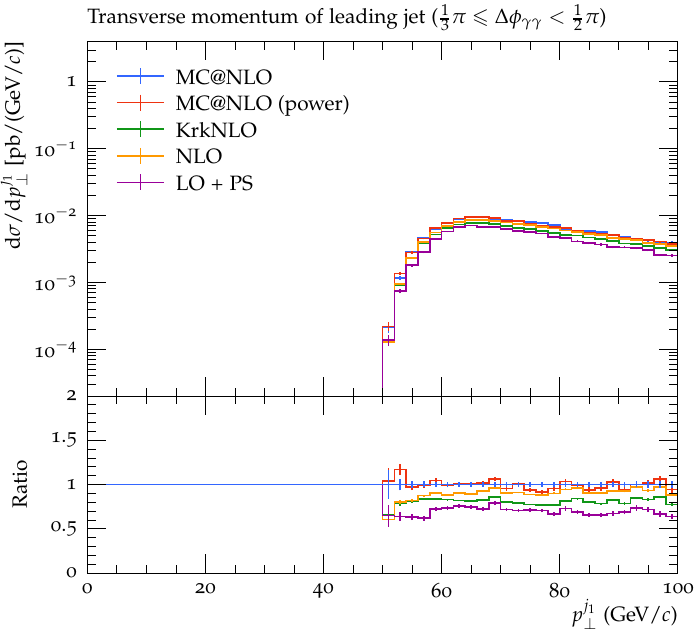}
  }
\makebox[\textwidth][c]{
		\includegraphics[width=.36\textwidth]{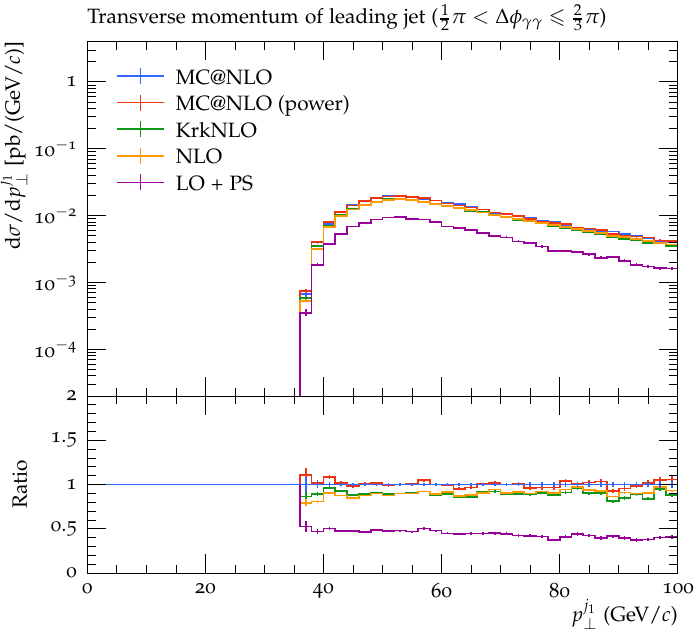}
		\includegraphics[width=.36\textwidth]{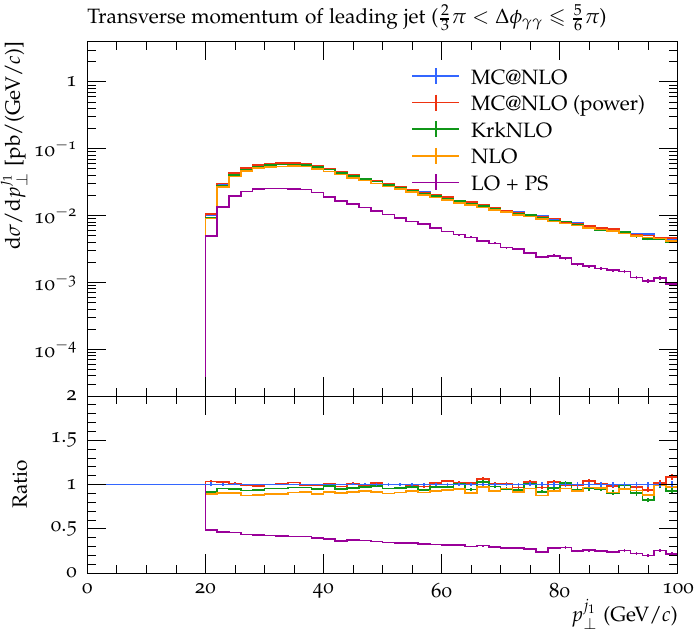}
		\includegraphics[width=.36\textwidth]{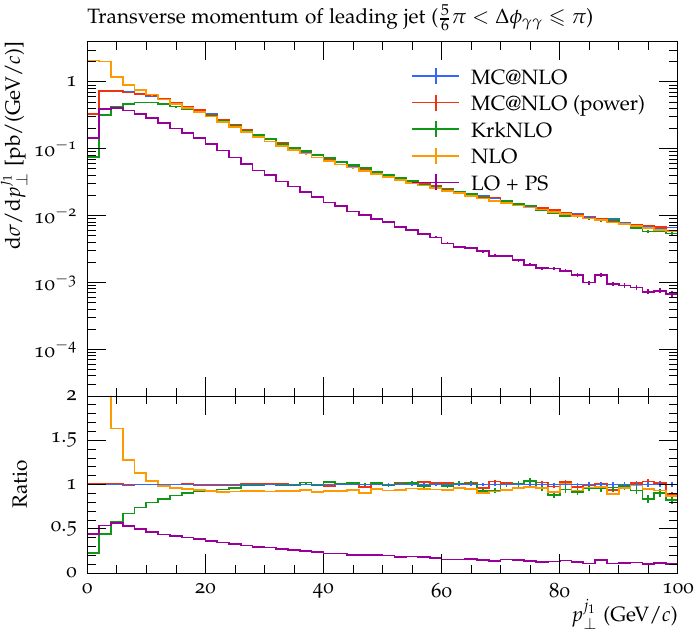}
	}
 }
	\vspace{1cm}
	\\
	\subcaptionbox{The invariant mass distribution of the diphoton pair, $\dd \sigma / \dd \Mgg$.
		\label{fig:matchingcomparisons_1em_mgg_dphi}}[\textwidth]{

\makebox[\textwidth][c]{
		\includegraphics[width=.36\textwidth]{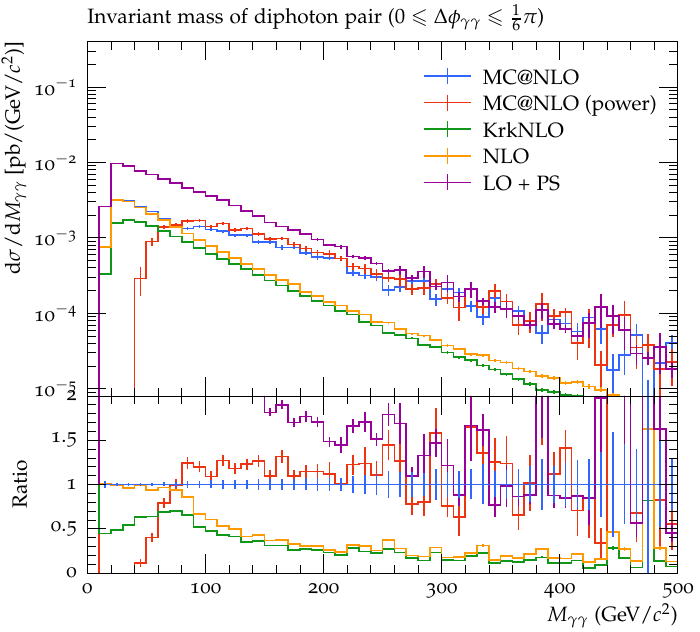}
		\includegraphics[width=.36\textwidth]{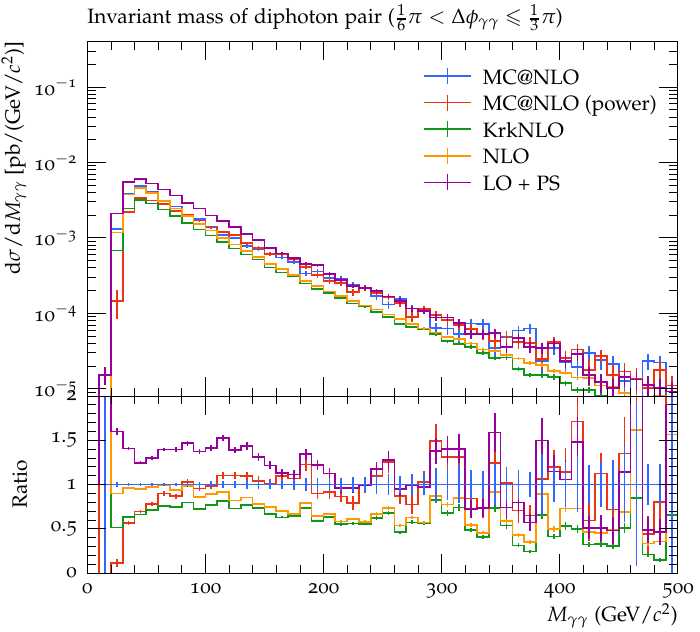}
		\includegraphics[width=.36\textwidth]{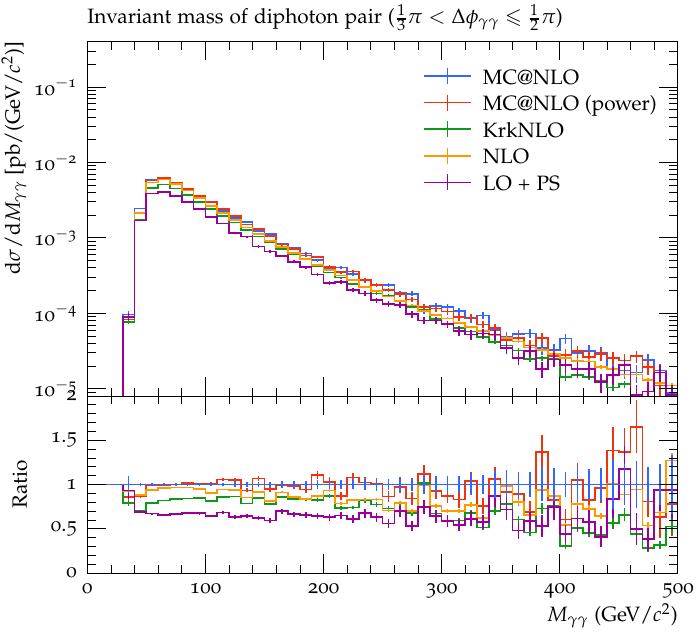}
  }
\makebox[\textwidth][c]{
		\includegraphics[width=.36\textwidth]{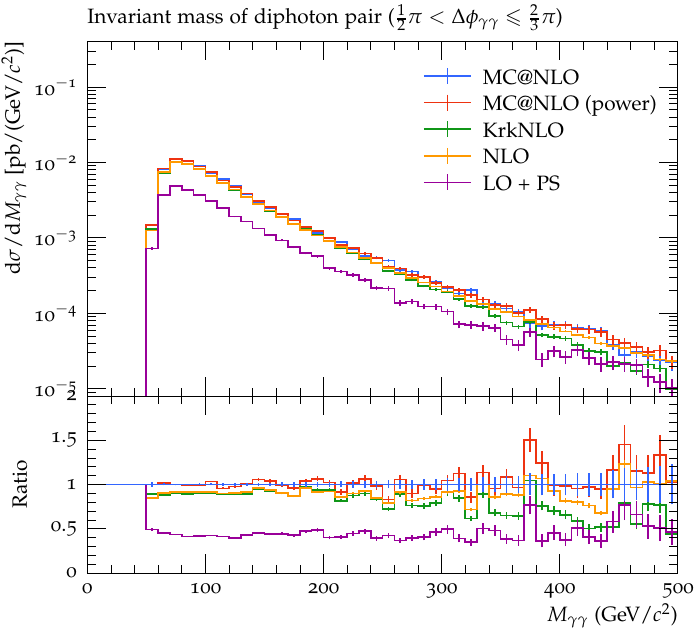}
		\includegraphics[width=.36\textwidth]{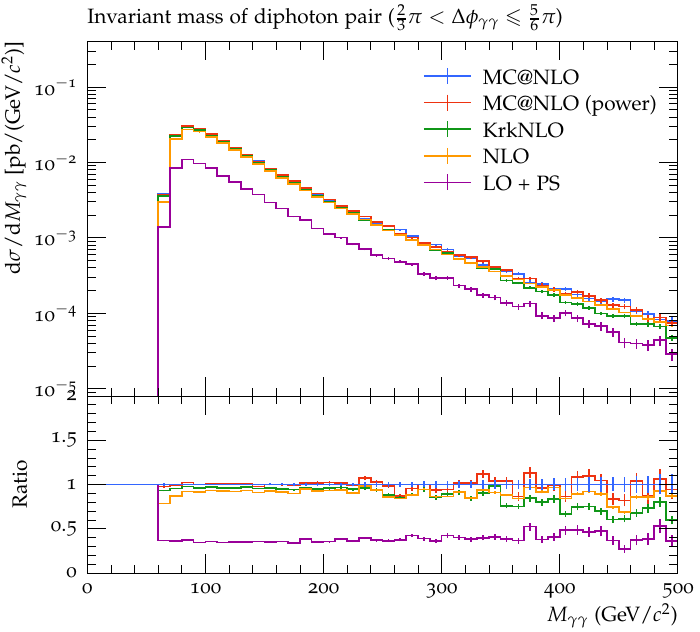}
		\includegraphics[width=.36\textwidth]{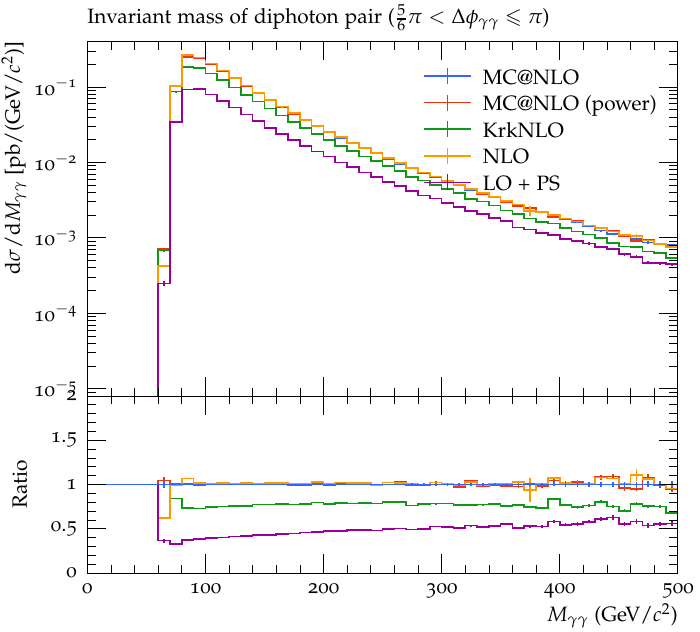}
  }
	}
	\caption{
		`Parton level' (first-emission) comparison of
		differential cross-sections divided into six equal bins of $\dphigg$,
		generated by \krknlo, \mcatnlo, NLO fixed-order,
		and the corresponding first-emission distributions generated by the
		parton shower from a leading-order calculation.
		\label{fig:matchingcomparisons_1em_ptj_mgg_dphi}}
\end{figure}

This effect is isolated double-differentially in \cref{fig:matchingcomparisons_1em_ptj_dphi},
broken into six equal slices of $\dphigg$.
Born-like kinematic configurations require $\dphigg = \pi$,
while small values of $\dphigg$ are only possible
if the diphoton system recoils against a hard jet.
The predictions can be seen to agree well in the central four $\dphigg$ bins.
The slight deviations at high-$\ptj{1}$ can be seen to arise from the
low-$\dphigg$ slices, in which the diphoton system recoils against a jet,
gradually becoming kinematically possible; within each slice the relationship
between the predictions is fairly stable, but the compositional effect 
changes the overall ratio in \cref{fig:matchingcomparisons_1em}.
The low-$\ptj{1}$ Sudakov suppression shown in \cref{fig:matchingcomparisons_1em} arises solely from the close-to-Born configuration in the final $\dphigg$ slice, which is the dominant contribution and in which the photons are approximately back-to-back.
This is the only slice in which soft gluon radiation is kinematically permitted
and it is in this slice that the behaviour of the Sudakov factor at low-$\ptj{1}$ drives the deviation between \krknlo and \mcatnlo.

A similar pattern can be seen in \cref{fig:matchingcomparisons_1em_mgg_dphi}, in which
once again there is generally good agreement between the methods in the central four $\dphigg$ bins.
In this case the dominant difference arising from the Sudakov factor in the low-$\ptj1$ region 
for $\dphigg \approx \pi$
is smeared across the $\dd \sigma / \dd \Mgg$ distribution,
leading to an apparent normalisation difference whose
origin is nevertheless again the Sudakov factor which
accompanies $\rR$ in the \krknlo method.
The differences between the \mcatnlo predictions arising from the change
in shower starting-scale $Q(\Phi_m)$ can also be seen in the lowest $\dphigg$ bin;
the effect arises through both the emission-scale theta-function and the Sudakov factor in 
\cref{eq:nlomatching_notationdefinitions_nlomatching_dsigmaqqb_mcatnlo_1_singlet,eq:nlomatching_notationdefinitions_nlomatching_dsigmaqg_mcatnlo_1_singlet}. This discrepancy is sensitive to the choice of generator cuts in \cref{eqn:rungeneratorcuts} and may in practice be tamed by the use of more restrictive generator cuts.
 
This illustrates the general pattern, which is that for inclusive single-differential distributions, the differing behaviour of the predictions for low
$\ptof{1}$ will not be directly observed, manifesting instead as a normalisation difference in regions of phase-space 
which correspond to a close-to-Born configuration if one exists, or a global
normalisation difference if not.
This can be seen in the remaining distributions of \cref{fig:matchingcomparisons_1em}, in particular
in the $\dd \sigma / \dd y_{j_1}$ distribution.

\subsection{Full shower}
\label{sec:matchingcomparisons_fullshower}

We turn now to the case of direct relevance to phenomenology, in which the parton shower is allowed
to run to completion and populate the full phase-space.
We again
consider the same `power-shower', `default' and `DGLAP'
showers as described above in the \mcatnlo
case, to understand how the matching uncertainty
between the alternative matching schemes
relates to the matching uncertainty
arising within \mcatnlo
due to the choice of the shower starting-scale $Q$.

\begin{figure}[p]
	\centering
	\includegraphics[width=.4\textwidth]{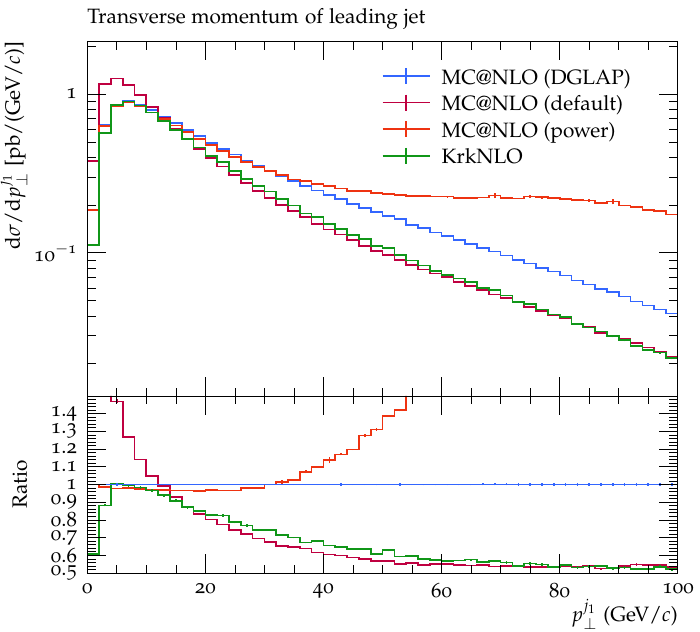}
	\qquad
	\includegraphics[width=.4\textwidth]{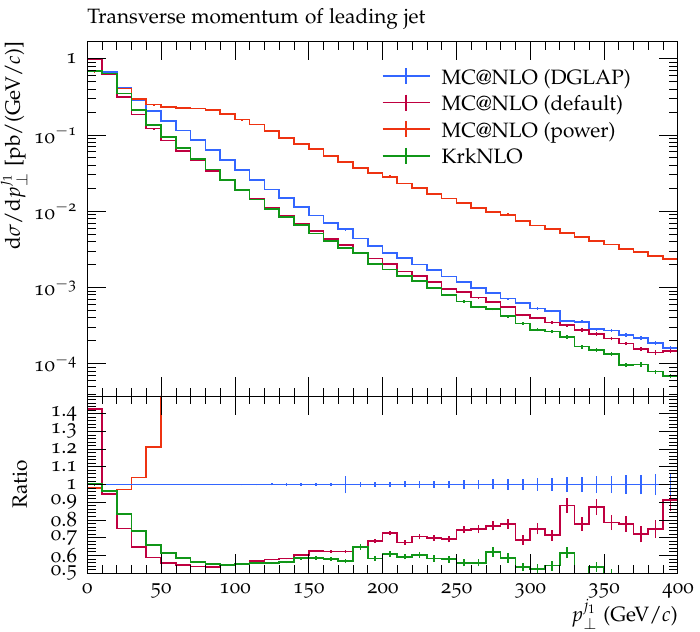}
	\includegraphics[width=.4\textwidth]{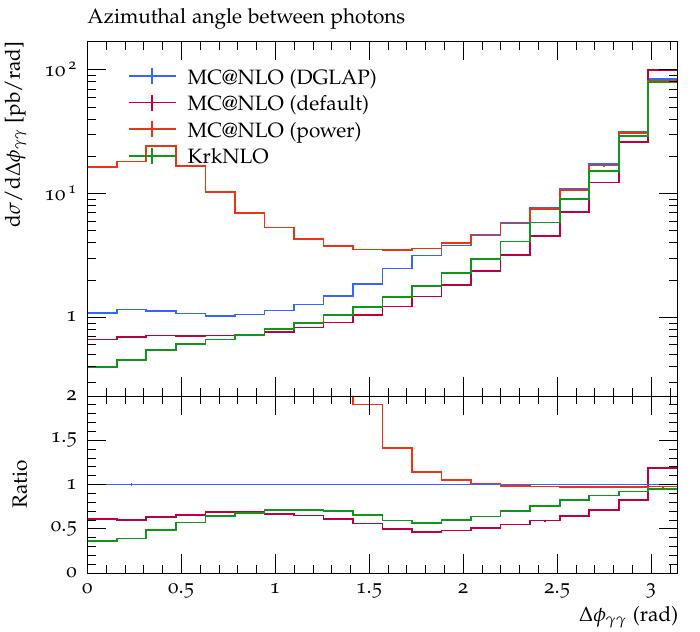}
	\qquad	\includegraphics[width=.4\textwidth]{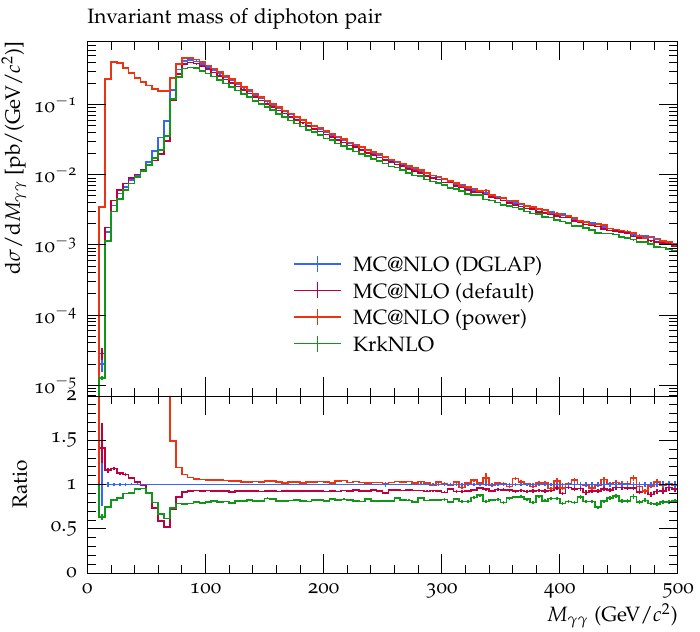}
	\includegraphics[width=.4\textwidth]{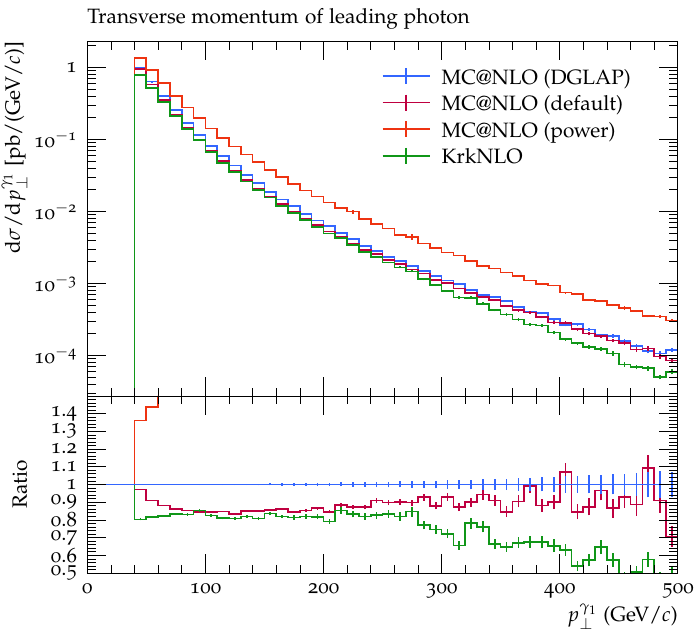}
	\qquad
	\includegraphics[width=.4\textwidth]{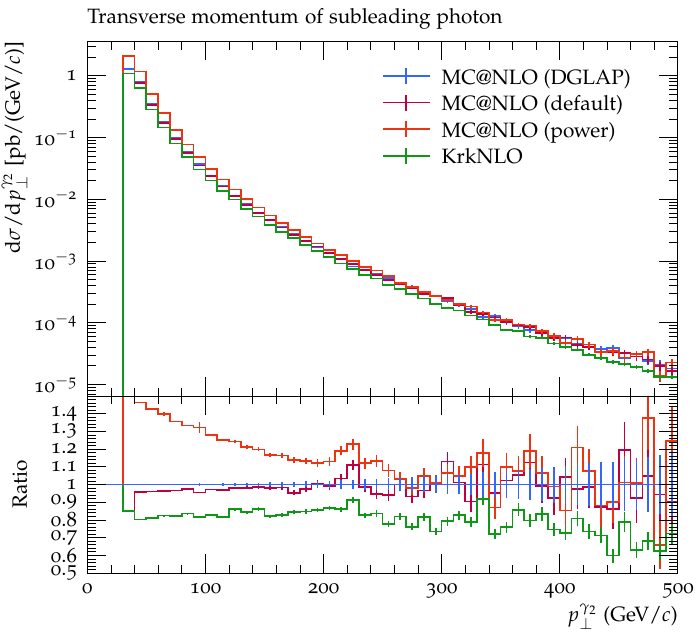}
	\includegraphics[width=.4\textwidth]{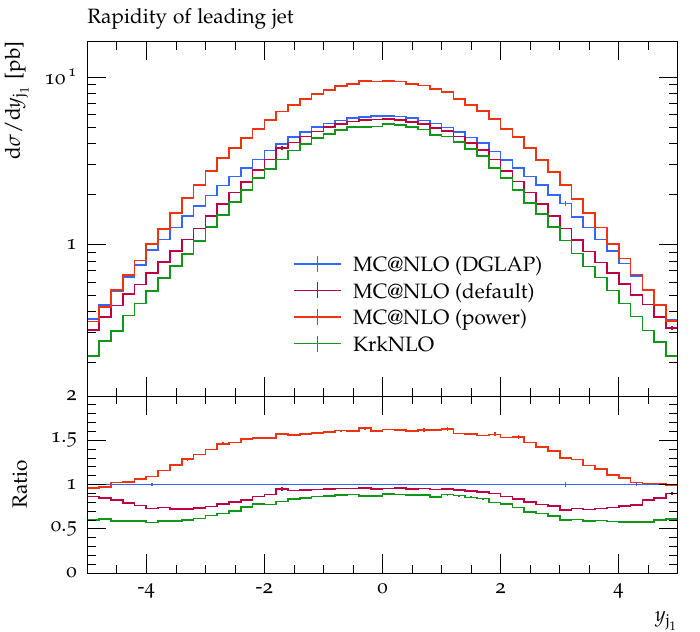}
	\qquad
	\includegraphics[width=.4\textwidth]{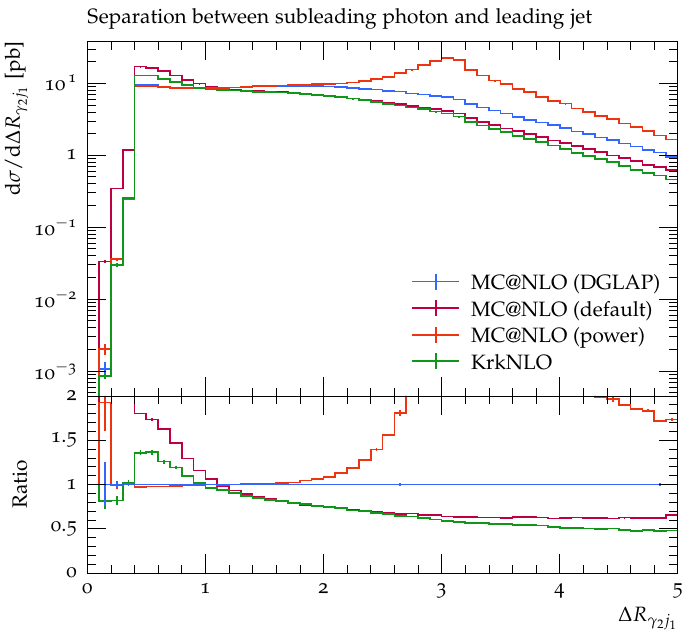}
	\caption{
		Comparison of matched
		differential cross-sections 
		generated by \krknlo and \mcatnlo with the `default', `power'-shower and `DGLAP' starting-scales.
		\label{fig:matchingcomparisons_fullshower}}
\end{figure}

In \cref{fig:matchingcomparisons_fullshower}
we see that the \krknlo prediction lies close to the
`default' \mcatnlo prediction.
At high-$\ptj1$ there is a large sensitivity to $Q(\Phi_{m+1})$,
but the \krknlo and `default' \mcatnlo choices have qualitatively
the same shape and agree moderately well.
The `power' shower diverges strongly from the other three
choices as has previously been observed in 
\cite{Bellm:2016rhh},
substantially favouring the emission of a hard jet relative 
to the other matching methods.
At lower-$\ptj{1}$ the \krknlo method again largely lies between the
\mcatnlo alternatives.

\begin{figure}[p]
	\centering
	\subcaptionbox{The transverse-momentum distribution of the hardest jet, $\dd \sigma / \dd \ptj{1}$.
		\label{fig:matchingcomparisons_fullshower_ptj_dphi}}[\textwidth]{
\makebox[\textwidth][c]{
		\includegraphics[width=.36\textwidth]{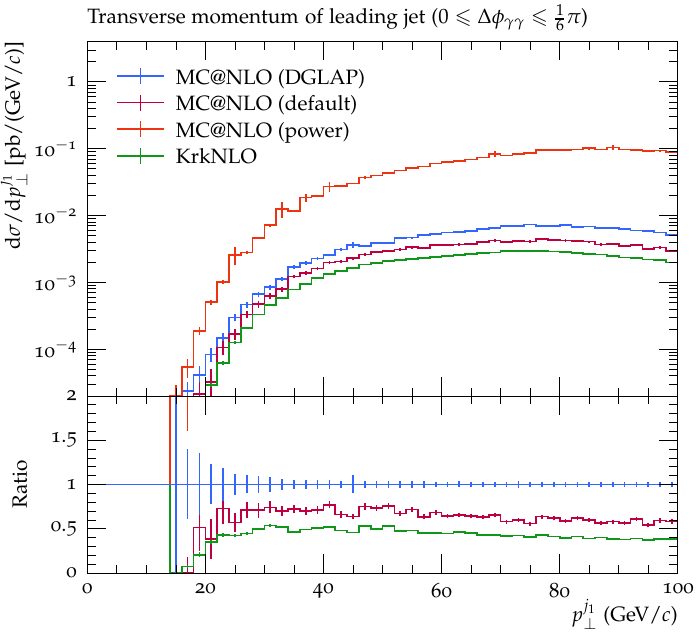}
		\includegraphics[width=.36\textwidth]{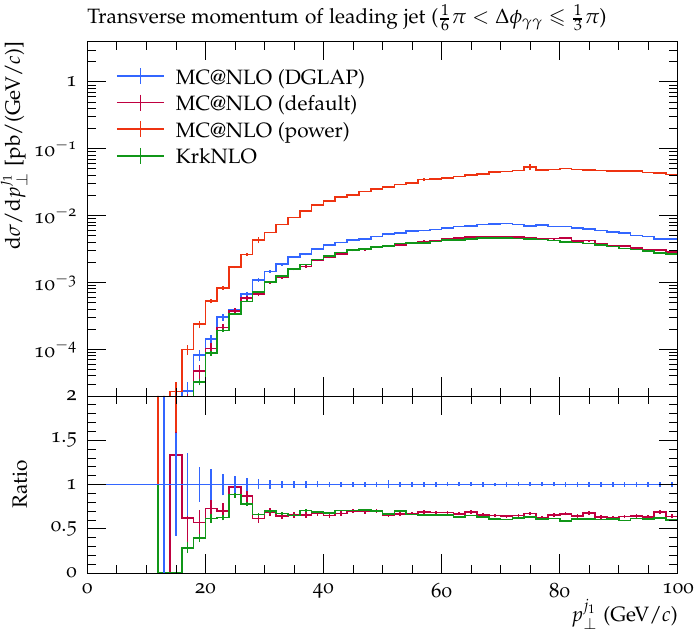}
		\includegraphics[width=.36\textwidth]{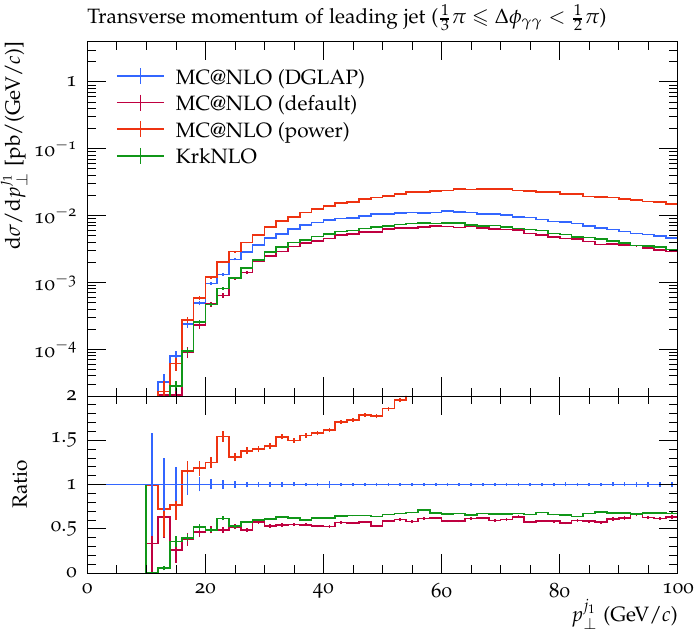}
  }
\makebox[\textwidth][c]{
		\includegraphics[width=.36\textwidth]{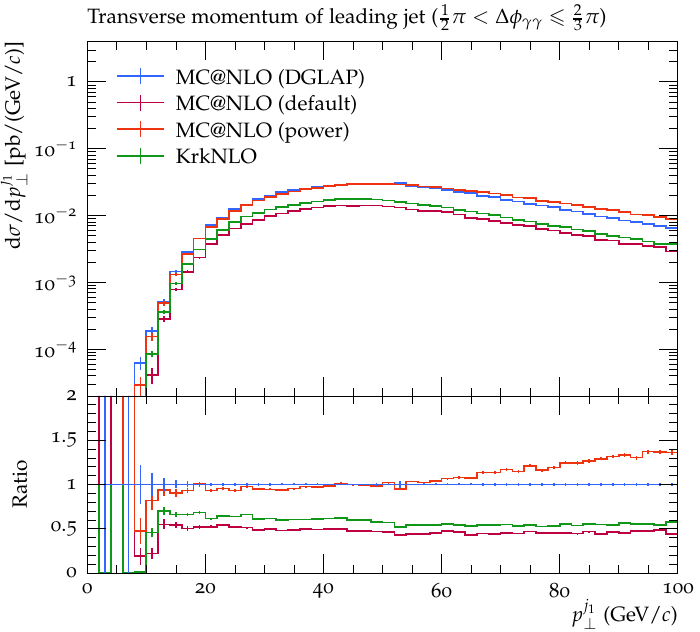}
		\includegraphics[width=.36\textwidth]{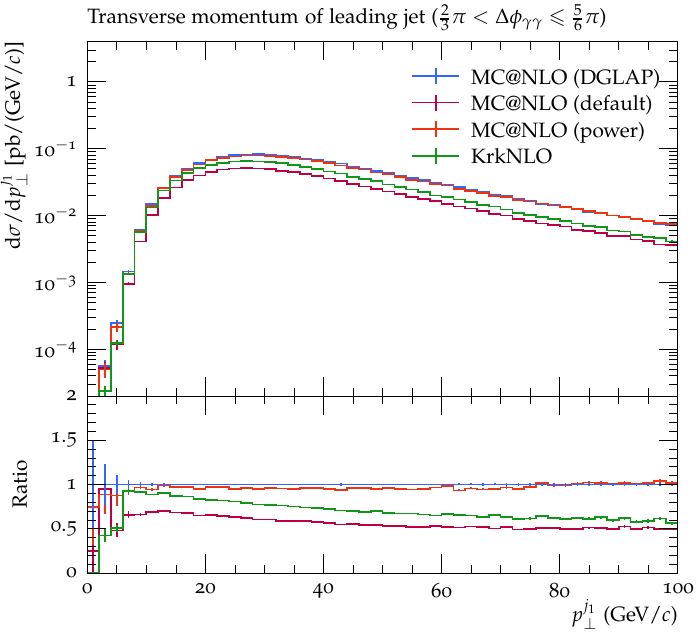}
		\includegraphics[width=.36\textwidth]{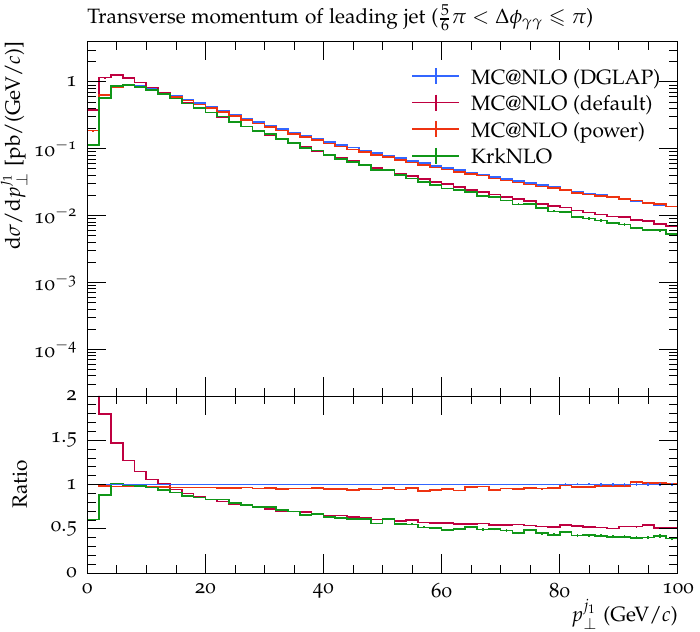}
  }
	}
	\vspace{1cm}
	\\
	\subcaptionbox{The invariant mass distribution of the diphoton pair, $\dd \sigma / \dd \Mgg$.
		\label{fig:matchingcomparisons_fullshower_mgg_dphi}}[\textwidth]{
\makebox[\textwidth][c]{
		\includegraphics[width=.36\textwidth]{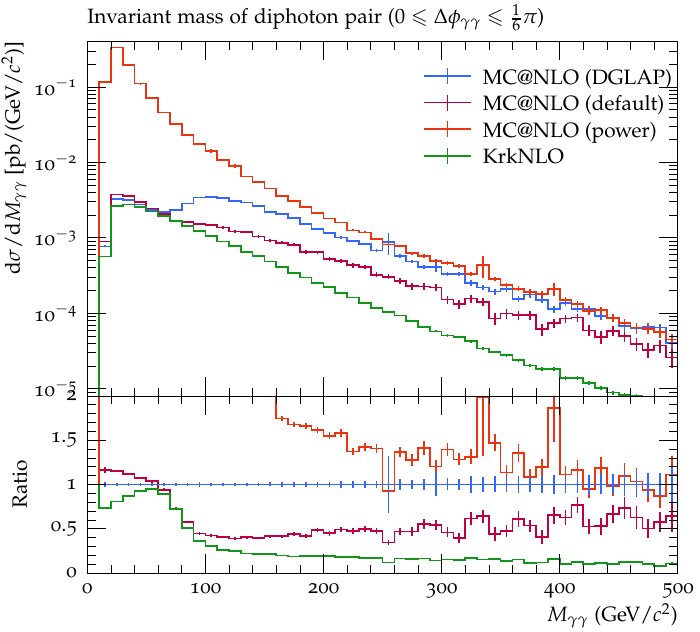}
		\includegraphics[width=.36\textwidth]{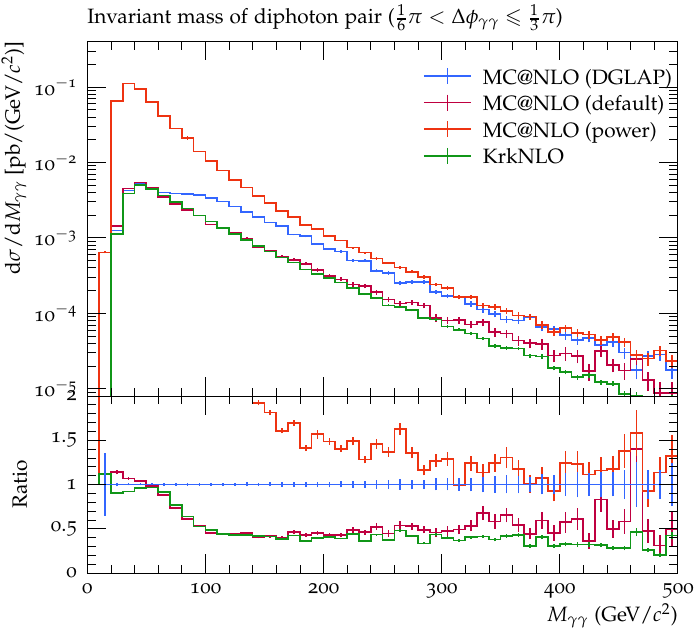}
		\includegraphics[width=.36\textwidth]{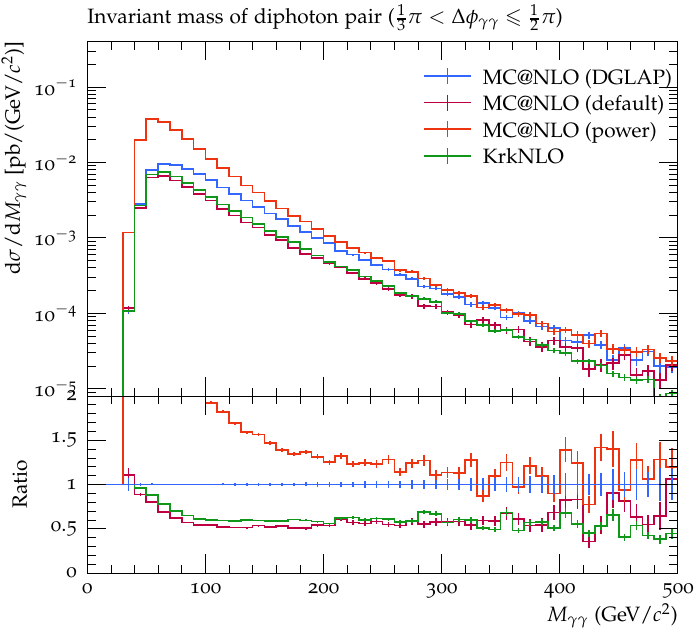}
  }
\makebox[\textwidth][c]{
		\includegraphics[width=.36\textwidth]{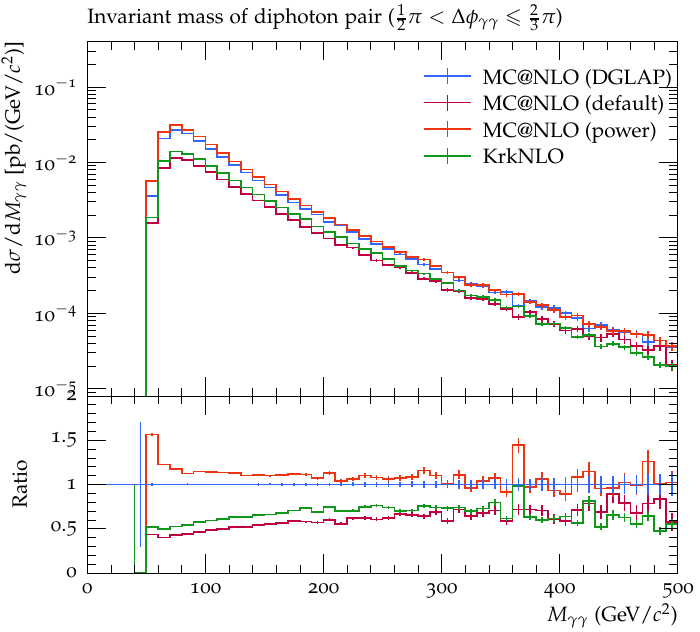}
		\includegraphics[width=.36\textwidth]{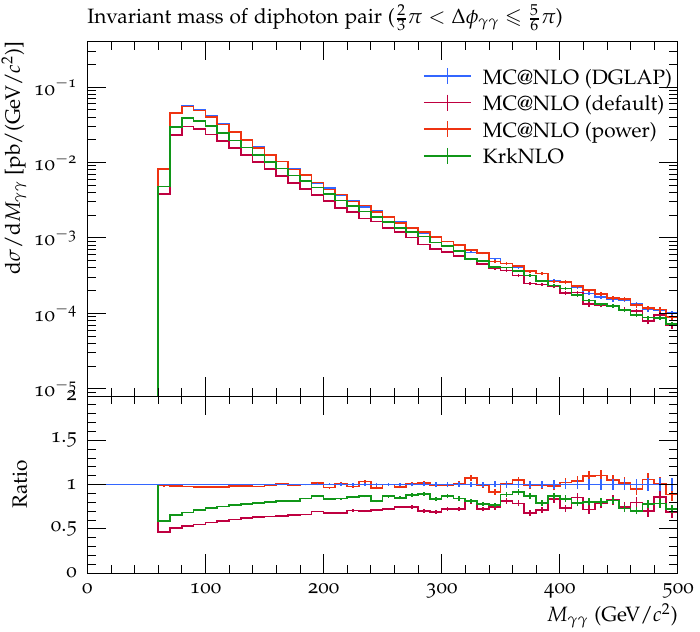}
		\includegraphics[width=.36\textwidth]{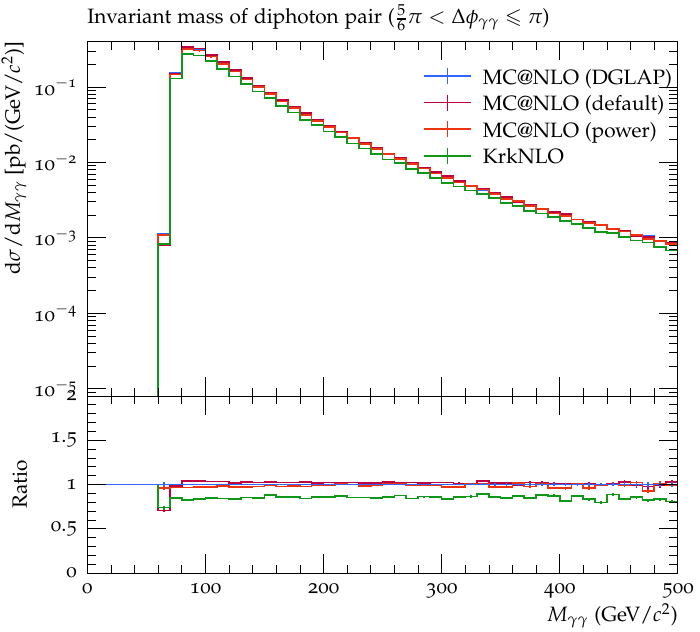}
  }
	}
	\caption{
		Comparison of matched
		differential cross-sections 
		divided into six equal bins of $\dphigg$,
		generated by \krknlo and \mcatnlo with the `default', `power'-shower and `DGLAP' starting-scales.
		\label{fig:matchingcomparisons_fullshower_ptj_mgg_dphi}
	}
\end{figure}

This region can be studied in further detail in \cref{fig:matchingcomparisons_fullshower_ptj_dphi}
where, as in \cref{sec:matchingcomparisons_1em},
it is further subdivided according to the value of $\dphigg$
into six equal slices.
As in the one-emission case, at low-$\ptj{1}$ the overall distribution
is dominated by the final slice $\dphigg \sim \pi$,
in which the two photons are approximately back-to-back.
In this region we again see the low-$\ptof{1}$ effect of the Sudakov factor
applied to $\rR$ in the \krknlo method, as in \cref{fig:matchingcomparisons_1em_ptj_mgg_dphi}.
In the \mcatnlo method, the Sudakov suppression of $\rR$ 
at low-$\ptj{1}$ is dominated by the 
Sudakov factor associated with the generation of the second emission,
$\Delta \bigr\vert_{\ptcut}^{Q({\Phi}_{m+1})}$,
as shown in \cref{eq:nlomatching_notationdefinitions_nlomatching_dsigmaqqb_mcatnlo_1_singlet,eq:nlomatching_notationdefinitions_nlomatching_dsigmaqg_mcatnlo_1_singlet}
and is therefore linked by unitarity to the phase-space available
for the generation of the second emission.
In this low-$\ptof{1}$ region, for the `default' scale, $ Q(\Phi_{m+1}) = \ptj{1} \sim \ptcut $ and so the Sudakov suppression of $\rR$
is negligible, whilst for the `power' and `DGLAP' choices, $ Q(\Phi_{m+1}) \gg \ptcut $ generating a Sudakov suppression
comparable to that of the \krknlo method arising instead from the 
generation of the first shower-emission from the underlying Born.
In the `default' case the divergence of the real-emission matrix-element is partially exposed by the miscancellation
between
the shower-approximation
term from the first-emission of the `S'-event,
which is suppressed by a factor of the Born Sudakov, here
\begin{align}
	\Delta \bigr\vert_{\ptof{1}}^{Q(\tilde{\Phi}_m^{\II_i})}
	(\tilde{\Phi}_m^{\II_i})
	 = \Delta \bigr\vert_{\ptof{1}}^{\Mgg}(\tilde{\Phi}_m^{\II_i}),
\end{align}
and therefore large, and the dipole-subtraction term from the `H'-event,
which has no such suppression.
Viewed differently, this limit exposes a sensitivity to the finite/non-singular
contributions from the dipole terms, which is rendered small when the `H'- and `S'-event Sudakov factors are similar in magnitude.

In other slices of \dphigg we see the only identifiable region of phase-space for which 
the \krknlo prediction lies outwith the uncertainty envelope
implied by the \mcatnlo predictions, $\dphigg \in [0, \frac{1}{6}\pi]$.

Inclusive observables such as $\Mgg$ are described
to NLO accuracy by all the matching set-ups, and might therefore
be expected to exhibit reduced matching uncertainty
relative to distributions described perturbatively only to leading-order.
Invariant-mass-type observables between colour-singlet particles are further
privileged by the momentum mappings used within the dipole
shower, which leave them unchanged.
Each event therefore retains the $\Mgg$ of the underlying phase-space point of its initial generation (whether
$\Phi_m$ or $\Phi_{m+1}$)
throughout the shower evolution.
The differences between the alternatives
seen inclusively, as in \cref{fig:matchingcomparisons_fullshower},
are unsurprisingly small.
In \cref{fig:matchingcomparisons_fullshower_mgg_dphi} we see the overall distribution is dominated by the final
$\dphigg$ slice in which the distributions agree well.
Viewed double-differentially in slices of $\dphigg$,
we can see the effect of the recoil from the additional radiation,
which distributes events to slices of $\dphigg$ according
to their subsequent shower evolution.
For soft radiation $\dphigg$ remains unchanged from its Born value,
$\dphigg \approx \pi$, and the alternative schemes agree.
Successively harder radiation probes the upper boundary
of the shower phase-space, 
in particular the difference between the `DGLAP' and `power'-shower alternatives which emerges only for $\dphigg \leqslant \frac{1}{2}\pi$.
Intuitively, the additional very-hard-radiation permitted by
the power-shower pushes more events into the lowest
$\dphigg$ bins than the alternatives.
Once again the \krknlo prediction lies within the 
uncertainty envelope of the various \mcatnlo predictions,
across phase-space, save for the lowest bin of $\dphigg$.

The distribution with respect to 
the transverse momenta of the two photons are shown in
\cref{fig:matchingcomparisons_fullshower}
and are also inclusive distributions, i.e.\@ are calculated 
to NLO accuracy.%
\footnote{With the exception of the region $\ptg{2}\in[30,40]\,\GeV$ which is kinematically prohibited
	at LO by the 2--2 kinematic constraint $\ptg1 = \ptg2$ and the 
$\ptg{1} > 40 \, \GeV$ cut on the hardest photon.}
Unlike \Mgg, these are sensitive to the transverse momentum of the
jet radiation, and like the jet the `power'-shower again favours larger
values for $\ptg{1}$ (against which the jet recoils in a photon-plus-jet type configuration).
Again, the \krknlo prediction lies close to the `default'-
and `DGLAP' \mcatnlo variants.
The matching uncertainty for $\dd \sigma / \dd \ptg{2}$ is considerably smaller,
as might be expected.

\begin{figure}[t]
	\centering
 
\makebox[\textwidth][c]{
	\includegraphics[width=.36\textwidth]{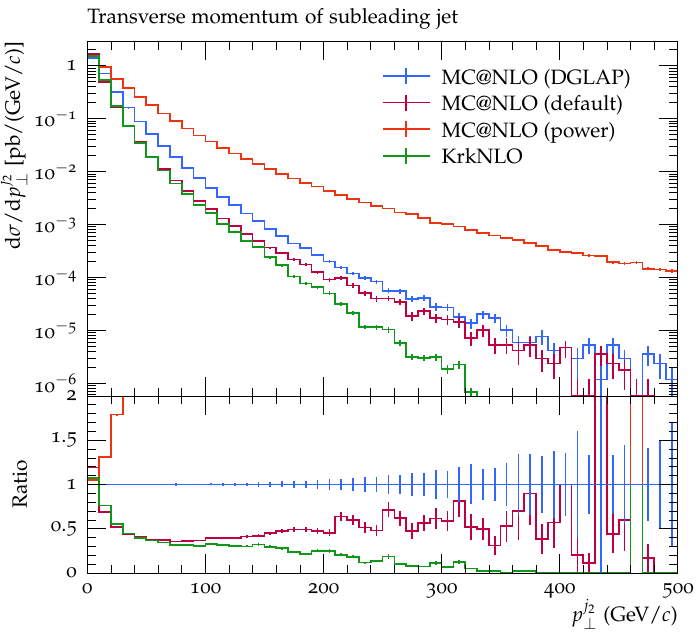}
	\includegraphics[width=.36\textwidth]{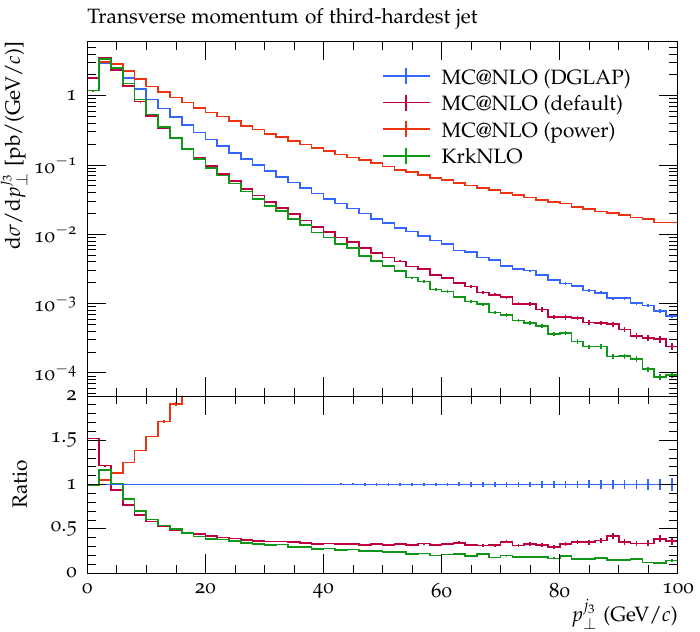}
	\includegraphics[width=.36\textwidth]{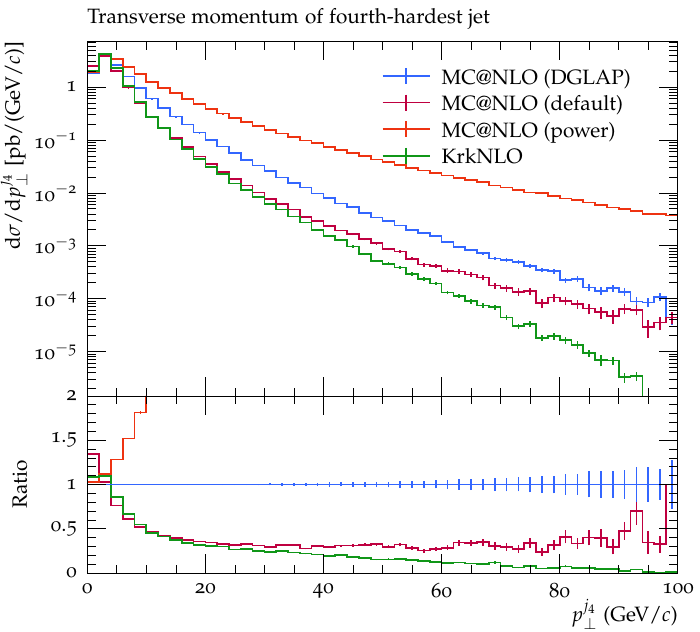}
}
\makebox[\textwidth][c]{
	\includegraphics[width=.36\textwidth]{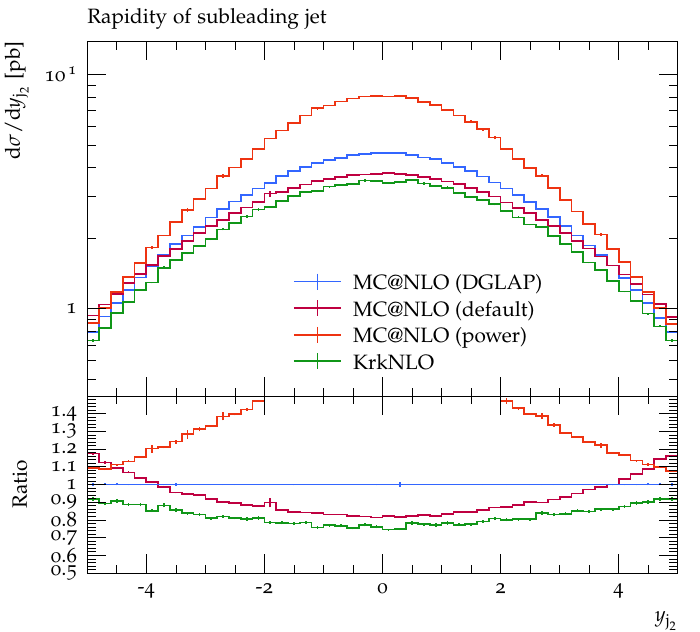}
	\includegraphics[width=.36\textwidth]{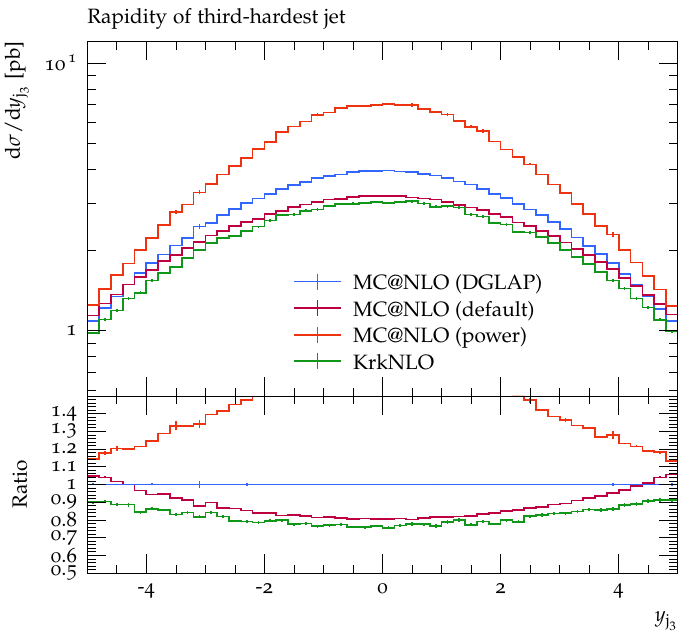}
	\includegraphics[width=.36\textwidth]{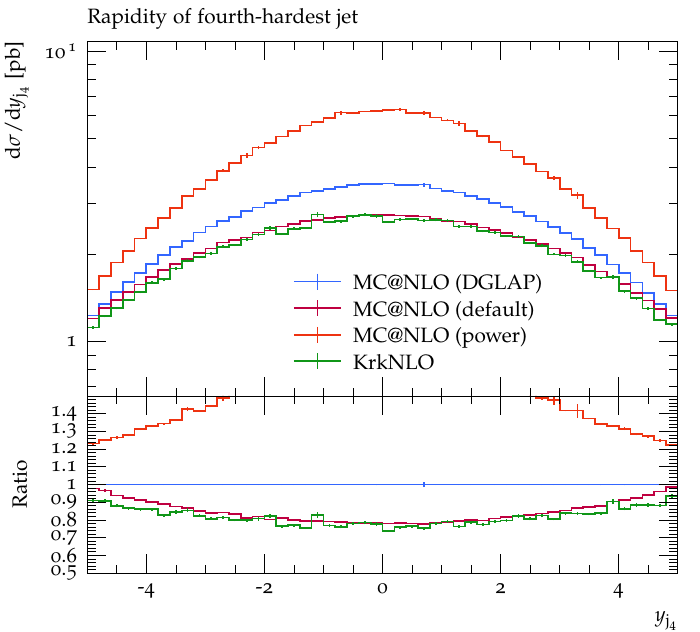}
 }
	\caption{
		Comparison of matched
		differential cross-sections 
		with respect to jet transverse momentum and rapidity,
		generated by \krknlo and \mcatnlo with the `default', `power'-shower and `DGLAP' starting-scales.
		\label{fig:matchingcomparisons_fullshower_jets}}
\end{figure}

Finally, distributions of transverse-momentum and rapidity for the additional identified jets are 
shown in \cref{fig:matchingcomparisons_fullshower_jets}.
These are not described to any order in perturbation theory and
are purely generated by the parton-shower evolution
from the initial-conditions provided by the matching method.
Again the `power'-shower \mcatnlo configuration favours harder
jets due to the unrestricted $Q(\Phi_{m+1})$,
while the \krknlo and `default' methods agree well, as expected
from their similar underlying scale choices.

\begin{figure}[htbp]
\centering
\makebox[\textwidth][c]{
\includegraphics[width=.36\textwidth]{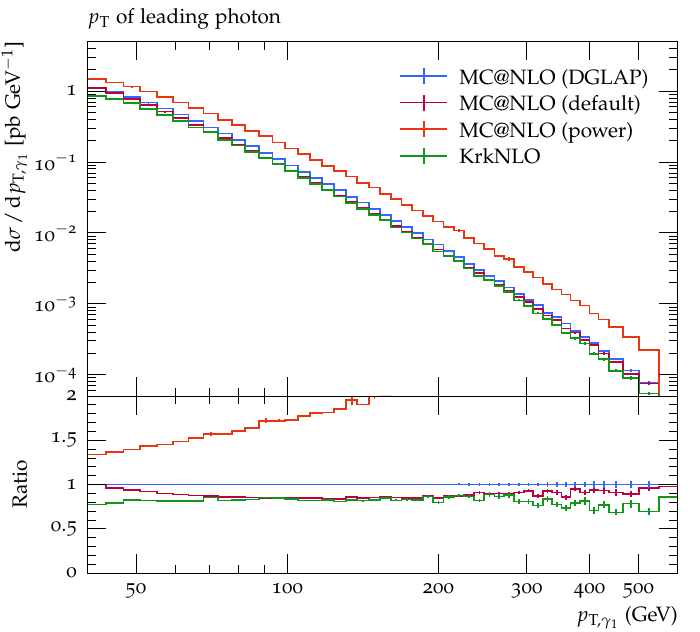}
\includegraphics[width=.36\textwidth]{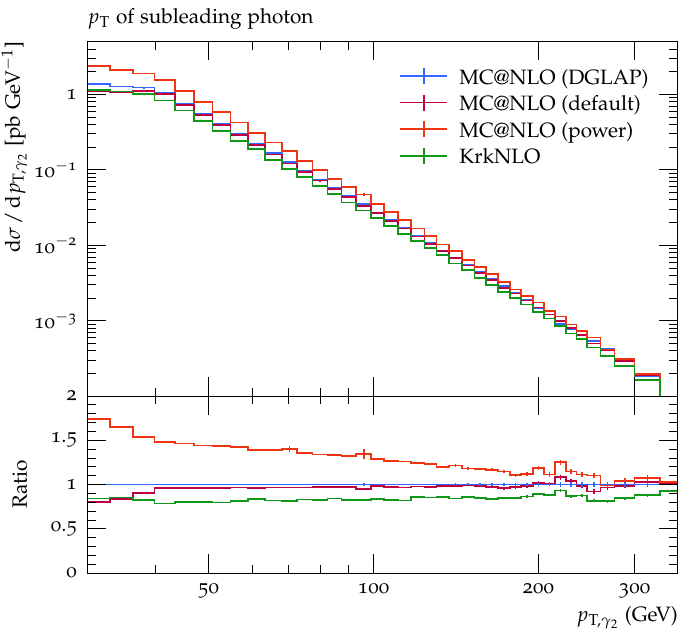}
\includegraphics[width=.36\textwidth]{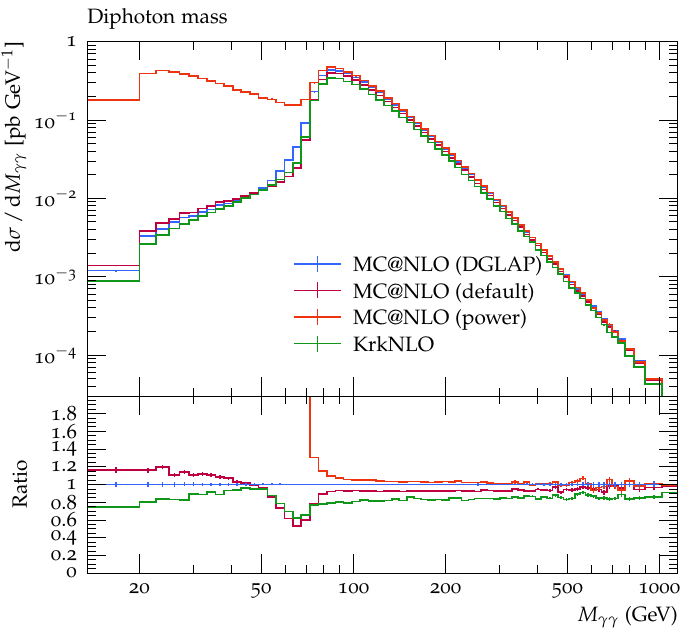}
}
\caption{
Comparison of matched NLO-plus-parton-shower differential cross-sections
as generated by \krknlo and \mcatnlo.
Refer to \cref{fig:pheno_fullshower_ATLAS}
for comparable plots also including \atlas data.
\label{fig:matchingcomparisons_fullshower_ATLAS_nodata}}
\end{figure}

\section{Results and phenomenology}
\label{sec:pheno}

In this section we augment the comparison of \cref{sec:matchingcomparisons}
with a comparison of the
same predictions against \atlas 13 TeV data \cite{ATLAS:2021mbt},
to assess the implications of matching uncertainty for comparisons to data.
We use the \atlas fiducial cuts
\begin{subequations}
	\label{eqn:ATLAScuts}
	\begin{align}
		\ptg{1} &> 40 \;\GeV \,, 	&	\ptg{2} &> 30 \;\GeV \,,  \\
		\dRgg &> 0.4 \,, & \absyg &\in \left[ 0, 1.37 \right) \cup \left( 1.52, 2.37 \right), \\
		\Etisopart &< 0.09 \; \ptg{}  & \text{within cone}\enspace  \dR &\leqslant 0.2.
	\end{align}
\end{subequations}
implemented in the
{\texttt{ATLAS\_2021\_I1887997}} analysis within \rivet \cite{Bierlich:2019rhm},
where $\Etisopart$ is the particle-level transverse isolation energy,%
\footnote{$\Etisopart$ is calculated
as the excess in the transverse magnitude of the total momentum of all
non-photon particles within the isolation cone,
over
an estimate of the average transverse momentum deposited in a cone of that area
due to the underlying event.
For more details consult \cite{ATLAS:2021mbt} or the corresponding \rivet analysis, {\texttt{ATLAS\_2021\_I1887997}}.}
and again with generator cuts
\begin{align}
	\hphantom{\Etisopart} \negphantom{\ptg{}} \ptg{} &> 5 \;\GeV \, , & \hphantom{\text{within cone}} \absyg &< 25 \negphantom{25}\hphantom{\left[ 0, 1.37 \right) \cup \left( 1.52, 2.37 \right)}
\end{align}
consistently for all predictions.

As explained in \cref{sec:matchingcomparisons}, the diphoton production process
was chosen among the possible Standard Model colour-singlet processes
due to its relative kinematic simplicity and absence of intrinsic mass-scales.
It is also, however, a process with large NLO and NNLO $K$-factors,
which has been found to require 
NNLO contributions to provide a good description of data throughout phase-space
\cite{Catani:2011qz, Campbell:2016yrh,Gehrmann:2020oec}
as well as NNLO contributions to the diphoton-plus-jet process
\cite{Chawdhry:2021hkp},
a \minnlops calculation \cite{Gavardi:2022ixt},
or the inclusion of the corresponding 
multiple-real-emission
matrix elements via parton-shower merging methods
\cite{Hoeche:2012yf}
to give a good description of observables related to the jet-recoil against the diphoton system.

Accordingly, of the \atlas observables measured in \cite{ATLAS:2021mbt} we here present only those
which are non-zero at LO and which we can therefore calculate to NLO using \krknlo.
These are nevertheless provided primarily for context and not as an attempt to produce a state-of-the-art prediction.

To improve the numerical comparison with the data, the
 gluon-gluon box contribution, formally an NNLO contribution relative to the $q\qbar$ Born partonic subprocess,
has been calculated with
\herwigseven and \matchbox using 
the one-loop provider
\gosam 2.1.1 \cite{Cullen:2011ac,GoSam:2014iqq}
and is in this section included consistently in all calculations.

\begin{figure}[tbp]
\centering
\makebox[\textwidth][c]{
\includegraphics[width=.36\textwidth]{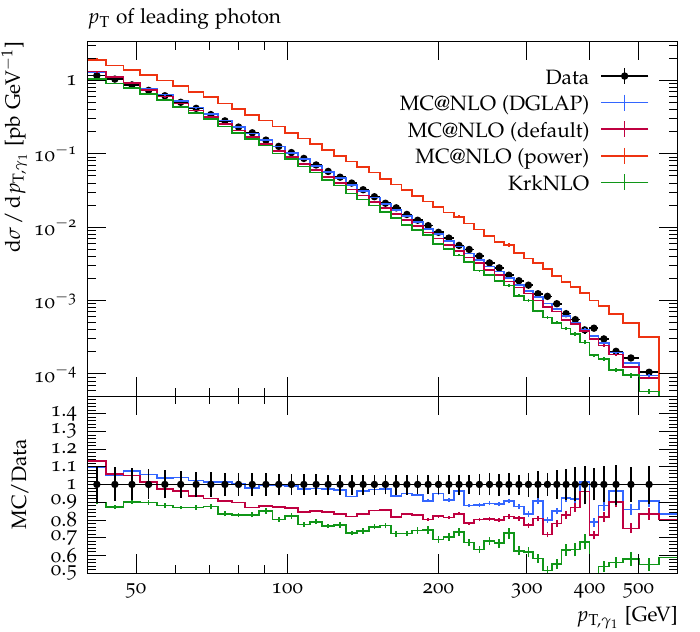}
\includegraphics[width=.36\textwidth]{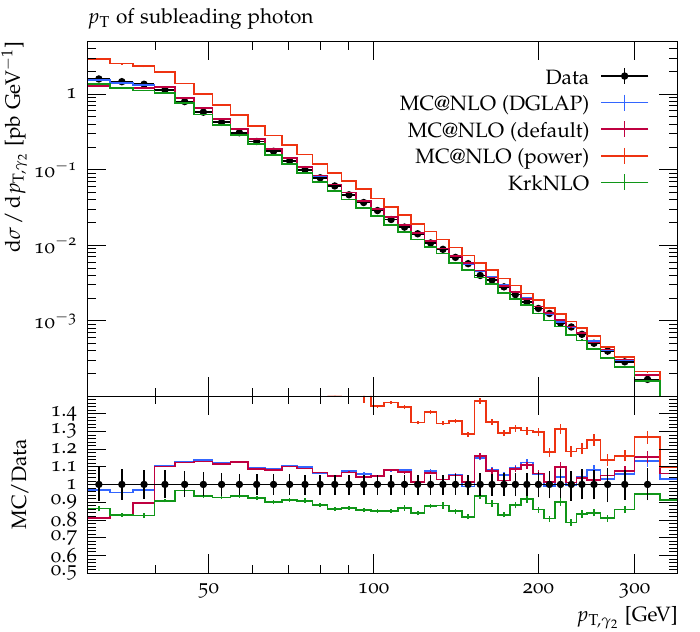}
\includegraphics[width=.36\textwidth]{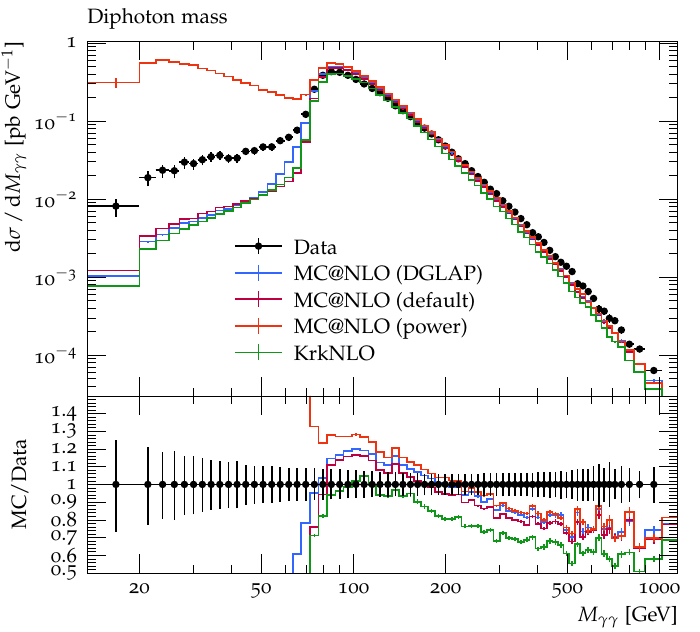}
}
\caption{
Comparison of matched NLO-plus-parton-shower differential cross-sections
as generated by \krknlo and \mcatnlo,
to \atlas data from \cite{ATLAS:2021mbt}.
The gluon-gluon box contribution (formally NNLO) has been
consistently included in all three predictions.
The remaining NNLO contributions are significant
and their inclusion is required to better describe the data.
Refer to \cref{fig:matchingcomparisons_fullshower_ATLAS_nodata}
for comparable plots without \atlas data.
\label{fig:pheno_fullshower_ATLAS}}
\end{figure}

These results are shown in \cref{fig:pheno_fullshower_ATLAS} (and in \cref{fig:matchingcomparisons_fullshower_ATLAS_nodata} without the data).  As seen in \cref{sec:matchingcomparisons_fullshower},
the matching uncertainty arising within \mcatnlo from the shower-scale variation is substantial for the $\ptg{1}$- and $\ptg{2}$-distributions, and much smaller for the $\Mgg$ distribution.

The latter is poorly-described by the available predictions for most of the available kinematic range, as anticipated by the known
requirement for NNLO calculations.

The former largely lie within the matching uncertainties of the methods, with good agreement in particular for $\dd \sigma / \dd \ptg{2}$ for the `DGLAP' variant of \mcatnlo.
However, all the predictions give an acceptable level of agreement considering their limited perturbative accuracy.

\section{Conclusion}
\label{sec:conclusion}

In this paper we have recapitulated the `\krknlo' method for the matching of NLO-accurate
perturbative calculations of the production of colour-singlet final-states in hadronic collisions
to a parton shower,
and systematically compared it to one of the main methods in use for LHC
phenomenology, the \mcatnlo method.
Building upon previous work on the \krknlo method,
the formulation presented here is process-independent
and can readily be applied to any colour-singlet final-state which proceeds via quark-antiquark annihilation at leading-order.

In comparison to \mcatnlo, \krknlo has the advantage of being fully and unambiguously
defined without the need to choose (unphysical) shower starting-scales,
the variation of which
has been found to lead to large uncertainties
\cite{Hoeche:2012fm,Dittmaier:2012vm}.
Although heuristics for scale choice have been formulated to reduce
this uncertainty
\cite{Alwall:2014hca,Harlander:2014uea,Bagnaschi:2015qta,Bagnaschi:2015bop,Frederix:2020trv},
it remains a potential obstacle to precision phenomenology.
The \krknlo method has no such parameters and therefore no
such uncertainty.
Despite the very different matching methods used, 
and the wide range of predictions which may be generated within the \mcatnlo
method using alternative scale choices,
when used with a common choice (and the \herwig default) the \mcatnlo
and \krknlo methods are generally found to be in good agreement.

The other method for NLO matching in widespread use, \powheg, like \krknlo uses
an unrestricted phase-space for the first parton-shower emission,
and so avoids \mcatnlo's dependence on an unphysical choice of shower starting-scale.
However, in \powheg this has been found to lead to 
a transverse-momentum distribution for the first jet that is unreasonably large and is
far from that implied by the key NLO ingredient, the real-emission matrix element
\cite{Alioli:2008tz,Nason:2009ai,Nason:2012pr,Bagnaschi:2015qta}.
In \cite{Hoeche:2011fd}
this was attributed to the exponentiation of the $\rR/\rB$ contribution throughout the emission phase-space,
rather than solely in the singular region in which the universal factorisation behaviour
holds.
For practical phenomenology, a heuristic has been adopted to fix this in the form of
a `damping' function \cite{Alioli:2008tz}
which can be tuned to suppress the exponentiated contribution from the non-singular region and recover the real-emission $\pt$-distribution.
However, this introduces its own intrinsic uncertainty via the choice of functional
form for the damping function and its chosen parameters 
\cite{Heinrich:2017kxx,FebresCordero:2021kcc,Banfi:2023mhz,ATLAS:2015wxj,ATLAS:2021pyq}%
.
We emphasise that the \krknlo method also avoids this uncertainty, despite also using a power-shower.
Similarly to \mcatnlo, it may be possible to tune the
functional form and parameters of the
\powheg damping-function to closely match \krknlo predictions.

By presenting a formulation of the \krknlo method
suitable for general colour-singlet processes we demonstrate
that it is a novel matching method ready for practical LHC 
phenomenology,
with several features making it an interesting and useful complement
to \mcatnlo and \powheg.
We are currently validating the \krknlo method for a range of
other colour-singlet processes and
expect to present further matching comparisons for other processes in future work.
We anticipate the public availability of the \krknlo
code within an upcoming release of \herwigseven.
Beyond colour-singlets at NLO we remain optimistic about the applicability
of a variant of the \krknlo method to more complex processes and higher-orders.

\acknowledgments

The authors wish to thank 
Wiesław Płaczek
and 
the late Stanisław Jadach
for their work on, and for many fruitful discussions about, the KrkNLO method, and 
Simon Plätzer
for productive discussions about NLO matching, and for his insight into and assistance with \herwig and \matchbox.
JW and AS thank Olek Kusina and Stéphane Delorme for valuable discussions about
PDF convolutions and factorisation schemes.
We are grateful to 
Wiesław Płaczek,
Mike Seymour
and
Simon Pl\"atzer for their thoughtful comments on the manuscript.

This work was supported by 
grant 2019/34/E/ST2/00457 of the National Science Centre, Poland.
AS is also supported by the Priority Research Area Digiworld under the program 
`Excellence Initiative -- Research University'
at the Jagiellonian University in Krakow.
We gratefully acknowledge the Polish high-performance computing infrastructure PLGrid (HPC Centre: ACK Cyfronet AGH) for providing computing facilities and support within computational grant PLG/2023/016494.

\appendix

\section{Summary of Catani--Seymour insertion operators}
\label{sec:appendix_cssummary}

For ease of reference, we briefly recap the initial-state insertion operators of Catani--Seymour dipole
subtraction that are used for the construction of the \krknlo matching expressions, largely following the notation of \cite{Catani:1996vz}.

Parton density functions in the factorisation scheme $\FS$ are related to those in the \msbar scheme by the transformation
\begin{align}
\label{eq:app_CS_faFSvec}
\mathbf{f}^{\rFS}
&=
\mathbb{K}^{\msbar \to \rFS}
\otimes
\mathbf{f}^{\msbar},
\end{align}
where explicitly `input' $\msbar$ PDFs are convolved, locally in $\muf$, with a kernel,
\begin{align}
\label{eq:app_CS_faFS}
f^{\rFS}_a (x, \muf)
&=
\sum_b
\int_x^1
\frac{\dd z}{z}
\;
\mathbb{K}^{\msbar \to \rFS}_{ab} \left(z, \muf\right) \ 
f_b^{\msbar} \left(\frac{x}{z}, \muf\right)
\end{align}
and the matrix of convolution kernels $\mathbb{K}^{\msbar \to \rFS}_{ab}$ has perturbative expansion
\begin{align}
\mathbb{K}^{\msbar \to \rFS}_{ab} \left(z, \mu\right)
=
\delta_{ab} \; \delta(1-z)
+
\frac{\alphas(\mu)}{2\pi}
\;
\rK^{\msbar \to \rFS}_{ab}(z) + \order{\alphas^2}.
\label{eq:FS_pertexp}
\end{align}
The corresponding insertion operators for processes with two initial-state hadrons are given, for a colour-singlet final-state, by 
\cite[][eq.\ C.29]{Catani:1996vz}
\begin{align}
\label{eq:app_CS_Pop}
\rP_{ab} (x, \mu)
=
-
P_{ab}(x)
\log \frac{\mu^2}{\hat{s}_{12}}
\end{align}
where $P_{ab}$ are the regularised DGLAP splitting kernels,
and \cite[][eq.\ C.33]{Catani:1996vz}
\begin{align}
\rK_{ab} (x, \mu)
=
\bar{K}_{ab} (x)
+ \tilde{K}_{ab} (x)
- K_{ab}^{\FS} (x),
\end{align}
where full expressions for $\bar{K}_{ab}$ and $\tilde{K}_{ab}$ may be found in Appendix C of \cite{Catani:1996vz}.

For colour-singlet processes, the relevant combinations of insertion operators simplify to
\begin{alignat}{3}
\label{eq:app_CS_Pop_PKqq}
\left(
\rP(\mu) + \rK
\right)^{\rFS}_{qq} \negthick (x)
=
\cf
\Bigg[
& 4 \lnomxoomxplus{}
- 2 (1 + x) \log (1-x)
- \frac{1+x^2}{1-x} \log x
\\ \notag & {} +
1-x
+
\left( \frac{2}{3}\pi^2-5 \right)
\delta(1-x)
-
p_{qq} (x) \log \frac{\mu^2}{\hat{s}_{12}}
-
\cf^{-1}
K_{qq}^{\rFS} (x)
\Bigg]
\\ 
\label{eq:app_CS_Pop_PKgq}
\left(
\rP(\mu) + \rK
\right)^{\rFS}_{gq} \negthick (x)
=
\tR
\Bigg[
& 
p_{gq} (x) \log \frac{(1-x)^2}{x}
\\ \notag & {} +
2 x (1-x)
-
p_{gq} (x) \log \frac{\mu^2}{\hat{s}_{12}}
-
\tR^{-1}
K_{qg}^{\rFS} (x)
\Bigg],
\end{alignat}
where the leading-order colour-factor-stripped regularised DGLAP splitting functions are given in four dimensions by
\begin{align}
    p_{qq} (x) &= \left(\frac{1+x^2}{1-x}\right)_+ \equiv 2\left(\frac{1}{1-x}\right)_+ - (1 + x) + \frac{3}{2} \, \delta(1-x)
    \\
    p_{qg} (x) &= \frac{1 + (1-x)^2}{x}
    \\
    p_{gq} (x) &= x^2 + (1-x)^2
    \\
    p_{gg} (x) &= 2 \left[ 
\left(\frac{1}{1-x}\right)_+ + \frac{1-x}{x} -1 + x(1-x) \right] + \frac{b_0}{\ca}\delta(1-x),
\end{align}
and
\begin{align}
    b_0 = \frac{11}{6} \ca -\frac{2}{3} \nf \tR.
\end{align}

The factorisation-scheme-dependent contributions $K_{ab}^{\FS}(x)$, which are also the 
transition kernels of \cref{eq:FS_pertexp}
(transposed%
\footnote{The transposed indices arise from the difference between indexing the
	kernels for matrix-multiplication on a column vector of PDFs as in \cref{eq:app_CS_faFSvec,eq:app_CS_faFS}
	and on a matrix of partonic cross-sections as in \cite{Catani:1996vz}.
	The same considerations apply to the splitting kernels.
}%
)
then allow us to write the insertion operators in
any factorisation scheme in terms of the \msbar operators
and the transition kernel,
\begin{align}
	\left(
	\rP(\mu) + \rK
	\right)^{\rFS}_{aq} \negthick (x)
	=
	\left(
	\rP(\mu) + \rK
	\right)^{\msbar}_{aq} \negthick (x)
	-
	\rK^{\msbar \to \rFS}_{qa} (x).
\end{align}
The factorisation scheme kernels are, by definition,
\begin{align}
    \rK_{ab}^{\msbar\to\msbar} (x) &\equiv 0
    \equiv K_{ba}^{\msbar} (x)
\end{align}
for the \msbar scheme and
\begin{alignat}{3}
    & \rK_{qq}^{\msbar\to\dis} (x) \equiv K_{qq}^{\dis} (x) 
    & {} = {} & \cf \left[
    \frac{1+x^2}{1-x}\left(\log \frac{1-x}{x} - \frac{3}{4}\right)
    + \frac{1}{4}(5x + 9)
    \right]_+
    \\
    & \rK_{qg}^{\msbar\to\dis} (x) \equiv K_{gq}^{\dis} (x) 
    & {} = {} & \tR \left[
    p_{gq}(x) \, \log\frac{1-x}{x} + 8x(1-x)-1
    \right]
    \\
    & \rK_{gq}^{\msbar\to\dis} (x) \equiv K_{qg}^{\dis} (x) 
    & {} = {} & - \rK_{qq}^{\msbar\to\dis}
    \\
    & 
    \rK_{gg}^{\msbar\to\dis} (x) \equiv K_{gg}^{\dis} (x) 
    & {} = {} & - 2 \nf \rK_{gq}^{\msbar\to\dis}
\end{alignat}
for the DIS scheme%
\footnote{By convention, in the DIS scheme $\rK_{gq}^{\msbar\to\dis}$ and $\rK_{gg}^{\msbar\to\dis}$ are fixed by the requirement that the DIS PDFs
satisfy momentum conservation locally in $x$-space
(equivalently, by extending the constraint on the second Mellin moment
implied by momentum conservation to all Mellin moments)
\cite{Altarelli:1979ub,Diemoz:1987xu,Catani:1994sq,Martin:1998np}
.}
\cite{Altarelli:1979ub}.
In the \krk scheme,
following from
\cref{eq:nlomatching_krkfs_Deltaeq_Kqq,eq:nlomatching_krkfs_Deltaeq_Kqg,eq:app_CS_Pop_PKqq,eq:app_CS_Pop_PKgq},
\begin{alignat}{3}
	\label{eq:app_CS_Kqq}
	& \rK_{qq}^{\msbar\to\krk} (x) \equiv K_{qq}^{\krk} (x) 
	& {} = {} & 
	\cf \Biggl[
	\biggl[ \frac{1+x^2}{1-x} \log \frac{(1-x)^2}{x} + 1 - x \biggr]_+
	- \frac{3}{2} \delta(1-x)
	\Biggr]
	\\ & \notag
	& {} = {} & \cf \Biggl[ 
	4 \left[\frac{\log(1-x)}{1-x}\right]_+ 
	- 2 (1 + x) \log (1-x)
	- \frac{1+x^2}{1-x} \log x
	\\ &&& \notag
	\qquad + 1-x
	- \left(\frac{\pi^2}{3} + \frac{17}{4}\right)\delta(1-x) \Biggr]
	\\
	& \rK_{qg}^{\msbar\to\krk} (x) \equiv K_{gq}^{\krk} (x) 
	& {} = {} & \tR \Biggl[
	2 p_{gq}(x) \log (1-x) - p_{gq}(x) \log x - p_{gq}(x) + 1 
	\Biggr]
\end{alignat}
and so
\begin{alignat}{1}
	\left(
	\rP(\mu) + \rK
	\right)^{\krk}_{qq} \negthick (x)
	&{} =
	\rP_{qq} (x,\mu) 
	+ \rK_{qq}^{\msbar}(x)
	- \rK_{qq}^{\msbar \to \krk} (x)
	\\
	&{} =
	-
	\cf \;
	p_{qq} (x) \log \frac{\mu^2}{\hat{s}_{12}}
	+ \frac{1}{2} \, \Delta_0^{\krk} \, \delta(1-x)
	\\
	\left(
	\rP(\mu) + \rK
	\right)^{\krk}_{gq} \negthick (x)
	&{} =
	\rP_{gq} (x,\mu) 
	+ \rK_{gq}^{\msbar}(x)
	- \rK_{qg}^{\msbar \to \krk} (x)
	\\
	&{}
	=
	- \tR \;
	p_{gq} (x) \log \frac{\mu^2}{\hat{s}_{12}},
\end{alignat}
again illustrating the justified exclusion of explicit
collinear terms from the perturbative cross-section,
with
\begin{align}
	\Delta_0^{\krk}
	=
	\cf \left( 2 \pi^2 - \frac{3}{2} \right)
\end{align}
following directly from the difference between \cref{eq:app_CS_Pop_PKqq} and \cref{eq:app_CS_Kqq}.

\section{PDFs in the Krk factorisation scheme}
\label{sec:appendix_krkpdfs}

\begin{figure}[p]
	\centering
	\includegraphics[width=\textwidth]{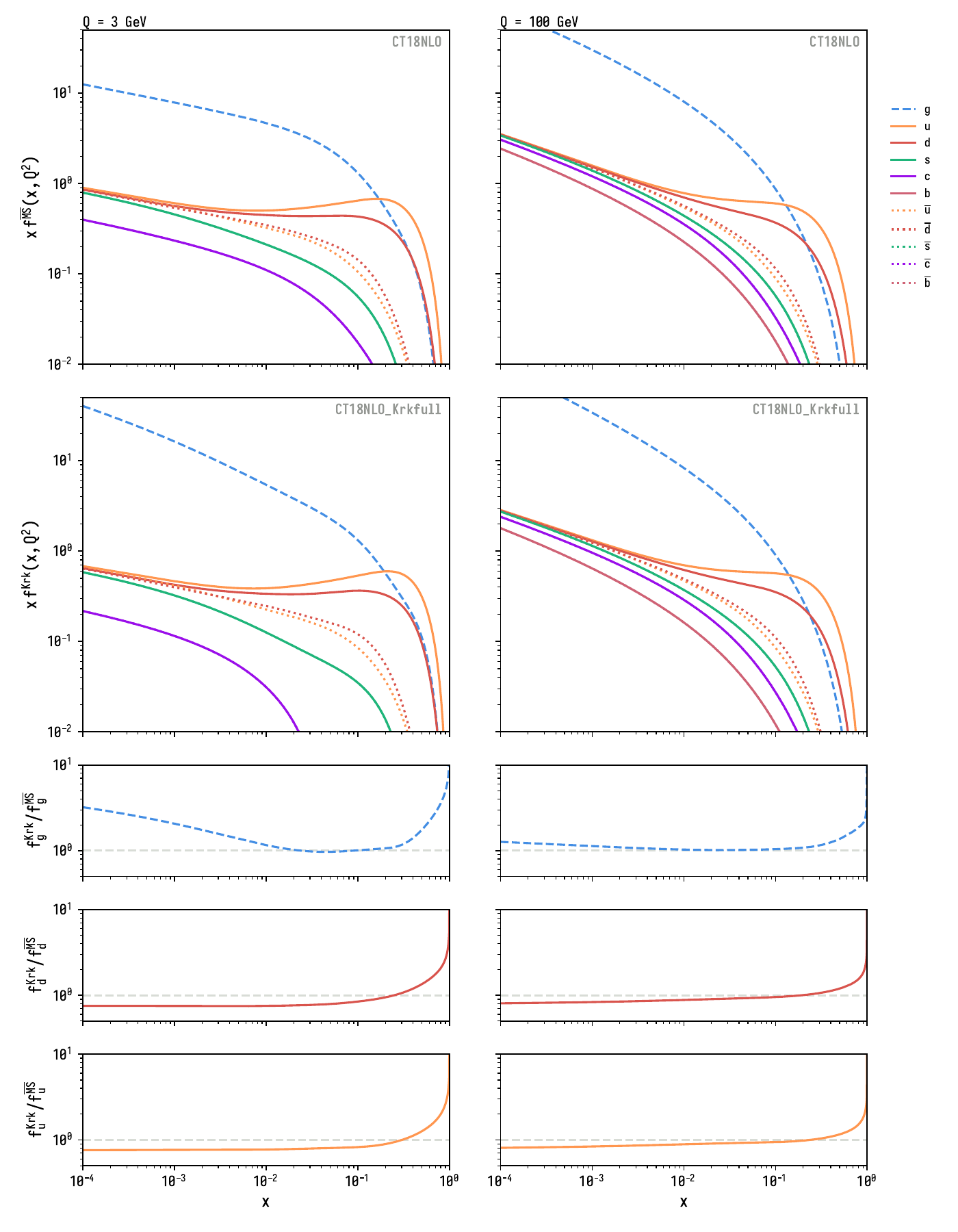}
	\caption{Comparison of the central \texttt{CT18NLO} \cite{Hou:2019efy} PDF set 
		in the default \msbar factorisation scheme against the same PDF set transformed into the (full) \krk scheme
		as described in \cref{sec:nlomatching_krkfs,sec:appendix_krkpdfs}.
		Left: $Q=3\;\GeV$ (for \texttt{CT18NLO}, $Q_0 = 1.3 \; \GeV$);
		right: $Q=100\;\GeV$.
		\label{fig:appendix_krkvsmsbarpdf}}
\end{figure}

Throughout we use the \texttt{CT18NLO} PDF \cite{Hou:2019efy},
transformed from the \msbar to the \krk factorisation scheme
with the \texttt{C++} code used for \cite{Jadach:2016acv}.
This interfaces with the LHAPDF 6 \cite{Buckley:2014ana}
library to perform the required convolution integrals on the interpolated
PDF objects, writing the transformed PDFs as LHAPDF grids with
the same $(x, Q^2)$ knots as the original input PDF grid.
\texttt{CT18NLO} PDFs in the \msbar and \krk schemes
at 3 and 100 GeV are shown in \cref{fig:appendix_krkvsmsbarpdf}.

The transformation kernels for the \krk scheme for the quark
and antiquark PDFs,
$\rK^{\msbar \to \krk}_{qa}$,
are determined by the 
Catani--Seymour dipoles as outlined in \cref{sec:appendix_cssummary}.
The \krk `DY' scheme \cite{Jadach:2015mza,Jadach:2016acv}
transforms only the quark PDFs and would be equally suitable for the processes
considered in this paper.
We use the `full' \krk scheme which also transforms the gluon PDF
as outlined in \cite{Jadach:2016acv,Jadach:2016qti}.

Concretely, after imposing the momentum sum rule of \cref{eq:nlomatching_krkfs_PDFmomsumrule} \cite{Collins:1981uw},
the transformation from the \msbar into the (full) \krk
scheme is given by:
\begin{align}
\label{eq:app_krkpdfs_krkconv}
f_q^\krk (x, \muf)
    = {}
    & f_q^\msbar (x, \muf)
    \\ \notag 
    & {} - \mathrlap{
    \frac{\alphas(\muf)}{2\pi} \;  \frac{3}{2} \cf 
    \; f_q^\msbar \left({x}, \muf\right)
    }
    \\ \notag 
    & {} + \mathrlap{
    \frac{\alphas(\muf)}{2\pi} \; \cf
    \int_x^1
    \frac{\dd z}{z}
    \biggl[\frac{1+z^2}{1-z} \log \frac{(1-z)^2}{z} + 1 - z \biggr]_+
    f_q^\msbar \left(\frac{x}{z}, \muf\right)
    }
    \\ \notag 
    & + \mathrlap{
    \frac{\alphas(\muf)}{2\pi} \;
    \tR \negphantom{\tR} \hphantom{\cf}
    \int_x^1
    \frac{\dd z}{z}
    \biggl[
    p_{gq}(z) \, \log \frac{(1-z)^2}{z} + 2z(1 - z)
    \biggr]
    \hphantom{{}_+}
    f_g^\msbar \left(\frac{x}{z}, \muf\right)
    }
    \\
f_g^\krk (x, \muf)
    = {}
    & f_g^\msbar (x, \muf)
    \\ \notag 
    & {} - \mathrlap{
    \frac{\alphas(\muf)}{2\pi} \; \ca
    \biggl[
    \frac{\pi^2}{3}+\frac{341}{72}-\frac{59}{36}\frac{\nf\tR}{\ca}
    \biggr]
    \hphantom{{}_+}
    f_g^\msbar \left({x}, \muf\right)
    }
    \\ \notag 
    & {} +
    \frac{\alphas(\muf)}{2\pi} \; \ca
    \int_x^1
    \frac{\dd z}{z}
    \Biggl[ \begin{aligned}[t]
        & 4 \lnomxoomxplus{}
                - 2 \frac{\log z}{1-z}
     \\ & {} + 2 \left(
                       \frac{1}{z} - 2 + z(1-z)
                 \right) \log\frac{(1-z)^2}{z}
         \Biggr]
            \hphantom{{}_+}
            f_g^\msbar \left(\frac{x}{z}, \muf\right)
     \end{aligned}
    \\ \notag 
    & + \mathrlap{
    \frac{\alphas(\muf)}{2\pi} \; \cf
    \sum_{q_f,\qbar_f}
    \int_x^1
    \frac{\dd z}{z}
    \left[ 
    p_{qg} (z) \, \log \frac{(1-z)^2}{z} + z
    \right]
    \hphantom{{}_+}
    f_{q}^\msbar \left(\frac{x}{z}, \muf\right)
    }
\end{align}
In the \texttt{C++} implementation the number of active light quark flavours in the above transformations is fixed at $\nf = 5$ for all scales $\mu$,
corresponding to using the fixed-flavour-number scheme (FFNS)
for the transformation kernels.

\section{Validation}
\label{sec:appendix_validation}

In this section we summarise the validation of the implementation of both
the diphoton process within the \krknlo code, and of the \krknlo method
more generally.

\subsection*{Real corrections}
\label{sec:appendix_validation_real}

As described in \cref{sec:imp_validation_R},
it is possible to choose settings within \krknlo which cause the \krknlo implementation 
to calculate the fixed-order distribution, rather than the matched calculation.
This is done by truncating the \herwig dipole shower to generate a single emission
and reweighting by the reciprocal of the Sudakov factor,
$1 / \Delta \bigr\vert_{\ptof{1}}^{Q(\Phi_m)} (\Phi_m)$,
as calculated numerically by integrating the
competing shower emission kernels over their respective phase-spaces.
We veto the no-emission events.

We perform the validation
for the fiducial cuts of \cref{eqn:pseudoATLAScuts}, with the addition of
a jet cut
\begin{align}
	\hphantom{\Etiso (r)} \negphantom{\ptj{}}
	\ptj{} &> 1 \;\GeV \, , 
	\hphantom{ 0.1 \, \ptg{} \; \chi(r; R) } \negphantom{30 \; \GeV \, ,} 
	& 
	\hphantom{\text{ within cone } r} \negphantom{\absyg}
	\absetaj &< 5 \negphantom{5} \hphantom{\left[ 0, 2.5 \right), \cup \left( 1.52, 2.37 \right)}
\end{align}
and with the generator cuts of \cref{eqn:rungeneratorcuts}.
Note that identified `jets' at this order consist of exactly one parton.

\begin{figure}[hp]
	\centering
	\begin{subfigure}[t]{\textwidth}
		\centering
\makebox[\textwidth][c]{
		\includegraphics[width=.32\textwidth]{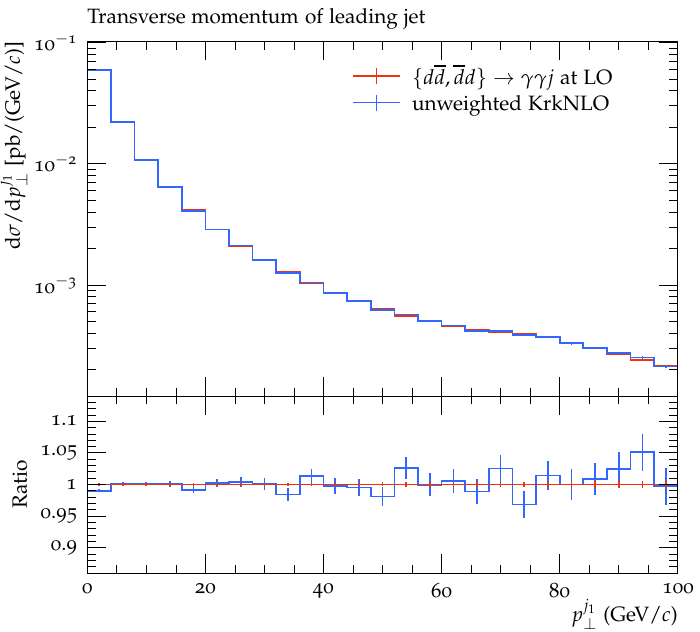}
		\includegraphics[width=.32\textwidth]{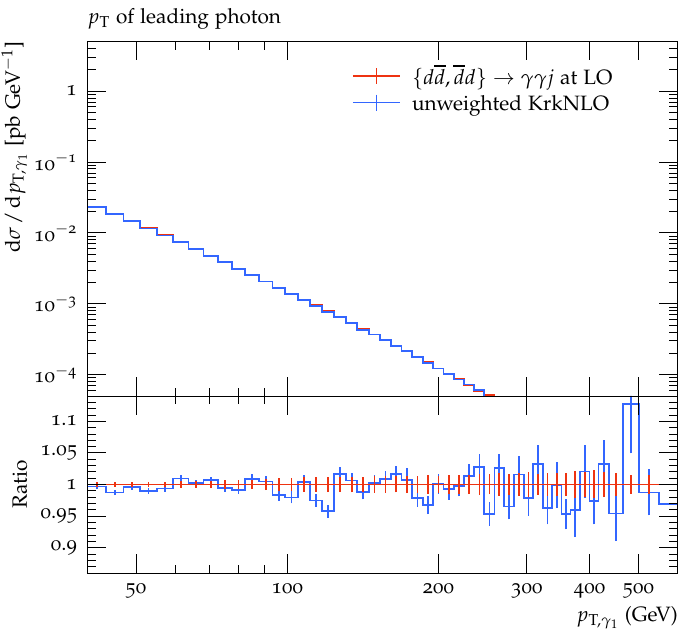}
		\includegraphics[width=.32\textwidth]{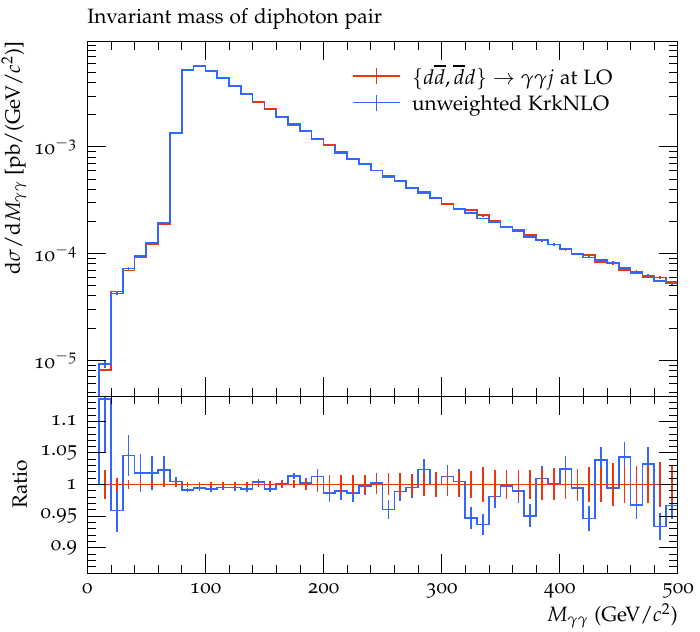}
  }
		\caption{ $\{d\overline{d}, \overline{d}d\} \to \gamma \gamma g$ \label{fig:validation_real_qqb}}
	\end{subfigure}
	\begin{subfigure}[t]{\textwidth}
		\centering
\makebox[\textwidth][c]{
		\includegraphics[width=.32\textwidth]{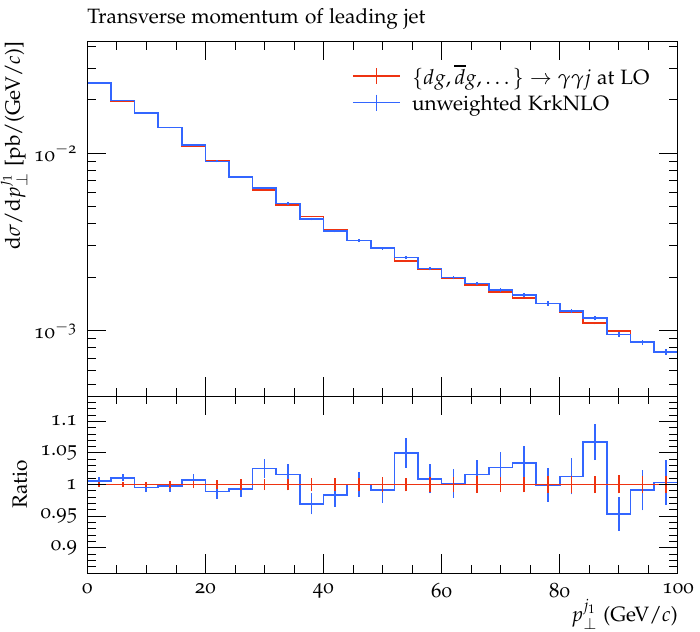}
		\includegraphics[width=.32\textwidth]{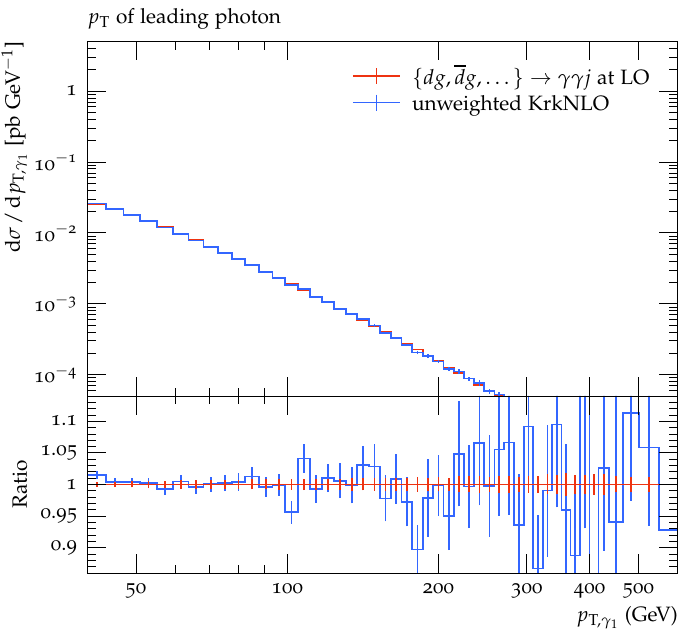}
		\includegraphics[width=.32\textwidth]{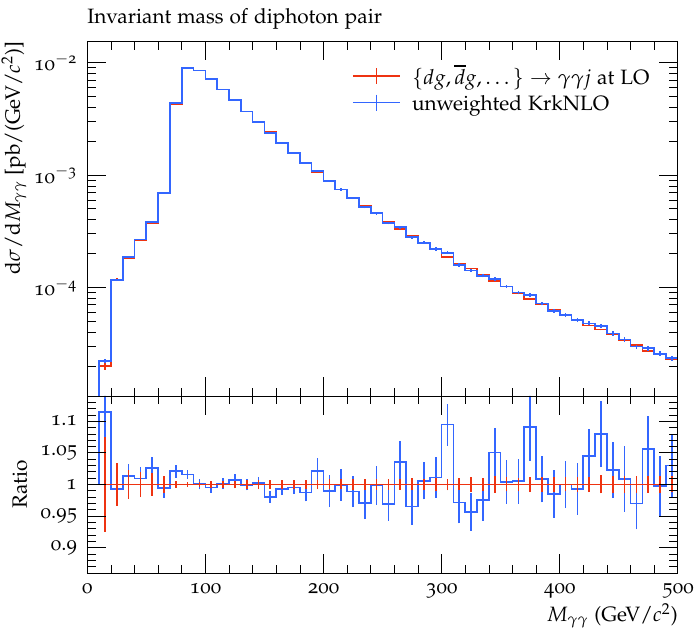}
  }
		\caption{$\{dg, gd \} \to \gamma \gamma d, \;
			\{\overline{d}g, g\overline{d}\} \to \gamma \gamma \overline{d}$ \label{fig:validation_real_qg}}
	\end{subfigure}
	\begin{subfigure}[t]{\textwidth}
		\centering
\makebox[\textwidth][c]{
		\includegraphics[width=.32\textwidth]{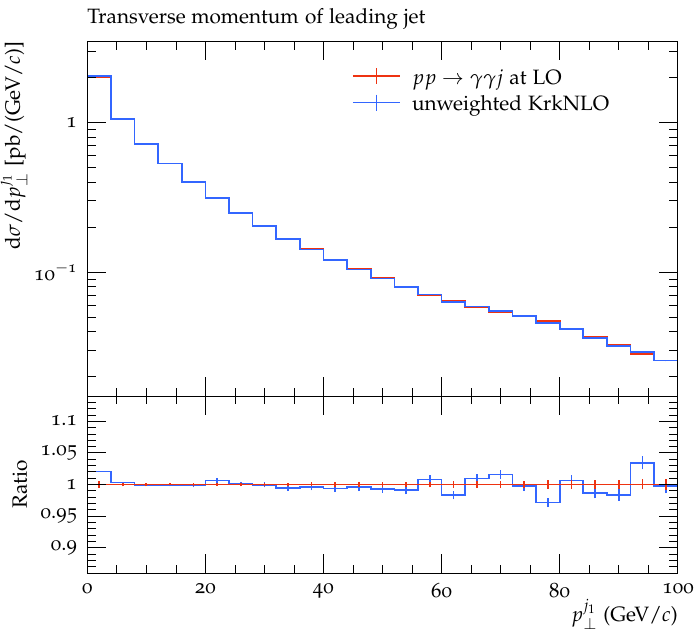}
		\includegraphics[width=.32\textwidth]{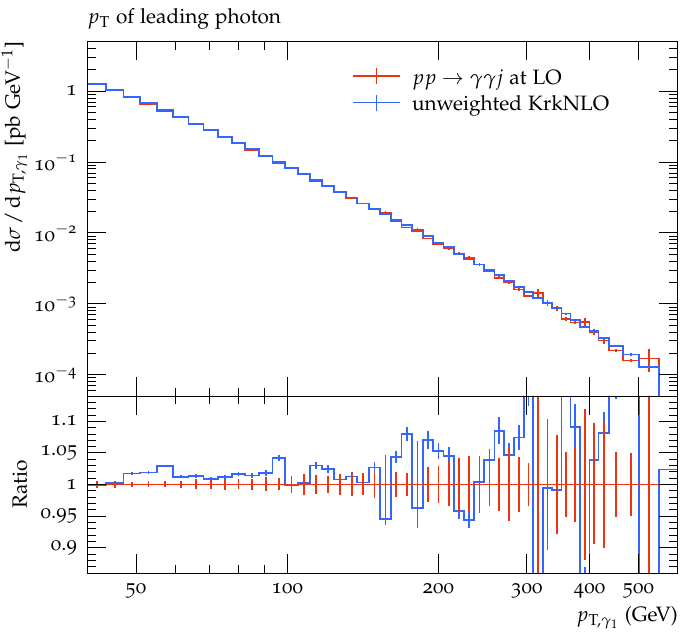}
		\includegraphics[width=.32\textwidth]{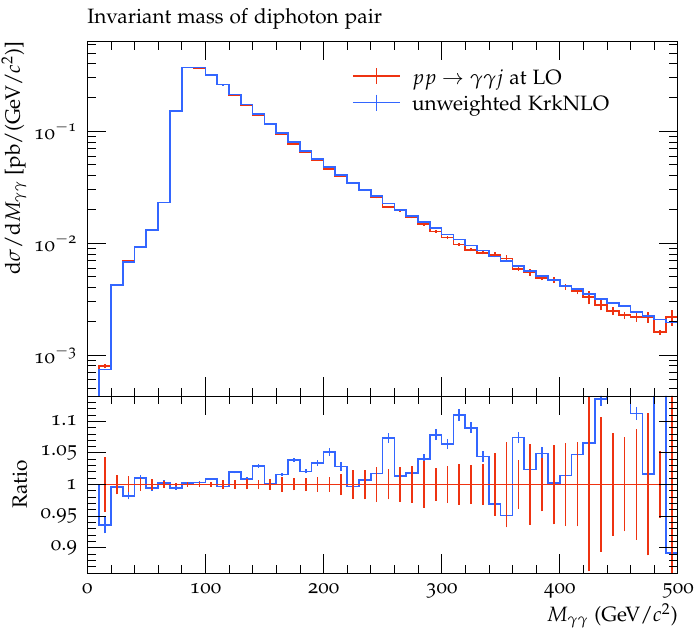}
  }
		\caption{$p p \to \gamma \gamma j$ \label{fig:validation_real_pp}}
	\end{subfigure}
	\caption{Validation of the real weight for the $q\qbar$-, $qg$- and $pp$-channels
	respectively by unweighting by the Sudakov factor 
	$\Delta \bigr\vert_{\ptof{1}}^{Q(\Phi_m)} (\Phi_m)$
	(independently calculated for each emission by numerical integration of the kernels over the radiative phase-space)
	to isolate the real matrix element within the \krknlo implementation,
	as described in \cref{sec:imp_validation_R}.
	\label{fig:validation_real}}
\end{figure}

The resulting distributions are shown in \cref{fig:validation_real}
and show
excellent agreement between the \krknlo code and the \matchbox calculation of
diphoton-plus-jet at leading order.
This is a non-trivial test of the real weight \cref{eq:nlomatching_notationdefinitions_nlomatching_krknlo_wR_sym} and 
of the consistency of the Sudakov integration and unweighting with the shower algorithm.

\subsection*{Virtual corrections}
\label{sec:appendix_validation_virt}

\begin{figure}[ht]
	\centering
\makebox[\textwidth][c]{
	\includegraphics[width=.36\textwidth]{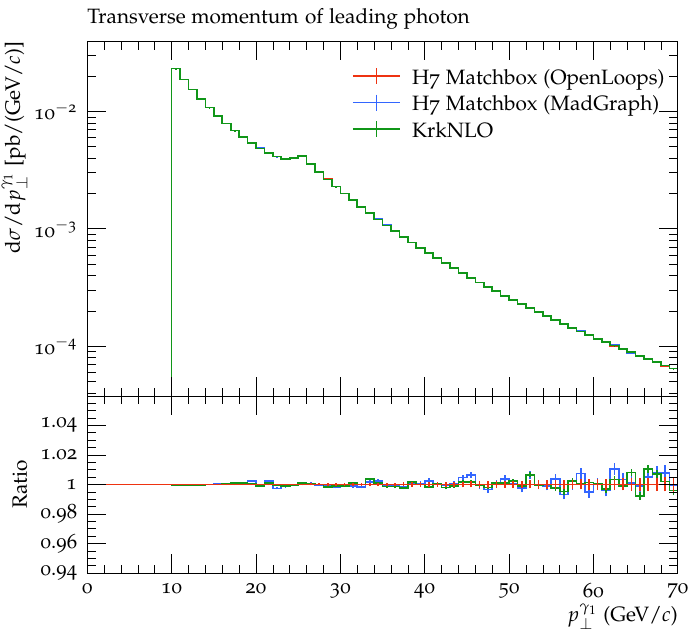}
	\includegraphics[width=.36\textwidth]{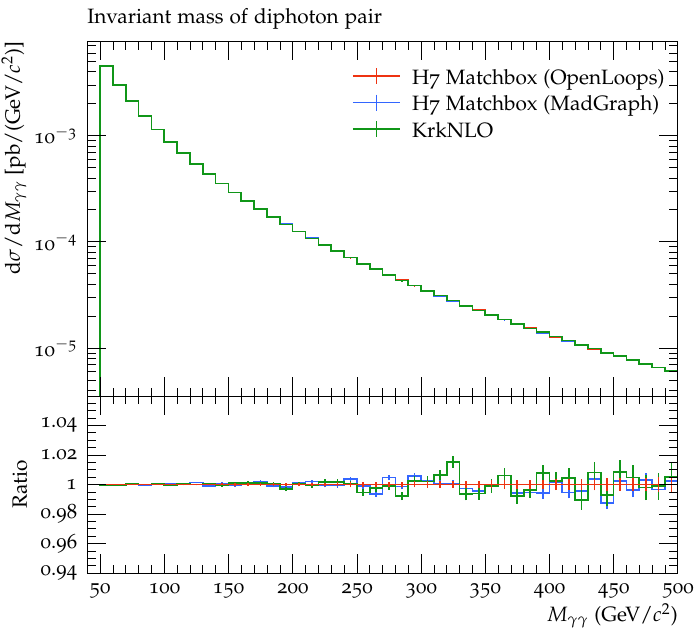}
	\includegraphics[width=.36\textwidth]{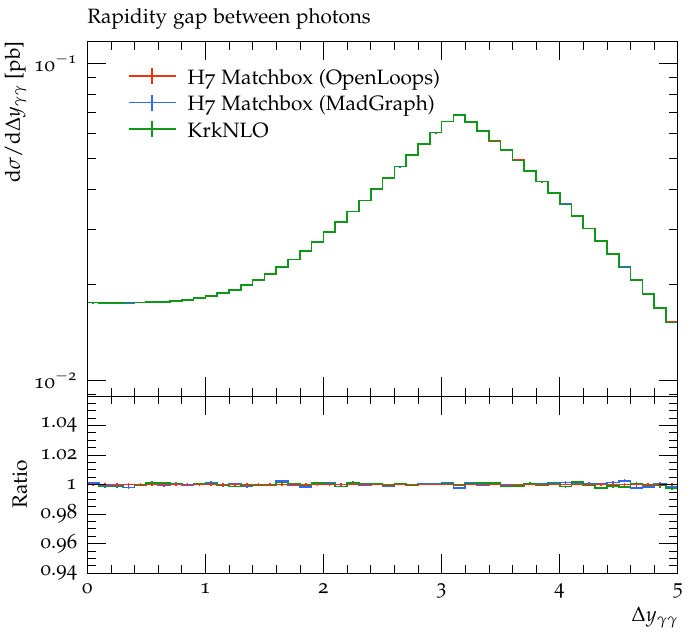}
 }
	\caption{Validation of the virtual weight $\rV + \rI$ for the $q\qbar$-channel
     		as described in \cref{sec:imp_validation_V}.
		    By setting the shower IR cutoff to guarantee $t_0 > Q(\Phi_m)$,
		    thus disabling the shower, and manually
		    disabling the Born and $\Delta_0^{\rFS}$ contributions within \krknlo,
		    the \krknlo implementation of the virtual terms can be compared with
		    those generated automatically by \matchbox within \herwigseven.
			\label{fig:validation_virtual}}
\end{figure}

As described in \cref{sec:imp_validation_V}, the virtual matrix elements 
within the \krknlo implementation
can be tested
by setting the shower cut-off $t_0$ to
ensure $t_0 > \max_{\Phi_m} Q(\Phi_m)$, prohibiting any parton-shower
radiation.
The shower algorithm is then bypassed and the Sudakov factor
identically equal to 1 for all events.
The components contributing to $\calO(\Phi_m)$ in 
\cref{eq:nlomatching_notationdefinitions_nlomatching_dsigmaqqb_krknlo_1_singlet}
can then be tested numerically
at the level of differential cross-sections
against reference distributions calculated using the automated \matchbox implementation within \herwig, as shown in \cref{fig:validation_virtual}.

\begin{figure}[hb]
	\centering
\makebox[\textwidth][c]{
	\includegraphics[width=.36\textwidth]{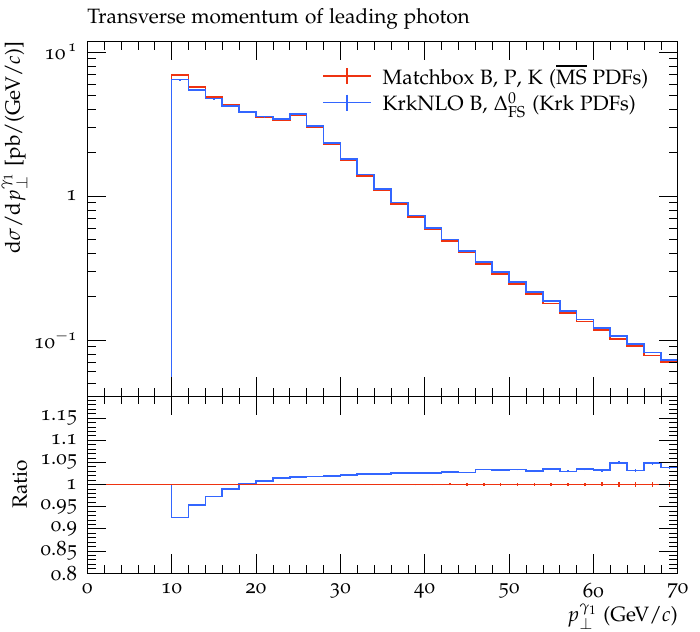}
	\includegraphics[width=.36\textwidth]{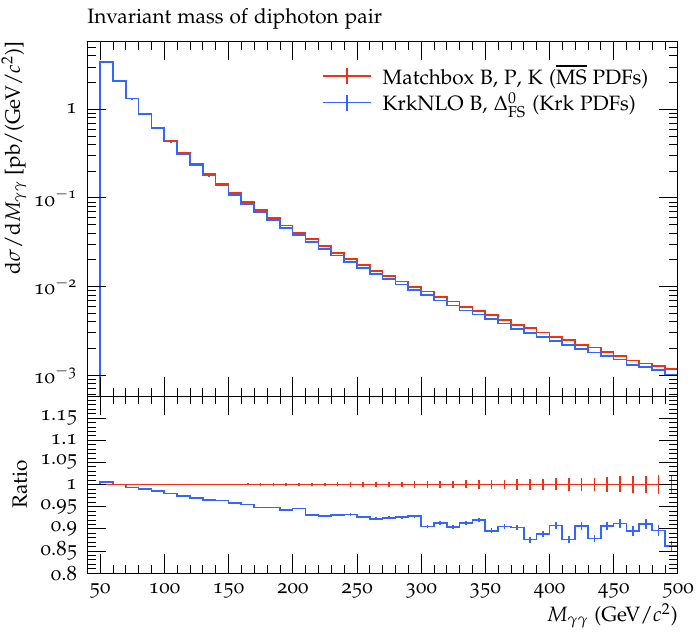}
	\includegraphics[width=.36\textwidth]{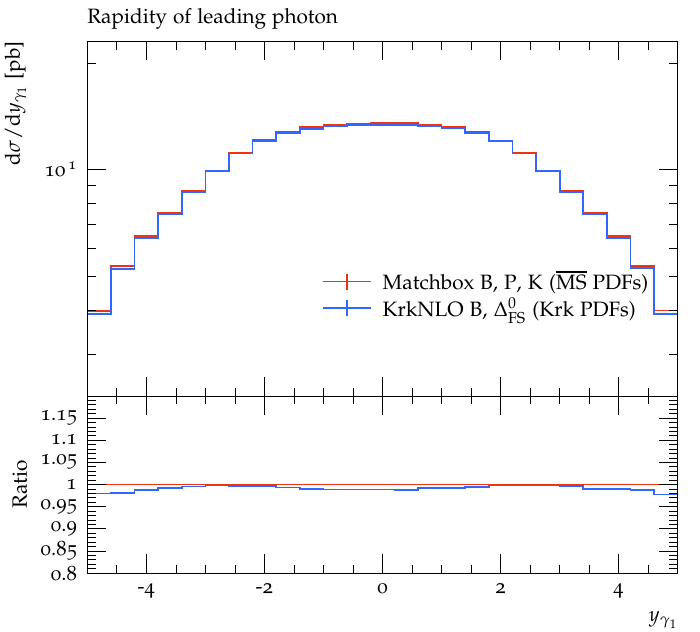}
 }
	\caption{Validation of the \krk PDF scheme transformation and associated 
		$\Delta_{0}^{\rFS}$ correction as described in \cref{sec:imp_validation_FS}.
		Plots shown for $pp \to \gamma \gamma$ at $\sqrt{s} = 8 \, \TeV$,
		for running $\alphas$ with $\alphas(M_Z)=0.118$
		and the \texttt{CT18NLO} PDF set.
		\label{fig:validation_PK}}
\end{figure}

\subsection*{Krk factorisation scheme}
\label{sec:appendix_validation_krkfs}

The PDF scheme transformation outlined in \cref{sec:nlomatching_krkfs,sec:appendix_krkpdfs}
and its compensating terms outlined in \cref{sec:nlomatching_krknlo,sec:appendix_cssummary}
ensure perturbative accuracy to NLO.  Numerically, however, the reorganisation of the
convolution between the hard process and the PDF is not guaranteed to give
a numerically-identical result.

We test this explicitly in \cref{fig:validation_PK} where we compare the \matchbox implementation
of the Catani--Seymour $\rP$ and $\rK$ collinear counterterms to their \krknlo counterparts,
where the convolution is performed upon the PDFs to transform them into the \krk scheme
and partially compensated by the $\Delta_{0}^{\krk}$ reweighting of the hard-process
(concretely, the left- and right-hand sides of \cref{eq:krkconvvalidation_krk}).

This is shown in \cref{fig:validation_PK} and illustrates that the higher-order terms
introduced to this contribution by the reorganisation of the convolution are
approximately 5--10\% of the total, varying primarily according to the rapidity-separation of the photon pair.
This difference should be regarded as a contribution to the overall `missing-higher-order'
perturbative uncertainty.

\bibliographystyle{JHEP}
\bibliography{main.bib}

\end{document}